\documentclass[12pt,reprint,floatfix,onecolumn,amssymb, secnumarabic, amsmath, tightenlines, nofootinbib, nobibnotes, aps, rmp]{revtex4}

\usepackage{fancyhdr}
\usepackage{graphicx}
\usepackage{dcolumn}
\usepackage{bm}

\begin{document}


\title{Canonical Models of Dielectric Response}

\thanks{Copyright (2018) D. Mark Riffe.  ALL RIGHTS RESERVED.}

\author{D.~Mark Riffe}
\homepage{http://www.physics.usu.edu/riffe/bio/index.htm}
\email{mark.riffe@usu.edu.}

\affiliation{Department of Physics, Utah State University, Logan, UT 84322-4415}


\date{\today}

\begin{abstract}
The interaction of electromagnetic fields with a solid is characterized by several interconnected response functions:  the dielectric function $\varepsilon(\omega)$, index of refraction $N(\omega)$, conductivity $\sigma(\omega)$, and optical impedance $Z(\omega)$.  Here we utilize three canonical models of dielectric response  -- the damped harmonic oscillator, Debye polarization response, and the Drude model -- to discuss these four optical response functions.  Special emphasis is devoted to the response of a Drude metal.  Our main focus is on electromagnetic wave propagation through a material.  We also discuss the relaxation of charge fluctuations within the context of the three canonical models of response.     
\end{abstract}



\maketitle

\newpage

\lhead{\small{D. M. Riffe}}
\rhead{\small{Canonical Models of Dielectric Response}}
\cfoot{\thepage}

\tableofcontents

\newpage

\section{Introduction}

What happens when sunlight strikes a window pane?  We all know that some of the light is transmitted and some is reflected.  Some of us may also be aware that a tiny fraction is absorbed within the pane.  Why do these things happen?  The answer to this question is not so straightforward.  Indeed, this question is the essential motivation for this monograph.

In general terms, the answer goes something like this.  The electric and magnetic fields associated with the sunlight interact with the charge in the glass, causing the charge to accelerate.  Of course, accelerating charge produces its own light.  So in addition to the sunlight, the light produced by each bit of accelerating charge also drives all of the other bits of charge.  The sunlight thus induces a rather intricate dance of charges and fields within the glass.  The impinging sunlight and the local-charge-produced light then interfere with each other to produce the total reflected and transmitted fields, which we observe as the transmitted and reflected sunlight.  

So why does a piece of metal react to sunlight rather differently than does a piece of glass?  It cannot be due to differences in the fundamental charge carriers, as these are identical for each material.   It could potentially be due to the density of charge carriers, and this does have some effect, but it is relatively minor.  Rather, the differences in response between a piece of glass and a piece of metal are largely attributable to the differences in internal forces that also act on the charge carriers while the sunlight is shining on the material.

Let's say that we wish to quantitatively determine the transmitted and reflected fields (as well as the fields inside the material) when sunlight hits the surface.  How do we go about figuring this out?  There are basically two approaches that we can use.  The first is to explicitly consider the interaction of an infinitesimal bit of charge with the sunlight and light generated by other bits of charge to develop integral equations for the total electric and magnetic fields at each point in space.  This approach has the advantage mimicking the above qualitative description of the physics.  However, it is somewhat cumbersome in that one must necessarily set up and then solve an integral equation to find the fields.  The more canonical method is to start with the differential form of Maxwell's equations and proceed from there.  In many cases one need only solve a (relatively well known) differential equation to find the fields.  This will be our approach.

No matter which approach one takes, one must deal with the motion of the charge in the solid under the influence of (i) the macroscopic electric and magnetic fields described by Maxwell's equations and (ii) the microscopic, internal forces inherent in the material of interest.  That is, one must develop (and then solve) equations of motion (either classical or quantum) that describe the response of the charge when the system is driven by external fields (such as those contained in sunlight).  In general, these solutions lead to the determination of material response functions, which characterize the macroscopic response of the material to electric and magnetic fields.  

Perhaps the most important of these response functions is the frequency-dependent dielectric function $\varepsilon(\omega)$, which relates the polarization of the charge to the electric field.  We shall spend significant time discussing various classical equations of motion for the charge and the consequent response $\varepsilon(\omega)$.  We also look at several other important response functions, which include the complex index of refraction $N(\omega)$, optical impedance $Z(\omega)$, and conductivity $\sigma(\omega)$.  Although in many cases the four response functions have simple relationships among them, they are all important because each one tells us something different about the interactions of the EM fields with the solid.  The index of refraction provides information on the spatial nature of the fields, the optical impedance relates the magnetic and electric fields in the material to each other, and the conductivity connects the current density to the electric field.

\section{Basic Field Equations}

In their fundamental form (written in terms of the ${\rm {\bf E}}$ and ${\rm {\bf
B}}$ fields only) Maxwell's equations for the averaged electromagnetic fields in matter can be expressed as
\begin{equation}
\label{eq1}
\nabla \cdot {\rm {\bf E}} = \frac{\rho}{\varepsilon _0 },
\end{equation}

\begin{equation}
\label{eq2}
\nabla \cdot {\rm {\bf B}} = 0,
\end{equation}

\begin{equation}
\label{eq3}
\nabla \times {\rm {\bf E}} = - \frac{\partial {\rm {\bf B}}}{\partial t},
\end{equation}

\noindent and
\begin{equation}
\label{eq4}
\nabla \times {\rm {\bf B}} = \mu _0 \, {\rm {\bf j}} + \mu _0 \varepsilon _0 \frac{\partial {\rm {\bf E}}}{\partial t}.
\end{equation}

\noindent Respectively, Eqs.~(\ref{eq1}), (\ref{eq3}), and (\ref{eq4}) are known as Gauss' law, Faraday's law, and the Amp\`{e}re/Maxwell (A/M) law.  Equation (\ref{eq2}) has no special name attached to it.  Written this way the \textbf{charge and current densities} $\rho({\rm \bf r})$ and ${\rm \bf j}({\rm \bf r})$ that appear in Gauss' and the A/M law are due to \textit{all} of the charge that may be present.

While Eqs.~(\ref{eq1}) -- (\ref{eq4}) are the most fundamental form of the electromagnetic field equations in matter, they are usually not particularly useful.  This is because one cannot independently specify $\rho({\rm \bf r})$ and ${\rm \bf j}({\rm \bf r})$.  Indeed, it doesn't take much thought to realize that these two densities are influenced by the fields ${\rm \bf E}$ and ${\rm \bf B}$.  

Four other equations are worth noting at this point.  First, charge conservation,  
\begin{equation}
\label{eq5}
\frac{\partial \rho }{\partial t} + \nabla \cdot {\rm {\bf j}} = 0,
\end{equation}

\noindent is implied by Maxwell's equations.\marginpar{\footnotesize{$\mathbb{E  X} \,$\ref{E1}}} Second, the (Lorentz) force on a particle of charge $q$ and velocity \textbf{v} is given by
\begin{equation}
\label{eq6}
{\rm {\bf F}} = q \, ( {\rm \bf E} + {\rm \bf v} \times {\rm \bf B} ).
\end{equation}

\noindent And lastly, the two fundamental constants of electromagnetism [the free space (or vacuum) \textbf{permeability} $\mu_0 = 4\pi \times 10^{-7}\,$T$\,$m/A and \textbf{permittivity} $\varepsilon_0 = 8.8542 \times 10^{-12}$ C$^2$/(N$\,$m$^2$)] can be used to define a two other constants,
\begin{equation}
\label{eq7}
c = \frac{1}{\sqrt{\mu_0 \varepsilon_0}} = 2.9979 \times 10^8 \, \textrm{m/s},
\end{equation}

\noindent the \textbf{speed of light} propagating in a free space, and
\begin{equation}
\label{eq7a}
Z_0 = \sqrt{\frac{\mu_0}{ \varepsilon_0}} = 376.73 \, \Omega,
\end{equation}  

\noindent which is known as the \textbf{impedance of free space}.

\section{Canonical Forms of Gauss and Amp\`{e}re/Maxwell}

\subsection{Charge and Current Densities}

Gauss' law and the A/M law are commonly written in forms where contributions to the charge and current densities have been divided into several components.  The assignment of particular types of charge to particular components is (at a fundamental level) completely arbitrary. There are, however, several conventions that are typically used in certain situations.

At the introductory-physics level the charge density is often separated into charge that is bound to atoms or molecules in the solid and charge that is free to move throughout the solid (such as the charge associated with
conduction electrons). In this case the total charge density is divided as $\rho = \rho _b + \rho _f$.

For a dc electric field this division is quite useful. The free-charge density $\rho _f $ is associated with any current density \textbf{j} while the bound-charge density $\rho _b $ is associated with any polarization \textbf{P} of the material.  However, when one starts to think about an applied ac field (such as that due to an electromagnetic wave), the division of the charge density into bound and free components becomes less distinct because both types of charge contribute to the current density in the material.

At a more sophisticated level, then, the bound and free charge are not formally differentiated (although in any specific case one is generally cognizant of the existence of both types of charge).  In this case the convention is to lump the free and bound charge together in a single charge density $\rho_{\rm p}$ that is used to define the polarization \textbf{P} of the material.  If there is no other charge around, then this is the total charge density. That is, $\rho = \rho_{\rm p}$.  Charge conservation (expressed as the continuity equation) can then be used to find the current density \textbf{j}.

Sometimes, though, there is some other charge density of interest, which we denote as $\rho _{ext} $, where $ext$ represents external (although $\rho _{ext} $ may reside inside the material of interest).  This charge density is associated with charge that we specify in some manner, \textit{independent} of the ${\rm \bf E}$ and ${\rm \bf B}$ fields.  In this case one typically writes $\rho = \rho_{\rm p} + \rho _{ext} $, where, again, $\rho _{\rm p}$ includes both the bound and free charge of the material.

All three of these situations can be subsumed under the following general scheme. We divide the charge density into $\rho _{\rm p}$ and any other charge,
\begin{equation}
\label{eq8}
\rho = \rho _{\rm p} + \rho _{other},
\end{equation}

\noindent keeping in mind that the free charge $\rho_f$ may be associated with either $\rho _{\rm p} $ or $\rho _{other} $, depending upon the situation. A consequence of this is that the meanings of the polarization field \textbf{P} (which, again, we always associate with $\rho _{\rm p})$ and (consequently) the displacement field \textbf{D} (see below) depend upon the assignment of the free charge density.

At times it is desirable to divide the charge unconventionally.  For example, as we see below when discussing the relaxation of charge density fluctuations, it can be convenient to divide the bound charge itself so as to include only part in $\rho_{\rm p}$, with the remainder in $\rho_{other}$.  The major lesson to be taken away from this discussion is that a number of choices exist for describing the charge density, and the choice that is made depends upon the problem at hand.  \textit{Caveat emptor}!

Once the charge-density assignments are chosen, the current-density division is straightforward. It is divided into \textit{three} components, 
\begin{equation}
\label{eq8a}
{\rm {\bf j}} = {\rm {\bf j}}_{\rm p} + {\rm {\bf j}}_{other} + {\rm {\bf j}}_M.
\end{equation}

\noindent The first two current densities are associated with $\rho _{\rm p}$ and $\rho _{other} $, respectively, while the third component is related to the \textbf{magnetization} ${\rm {\bf M}}$ of the material
\begin{equation}
\label{eq9}
{\rm {\bf j}}_M = \nabla \times {\rm {\bf M}}.
\end{equation}

\noindent The magnetization \textbf{M} is produced by intrinsic spin and/or motional degrees of freedom associated with electrons in the solid.  In the materials that we shall discuss here the average values of the spin and orbital angular momentum are close enough to zero that \textbf{M} can be neglected.  We do note that Eq.~(\ref{eq9}) implies that any magnetization ${\rm {\bf M}}$ (even if it is only from intrinsic spin) has an associated current density ${\bf j}_M$.

\subsection{Gauss' Law}
 
To convert Gauss' law into canonical form we start by using the densities $\rho _{\rm p}$ and ${\rm{\bf j}}_{\rm p}$ to define the \textbf{polarization field} ${\rm {\bf P}}$ via 
\begin{equation}
\label{eq10}
\rho _{\rm p} = - \nabla \cdot {\rm {\bf P}}
\end{equation}

\noindent and
\begin{equation}
\label{eq11}
{\rm {\bf j}}_{\rm p} = \frac{\partial {\rm {\bf P}}}{\partial t}.
\end{equation}

\noindent (Notice that this definition of ${\rm {\bf P}}$ is explicitly consistent with $\rho _{\rm p}$ and ${\rm{\bf j}}_{\rm p}$ satisfying the continuity equation.) We now use Eq.~(\ref{eq10}) to replace $\rho_{\rm p}$ in Eq.~(\ref{eq8}) and then use that result in Eq.~(\ref{eq1}), which transforms Gauss' law into 
\begin{equation}
\label{eq12}
\nabla \cdot ( {\varepsilon _0 } \, {\rm {\bf E} +{\rm {\bf P}} }) = \rho_{other}.
\end{equation}

\noindent This last equation naturally leads to the definition of the \textbf{displacement} field
\begin{equation}
\label{eq13}
{\rm {\bf D}} = \varepsilon _0 \,{\rm {\bf E}} + {\rm {\bf P}},
\end{equation}

\noindent and the compact, canonical form of Gauss' law,
\begin{equation}
\label{eq14}
\nabla \cdot {\rm {\bf D}} = \rho _{other}.
\end{equation}

\noindent There is nothing new in this form of Gauss' law that was not in the original. All that has happened is that $\rho _{\rm p}$ has been tucked away in the displacement ${\rm {\bf D}}$.

In order to solve a typical problem involving Gauss' law the response of the charge density to the electric field ${\rm \bf E}$ must (at some level) be known.  In principle, this response can be obtained by solving an appropriate equation of motion for the charge density [where $q{\rm \bf E}$ is one of (typically) several forces acting on the charges].  Several examples of this approach are worked out below.  However, sometimes it suffices to simply assume some general property about the charge-density response.

The simplest assumption is to let the polarization field ${\rm {\bf P}}$ [which is related to the charge via Eqs.~(\ref{eq10}) and (\ref{eq11})] be proportional to the electric field ${\rm {\bf E}}$,
\begin{equation}
\label{eq15}
{\rm {\bf P}} = \varepsilon _0 \chi _e {\rm {\bf E}}.
\end{equation}

\noindent Such an assumption is generically referred to as \textbf{simple linear response}.  In Eq.~(\ref{eq15}) the quantity $\chi _e $ is known as the \textbf{electric susceptibility} of the material.  Under this linear-response ansatz we can combine Eqs.~(\ref{eq13}) and (\ref{eq15}) to see that the displacement vector is also proportional to the electric field,
\begin{equation}
\label{eq16}
{\rm {\bf D}} = \varepsilon _0 \,\left( {1 + \chi _e } \right)\,{\rm {\bf E}}.
\end{equation}

\noindent This last equation leads to the definition of another quantity, the (electric) \textbf{dielectric constant} of the material,
\begin{equation}
\label{eq17}
\varepsilon =  1 + \chi _e, 
\end{equation}

\noindent and the rewriting of Eq.~(\ref{eq16}) as
\begin{equation}
\label{eq18}
{\rm {\bf D}} = \varepsilon_0  \varepsilon \, {\rm {\bf E}}. 
\end{equation}

\noindent  The combination $\varepsilon_0  \varepsilon$ is known as the (electric) \textbf{permittivity} of the material.  Because $\varepsilon$ is the ratio of the permittivity of the material to the permittivity of free space, it is sometime referred to as the \textbf{relative permittivity} of the material.\footnote{To make matters worse, a modern SI convention is to use the symbol $\varepsilon$ to represent the permittivity $\varepsilon_0 \, \varepsilon$ [rather than the dielectric constant (or relative permittivity)].  However, due to the rather universal (and historic) use of $\varepsilon$ to represent the dielectric constant, we shall stick with this convention.}

\subsection{Amp\`{e}re/Maxwell Law}

To convert the A/M law into canonical form we first combine Eqs.~(\ref{eq8a}), (\ref{eq9}), and (\ref{eq11}) as
\begin{equation}
\label{eq21}
{\rm {\bf j}} = {\rm {\bf j}}_{other} + \frac{\partial {\rm {\bf P}}}{\partial t} + \nabla \times {\rm {\bf M}}
\end{equation}

\noindent
and then substitute this expression for ${\rm {\bf j}}$ into the A/M law [Eq.~(\ref{eq4})], which yields
\begin{equation}
\label{eq22}
\nabla \times \left( {\frac{{\rm {\bf B}}}{\mu _0 } - {\rm {\bf M}}} \right)
= {\rm {\bf j}}_{other} + \frac{\partial }{\partial t}\left( {\varepsilon _0
{\rm {\bf E}} + {\rm {\bf P}}} \right).
\end{equation}

\noindent This equation motivates the definition of the vector field
\begin{equation}
\label{eq23}
{\rm {\bf H}} = \frac{{\rm {\bf B}}}{\mu _0 } - {\rm {\bf M}},
\end{equation}

\noindent which, like ${\rm {\bf B}}$, is also often called the magnetic field.
However, it is probably best to simply refer to this field as the ${\rm {\bf H}}$ field.  With this definition the A/M law becomes
\begin{equation}
\label{eq24}
\nabla \times {\rm {\bf H}} = {\rm {\bf j}}_{other} + \frac{\partial {\rm {\bf D}}}{\partial t}.
\end{equation}

\noindent As in the case of the canonical form of Gauss' law, there is nothing new in this form of the A/M law. All that has happened is that the current densities ${\rm {\bf j}}_{\rm p}$ and ${\rm {\bf j}}_M $ have been hidden away in
the vectors ${\rm {\bf D}}$ and ${\rm {\bf H}}$, respectively.

Similar to the case of Gauss' law above, in order to solve a problem involving the A/M law, one must be able to connect the response of the system (in this case the current density ${\rm \bf j}_M$ hidden in the magnetization ${\rm \bf M}$) to the magnetic field ${\rm \bf B}$.  In this case simple linear response is traditionally introduced via  
\begin{equation}
\label{eq25}
{\rm {\bf M}} = \chi _m {\rm {\bf H}},
\end{equation}

\noindent where $\chi _m$ is known as the \textbf{magnetic susceptibility} of the material.\footnote{The astute reader will notice that the parallel with Gauss' law and the electric field is not quite exact.  If it were, the magnetic susceptibility $\chi _m$ would be defined in terms of the ${\rm \bf B}$ field rather than the ${\rm \bf H}$ field.}  With this linear-response assumption Eq.~(\ref{eq23}) for the ${\rm \bf H}$ field becomes
\begin{equation}
\label{eq26}
{\rm {\bf H}} = \frac{{\rm {\bf B}}}{\mu _0 \left( {1 + \chi _m} \right)},
\end{equation}

\noindent which leads to the definition of the \textbf{relative permeability} $\mu$ of the material
\begin{equation}
\label{eq27}
\mu =  1 + \chi _m.
\end{equation}

\noindent  With this definition Eq.~(\ref{eq26}) can be simply expressed as
\begin{equation}
\label{eq28}
{\rm {\bf H}} = \frac{{\rm {\bf B}}}{\mu_0  \mu},
\end{equation}

\noindent Analogous to the electric-field case, the combination $\mu_0 \, \mu$ is known as the (magnetic) \textbf{permeability} of the material.  For nonmagnetic materials that are our main interest here, the magnetic susceptibility $\chi_M$ is practically zero, making $\mu$ = 1.\footnote{Note that this is consistent with our earlier discussion where we pointed out that \textbf{M} and (thus) \textbf{j}$_M$ are zero for most materials.}  Thus, in these circumstances one simply has
\begin{equation}
\label{eq30}
{\rm {\bf H}} = \frac{{\rm {\bf B}}}{\mu _0 }.
\end{equation}

\subsection{All Together Now}

We now summarize our results for the canonical form of Maxwell's equations in matter.  With no assumptions regarding the response of the charge to the fields, the equations can be written as \marginpar{\footnotesize{$\mathbb{E  X} \,$\ref{E1b}}}
\begin{equation}
\label{eq31}
\nabla \cdot {\rm {\bf D}} = \rho _{other}.
\end{equation}

\begin{equation}
\label{eq32}
\nabla \cdot {\rm {\bf B}} = 0,
\end{equation}

\begin{equation}
\label{eq33}
\nabla \times {\rm {\bf E}} = - \frac{\partial {\rm {\bf B}}}{\partial t},
\end{equation}

\noindent and
\begin{equation}
\label{eq34}
\nabla \times {\rm {\bf H}} = {\rm {\bf j}}_{other} + \frac{\partial {\rm {\bf D}}}{\partial t},
\end{equation}

\noindent where \textbf{D} and \textbf{H} are given by Eqs.~(\ref{eq13}) and (\ref{eq23}).

When simple linear response (${\rm \bf D}$ = $\varepsilon_0  \varepsilon \, {\rm \bf E}$, ${\rm \bf B}$ = $\mu_0  \mu \, {\rm \bf H}$, and $\varepsilon$ and $\mu$ both constants) is assumed, these equations can be re-expressed in terms of the fundamental fields ${\rm {\bf E}}$ and ${\rm {\bf B}}$  as

\begin{equation}
\label{eq35}
\nabla \cdot {\rm \bf E} = \frac{\rho _{other}}{\varepsilon_0 \varepsilon}.
\end{equation}

\begin{equation}
\label{eq36}
\nabla \cdot {\rm {\bf B}} = 0,
\end{equation}

\begin{equation}
\label{eq37}
\nabla \times {\rm {\bf E}} = - \frac{\partial {\rm {\bf B}}}{\partial t},
\end{equation}

\noindent and
\begin{equation}
\label{eq38}
\nabla \times {\rm \bf B} = \mu_0 \mu \, {\rm {\bf j}}_{other} + \mu_0 \mu \, \varepsilon_0 \varepsilon \, \frac{\partial {\rm \bf E} }{\partial t}.
\end{equation}

\noindent Notice that these equations are identical to their original fundamental form [Eqs.~(\ref{eq1})--(\ref{eq4})] but with the total charge replaced by \textit{other} charge and the constants $\varepsilon_0$ and $\mu_0$ replaced by $\varepsilon_0 \varepsilon$ and $\mu_0 \mu$, respectively.  The beauty of expressing Maxwell's equations in this form is that when $\rho_{other}$ and ${\rm \bf j}_{other}$ are zero, the equations are \textit{homogeneous} in the fields \textbf{E} and \textbf{B}.

\section{Electromagnetic Waves}

\subsection{Constant $\varepsilon$ and $\mu$}

One of the coolest aspects of Maxwell's equations is that they admit traveling wave solutions for the electric and magnetic fields. These electromagnetic (EM) waves can travel through vacuum, and when the waves encounter a material they can continue to propagate through the material, but their propagation is modified by the presence of the charge within material. In the next several sections we discuss this propagation.  As we shall see as these sections are developed, the relative permittivity $\varepsilon$ (which will later become a function of frequency) is the key element in understanding EM waves in most materials.  The relative permeability can also be key, but because typically $\mu = 1$, its effects are usually not as interesting as those due to $\varepsilon$.

\subsubsection{Wave equation}

In order to derive this wave behavior of the fields, it is useful to derive uncoupled equations for the propagating electric and magnetic fields.  We start by assuming that (i) all free and bound charge in the material is described by $\rho _{\rm p}$  and ${\rm \bf j}_{\rm p}$, and (ii) there is no external charge (so that $\rho_{other} = 0$ and ${\rm \bf j}_{other}=0$).  Here we also assume simple linear response. That is, the solid is described by constants $\varepsilon$  and $\mu$.  With these assumptions the A/M law [Eq.~(\ref{eq38})] simplifies to

\begin{equation}
\label{eq40}
\nabla \times {\rm {\bf B}} = \mu _0 \mu \varepsilon_0  \varepsilon \frac{\partial {\rm {\bf E}}} {\partial t}.
\end{equation}

\noindent Taking the curl of Eq.~(\ref{eq37}) and using Eq.~(\ref{eq40}) to replace the
curl of ${\rm {\bf B}}$ gives
\begin{equation}
\label{eq41}
\nabla \times \nabla \times {\rm {\bf E}} = - \mu _0 \mu \varepsilon_0  \varepsilon \frac{\partial ^2 {{\rm {\bf E}}}}{\partial t^2}.
\end{equation}

\noindent This equation can be simplified by noting for any vector ${\rm {\bf
a}}$ that $\nabla \times \nabla \times {\rm {\bf a}} = \nabla \left( {\nabla \cdot {\rm {\bf a}}} \right) - \nabla ^2{\rm {\bf a}}$. This allows us to write
\begin{equation}
\label{eq42}
\nabla \left( {\nabla \cdot {\rm {\bf E}}} \right) - \nabla ^2{\rm {\bf E}} = - \mu _0 \mu \varepsilon_0 \varepsilon \frac{\partial ^2 {{\rm {\bf E}}}}{\partial t^2}.
\end{equation}

\noindent We now appeal to Gauss' law,
\begin{equation}
\label{eq43}
\nabla \cdot \left( \varepsilon_0 \varepsilon \,{\rm {\bf E}} \right) = 0,
\end{equation}

\noindent which, because $\varepsilon_0 \varepsilon $ is a nonzero constant, allows us to see that the electric field is described by the standard \textbf{wave equation}
\begin{equation}
\label{eq44}
\nabla ^2{\rm {\bf E}} = \mu _0 \mu \varepsilon_0 \varepsilon \frac{\partial ^2{\rm {\bf E}}}{\partial t^2}.
\end{equation}

\noindent An equation identical to Eq.~(\ref{eq44}) can be derived for the magnetic field.\marginpar{\footnotesize{$\mathbb{EX} \,$\ref{E2}}}

\subsubsection{Index of refraction}

Those familiar with the wave equation will immediately identify the constant $\mu _0 \mu \varepsilon_0 \varepsilon$ in Eq.~(\ref{eq44}) as 1/$v^2$, where $v$ is the \textbf{propagation speed} of waves governed by the wave equation.  And because the wave propagation speed in vacuum is given by 1/$\sqrt{\mu _0 \varepsilon_0}$ = $c$,\footnote{This result follows from Eq.~(\ref{eq44}) with $\mu = \varepsilon$ = 1.}  we have the result
\begin{equation}
\label{eq45}
v = \frac{c}{\sqrt{\mu \varepsilon}}.
\end{equation} 

When assuming simple linear response, one is usually interested in the case where $\mu = 1$ and $\varepsilon >1$.  Then $\sqrt{\mu \varepsilon} > 1$, making the wave speed $v$ less than the speed of light $c$.  Furthermore, insofar as the \textbf{index of refraction} $n$ of a material is defined as the ratio $c$/$v$, we also have
\begin{equation}
\label{eq46}
n = \sqrt{\mu \varepsilon}.
\end{equation}

For those less familiar with the properties of the wave equation we now review some of the relevant ideas associated with the propagation of waves described by Eq.~(\ref{eq44}).  In particular, we consider the \textbf{plane-wave} solution
\begin{equation}
\label{eq47}
{\rm {\bf E}}( {\rm {\bf r}},t ) = {\rm {\bf E}}_0 \,e^{i\left( {{\rm {\bf k}} \cdot {\rm {\bf r}} - \omega \,t} \right)}.
\end{equation}

\noindent Here ${\rm {\bf k}}$ is the \textbf{wave vector}, and $\omega $ is the \textbf{angular frequency}. This solution is a traveling wave that propagates in the direction of the wave vector ${\rm {\bf k}}$.  Although the electric field \textbf{amplitude} \textbf{E}$_0$ is not constrained by the wave equation, Gauss' law [Eq.~(\ref{eq43})] requires \textbf{E}$_0$ and \textbf{k} to be orthogonal.\marginpar{\footnotesize{$\mathbb{EX} \,$\ref{E3}}}  That is, the electric field is a \textbf{transverse} field.  Note that the \textbf{wavelength} $\lambda $ and \textbf{period} $T$ of the wave are related to the wave vector and angular frequency via $k = 2\pi / \lambda$ and $\omega = 2\pi / T$.$\,$\footnote{We use the convention that the magnitude of a vector quantity (such as ${\rm {\bf k}})$ is represented by the same symbol in italics ($k)$.}

If we substitute the plane-wave form of the electric field [Eq.~(\ref{eq47})] into the wave equation, we obtain the condition that relates the frequency to the wave vector,
\begin{equation}
\label{eq49}
\omega (k) = \frac{c}{\sqrt{\mu \varepsilon}} \, k = \frac{c}{n} k.
\end{equation}

\noindent Any equation that relates $\omega$ to ${\rm \bf k}$ is known as a \textbf{dispersion relation}. In general, the \textbf{phase velocity} and \textbf{group velocity} are respectively obtained from the dispersion relation via $v_p = \omega / k$ and $v_g = d\omega / dk$. Thus, the plane-wave solutions are characterized by
\begin{equation}
\label{eq50}
v_p = v_g = \frac{c}{ \sqrt{\mu \varepsilon} } = \frac{c}{n}.
\end{equation}

\noindent Notice that both of these velocities are the same as the wave speed previously identified by inspection of the wave equation.  Because both $v_p $ and $v_g $ are independent of frequency, (i) all plane-wave solutions propagate at the same speed, and (ii) localized solutions (such as a pulse) also travel at this same speed and do not change their shape as they propagate.

Because the index of refraction modifies the wave speed, the wavelength $\lambda$ of a wave in the material is modified from its vacuum value $\lambda_0$ (for a given frequency $\omega$).  To see this we note that $\omega / c = 2 \pi / \lambda_0$ allows us to re-express Eq.~(\ref{eq49}) as$\,$\footnote{$\omega / c = 2 \pi / \lambda_0$ is the dispersion relation for waves traveling in a vacuum.  If can be deduced from Eq.~(\ref{eq49}) with $n =1$ and the definition $k_0 = 2 \pi / \lambda_0$, where $k_0$ is the vacuum wave vector.}
\begin{equation}
\label{eq50b}
k = \frac{2 \pi}{\lambda_0} \, n.
\end{equation}

\noindent This equation along with $k = 2 \pi / \lambda$ gives us the relation between the wavelength in the material and the vacuum wavelength,
\begin{equation}
\label{eq50c}
\frac{\lambda}{\lambda_0} = \frac{1}{n}.
\end{equation}

\noindent  We further note that Eq.~(\ref{eq50b}) allows us to write Eq.~(\ref{eq47}) in terms of $\lambda_0$ and $n$ as
\begin{equation}
\label{eq50d}
{\rm {\bf E}}( \textbf{r},t ) = {\rm {\bf E}}_0 \, e^{i\left( 2 \pi / \lambda_0  \right) n \, \hat{\textbf{k}} \cdot \textbf{r}} e^{-i \omega  t },
\end{equation}

\noindent where $\hat{\textbf{k}} = \textbf{k} / k$ is the unit vector that points in the direction of \textbf{k}.

\subsubsection{Optical impedance}

In any EM wave there is not only a propagating electric field, but also an accompanying magnetic field;\footnote{Hence the term -- \textit{electromagnetic} wave.} in this section we consider this magnetic field and its relationship to the electric field.  Following tradition for describing magnetic fields in solids, we use the \textbf{H} field rather than the \textbf{B} field, although for simple linear response that is our current assumption, these fields are related via ${\bf B} = \mu_0 \mu {\bf H}$.  Starting with the electric field given by Eq.~(\ref{eq47}) and assuming that the accompanying \textbf{H} field is also a plane wave, it is not hard to show (using Maxwell's equations) that the \textbf{H} field is given by\marginpar{\footnotesize{$\mathbb{EX} \,$\ref{E4}}}
\begin{equation}
\label{eqZ1}
{\rm {\bf H}}( {\rm {\bf r}},t ) = {\rm {\bf H}}_0 \,e^{i\left( {{\rm {\bf k}} \cdot {\rm {\bf r}} - \omega \,t} \right)},
\end{equation}

\noindent where the \textbf{H}-field and \textbf{E}-field amplitudes are related via
\begin{equation}
\label{eqZ2}
Z \, \textbf{H}_0 = \hat{\textbf{k}} \times \textbf{E}_0.
\end{equation} 

\noindent The constant 
\begin{equation}
\label{eqZ3}
Z = \sqrt{\frac{\mu_0 \mu}{\varepsilon_0 \varepsilon}}
\end{equation} 

\noindent is known as the (optical) \textbf{impedance of the material}.\footnote{Recall that $Z_0 = \sqrt{\mu_0 / \varepsilon_0}$ is the impedance of free space.}  Equation (\ref{eqZ2}) tells us that \textbf{H} is orthogonal to both \textbf{k} and \textbf{E}, and from Gauss' law we already have that \textbf{k} and \textbf{E} are orthogonal; thus \textbf{E}, \textbf{H}, and \textbf{k} form an orthogonal set of vectors.  Furthermore, Eq.~(\ref{eqZ2}) tells us (for $Z > 0$) that  \textbf{E} $\times$ \textbf{H} points in the direction of \textbf{k}.  We can thus write
\begin{equation}
\label{eqZ4}
Z = \frac{E_0}{H_0}.
\end{equation} 

\noindent  The ratio $E_0 / H_0$ is known as the \textbf{wave impedance}; for the situation at hand it is obviously equal to the impedance of the material.  

 It is sometimes convenient to work with the \textbf{normalized impedance} of the material $\zeta = Z / Z_0$.  It should be obvious that
 \begin{equation}
\label{eqZ5}
\zeta = \sqrt{\frac{\mu}{\varepsilon}}.
\end{equation}

\noindent  We note that $\zeta \, n = \mu$ and so for nonmagnetic ($\mu = 1$) materials  $\zeta = 1 / n$.

\subsection{Frequency Dependent $\varepsilon$}

This result that all EM waves propagate at the same speed in a given material is an oversimplification. We know this because different frequencies of light have different angles of refraction in a material such as glass. Hence, the wave equation cannot truly describe EM wave propagation in a material.  If you carefully follow the reasoning in the last section you will discover that the wave equation is the result of having made the simple linear-response approximations ${\rm {\bf D}} = \varepsilon_0 \varepsilon \,{\rm {\bf E}}$ and ${\rm {\bf B}} = \mu_0 \mu \,{\rm {\bf H}}$.  In principle we might have to abandon both of these approximations.  However, for nonmagnetic materials is it sufficient to only abandon the simple-response approximation that connects \textbf{D} to \textbf{E}.

Cognizant of the fact that waves with different frequencies propagate with different speeds, we instead start with the assumption that all fields of interest oscillate harmonically in time, and so we write
\begin{equation}
\label{eq51}
{\rm {\bf F}}( {{\rm {\bf r}},t}) = {\rm {\bf \tilde {F}}}( {\rm {\bf r}} )\,e^{ - i\omega \,t},
\end{equation}

\noindent where \textbf{F} represents any of the fields \textbf{E}, \textbf{D}, \textbf{B}, or \textbf{H} (and consequently also \textbf{P} and \textbf{M}).  With this assumption the four equations of Maxwell can be expressed as \marginpar{\footnotesize{$\mathbb{EX} \,$\ref{E4b}}}
\begin{equation}
\label{eqM1}
\nabla \cdot {\rm {\bf \tilde D}} = 0,
\end{equation}

\begin{equation}
\label{eqM2}
\nabla \cdot {\rm {\bf \tilde B}} = 0,
\end{equation}

\begin{equation}
\label{eqM3}
\nabla \times {\rm {\bf \tilde E}} = i \omega {\rm {\bf \tilde B}},
\end{equation}

\noindent and
\begin{equation}
\label{eqM4}
\nabla \times {\rm {\bf \tilde H}} =   - i \omega {\rm {\bf \tilde D}}.
\end{equation}

\noindent  We continue to assume the homogeneous conditions $\rho_{other} = 0$ and ${\rm \bf j}_{other}=0$.  We refer to these last four equations as the \textbf{harmonic Maxwell's equations}.

We now derive a single equation for the electric field.  To do this we introduce a slightly more sophisticated form of linear response: we assume that the spatial parts of the ${\rm {\bf D}}$ and ${\rm {\bf E}}$ fields are linearly (and spatially locally) related, but that the relationship is frequency dependent,
\begin{equation}
\label{eq52}
{\rm {\bf \tilde {D}}}( {\rm {\bf r}}) = \varepsilon_0  \varepsilon ( \omega )\,{\rm {\bf \tilde {E}}}( {\rm {\bf r}}).
\end{equation}

\noindent We maintain the simple relationship between \textbf{B} and \textbf{H},
\begin{equation}
\label{eq53}
{\rm {\bf \tilde {B}}}( {\rm {\bf r}}) = \mu_0  \mu \,{\rm {\bf \tilde {H}}}( {\rm {\bf r}}).
\end{equation}

\noindent The term $\varepsilon ( \omega)$ is known as the (frequency-dependent) \textbf{dielectric function} of the material.\footnote{In general $\varepsilon$ and $\mu$ are functions of both ${\rm {\bf k}}$ and $\omega $. The dielectric function we are now working with is the ${\rm {\bf k}} \to 0$ limit of the more general function. That is, $\varepsilon \left( \omega \right) = \varepsilon \left( {{\rm {\bf k}} \to 0,\,\omega } \right)$.  Likewise, the permeability that we are working with is the ${\rm {\bf k}} \to 0$ and $\omega \to 0$ limit of the more general function.  That is, $\mu = \mu \left( {{\rm {\bf k}} \to 0,\,\omega \to 0 } \right)$.} With this new linear-response assumption Eqs.~(\ref{eqM3}) and (\ref{eqM4}) can be combined into one equation for the electric field,
\begin{equation}
\label{eq54}
\nabla ( {\nabla \cdot {\rm {\bf \tilde {E}}}} ) - \nabla ^2{\rm {\bf \tilde {E}}} = \mu _0 \mu \varepsilon_0 \varepsilon (\omega)\,\omega^2 \, {\rm {\bf \tilde {E}}}.
\end{equation}

\noindent There is one more bit of simplification.  In conjunction with Eq.~(\ref{eq52}), Gauss' law [Eq.~(\ref{eqM1})] tells us  $\varepsilon_0 \varepsilon(\omega) \nabla \cdot {\rm \bf \tilde E} = 0$.  For any actual material $\varepsilon(\omega) \ne 0$,\footnote{$\varepsilon (\omega) \ne 0$ is true for any real $\omega$, which is our interest at present.  In general, $\varepsilon (\omega)$ may have one or more complex roots, which can correspond to damped oscillations.} and so $\nabla \cdot {\rm \bf \tilde E} = 0$.  Thus, our equation for the electric field becomes
\begin{equation}
\label{eq55}
\nabla ^2{\rm {\bf \tilde {E}}} = - \mu _0 \mu \varepsilon_0 \varepsilon (\omega)\,\omega ^2 \, {\rm {\bf \tilde {E}}}.
\end{equation}

\noindent This is known as the \textbf{Helmholtz equation}.  Perhaps not surprisingly, an identical equation for the \textbf{H} field can be derived from the harmonic Maxwell's equations.

If we are interested in plane-wave solutions to Eq.~(\ref{eq55}), then
\begin{equation}
\label{eq56}
\tilde{\textbf{E}}(\textbf{r}) = {\rm {\bf E}}_0 \,e^{i \textbf{k} \cdot \textbf{r}}.
\end{equation}

\noindent Substituting this into Eq.~(\ref{eq55}) gives us the dispersion relation\footnote{As should be obvious, this is an implicit equation for $\omega (k)$.  In general, one must first find $\varepsilon (\omega)$, substitute it into Eq.~(\ref{eq57}), and then solve for $\omega (k)$.  On the other hand, as long as $\varepsilon$ has no $k$ dependence (as is assumed here), then solving for $k$ in terms of $\omega$ is clearly straightforward.  Thus, $k(\omega)$ is the more natural function.  However, it is more common to plot $\omega(k)$ vs $k$ when graphing a dispersion relation. Go figure.}
\begin{equation}
\label{eq57}
\omega (k) = \frac{c \, k}{\sqrt{\mu \, \varepsilon(\omega)}}.
\end{equation}

\noindent Because $\varepsilon ( \omega )$ is generally not constant, a key result of this more general dispersion relation is that the phase and group velocities are frequency dependent.  It would be difficult to overstate the importance of Eq.~(\ref{eq57}).  Keep in mind that nonmagnetic material are characterized by $\mu =1$.  In much of what follows we implicitly have assumed that $\mu = 1$.

The fun part now comes in figuring out $\varepsilon ( \omega )$ for a given material. In principle one should use quantum mechanics to calculate the dielectric function. However, semiclassical models are often sufficient. The general principle is to find an equation of motion for the charge of interest, and relate that motion to the polarization ${\rm {\bf P}}$, from which naturally arises  $\varepsilon ( \omega )$. We now look at several semiclassical models of various types of dielectric response.

\section{Model Dielectric Functions}

\subsection{Classical Harmonic Oscillator}

Let's think about an ideal crystalline insulator,\footnote{An insulator is sometimes called a dielectric.  As we shall see, metals (which are sometimes called conductors) also have a dielectric function, even though they are not dielectrics.  Physics isn't always so logical, is it?} where there is no free charge to move about the crystal.\footnote{Thus $\rho_{\rm p}$ has contribution solely from bound charge.} The valence electrons are localized on the atoms  and/or in bonds between atoms, while the core electrons are tightly bound to their respective nuclei. The nuclei contain charge that (on average) balances the electronic charge.  For any of this charge a displacement away from equilibrium (typically) results in a linear restoring force back towards equilibrium.  As in any such system of coupled particles, there exists a set of normal modes that describe the fundamental excitations of the system.  

For an insulator there are two distinct bands of frequencies associated with normal mode excitations.  At the lowest frequencies (starting at zero up through perhaps a hundred THz or so) there will be normal modes that primarily correspond to vibrational motion of the nuclei.\footnote{Of course, as the nuclei vibrate the surrounding electronic charge also rearranges itself in response.  Due to the relatively low frequencies involved, the electronic response is usually in phase with the nuclear motion.  Such adiabatic response corresponds to changes in energy levels of the electrons, but no excitation of the electrons from those energy levels.}  The quantized excitations of these coupled (mechanical) oscillations are known as \textbf{phonons}.  At frequencies above the phonon band is a gap devoid of fundamental excitations of the insulator.  This gap typically exists up through the ultraviolet part of the spectrum.  However, at frequencies in the ultraviolet through the x-ray region there is a set of excitations that are primarily electronic in nature.  These excitations correspond to electrons being excited from one electronic band to another electronic band, and are thus known as \textbf{interband transitions}.  Although it may not be not obvious, both vibrational and electronic excitations can often be  modeled effectively by a collection of harmonic-oscillator modes.    

Let's now assume that an EM wave is propagating through the solid.  Whether or not the wave couples to any given normal mode of the system depends upon two conditions.  First, the symmetry of the mode must be such that it can be excited by EM radiation.  To first order this has to do with whether or not the system has (electric) dipole moments that oscillate when the mode is excited. Second, the wave vector \textbf{k} of the EM radiation and the normal mode must match.  Otherwise, coherent excitation of the normal mode cannot take place.  If the frequency of the driving field is close to the natural frequency $\omega_0$ of the mode then we expect a large response from the system; if these two frequencies are disparate, then the system response will relatively small.

We can characterize the response of the system to EM radiation by the microscopic \textbf{dipole moment p} induced in each unit cell of the solid.  For the time being let's imagine that for a given \textbf{k} the system has just one normal mode that couples to the EM fields.  Under this condition the (effective) equation of motion for each moment \textbf{p} is simply that of a driven harmonic oscillator,\footnote{In Eq.~(\ref{eq61}) one should technically think of \textbf{r} as a discrete variable that labels the position of the unit cell of interest.} 
\begin{equation}
\label{eq61}
 \frac{d^2 \mathbf{p(r},t)}{dt^2} + \omega_0^2 \, \mathbf{p(r},t)   = \frac{q^2}{m^*} \mathbf{\tilde E(r)} \,e^{- i  \omega \, t }.
\end{equation}
 
\noindent The normal-mode natural frequency $\omega_0$ is obviously related to the local forces that are trying to restore equilibrium. For simplicity we assume that $\omega_0$ is independent of \textbf{k}.\footnote{In general there will be some dispersion in the system's normal modes.  However, because the slope of $\omega$ vs $k$ (=$c$) is so steep for EM radiation, a flat dispersion for the system's modes is an excellent first approximation.}  The effective mass $m^*$ is related to the masses of the particles involved in the moment \textbf{p} and the relative displacements of these particles.\footnote{For phonon normal modes $m^*$ is determined by the nuclear masses.  For modes that correspond to electronic excitations $m^*$ is close to the electron mass.}  The charge $q$ is the charge associated with the dipole moment \textbf{p}.

We now look for a solution to Eq.~(\ref{eq61}) that is harmonic at the same frequency as the electric field \textbf{E}.\footnote{In principle solutions to the homogeneous equation must be included in the full solution. Without justification, we ignore the homogeneous contribution.} We thus write
\begin{equation}
\label{eq62}
{\rm {\bf p(r}},t) = \mathbf{\tilde p(r)} \,e^{ - i\omega \,t}.
\end{equation}

\noindent Substituting this expression into Eq.~(\ref{eq61}) gives the amplitude of the polarization as
\begin{equation}
\label{eq63}
{\rm {\bf \tilde{p}(r)}} = \frac{q^2}{m^* \left( {\omega_0^2 - \omega ^2} \right)}{\rm {\bf \tilde{E} (r)}}.
\end{equation}

To obtain the relationship between the macroscopic polarization ${\rm {\bf P}}$ and the electric field ${\rm {\bf E}}$ we note that \textbf{P} = $N_c$\textbf{p}, where $N_c$ is the number density associated with the unit cells in the solid.  This gives us
\begin{equation}
\label{eq64}
{\rm {\bf \tilde {P}}}( {\rm {\bf r}} ) =  {\frac{N_c \,q^2}{m^* \left( {\omega_0^2 - \omega ^2} \right)}\;} {\rm {\bf \tilde {E}}}( {\rm {\bf r}}),
\end{equation}

\noindent There are two important results contained in Eq.~(\ref{eq64}).  First, we have a concrete example of linear response, which is due to the linear equation of motion for the dipole moment \textbf{p}.  Second, we see that this harmonic-oscillator model gives rise to response that is frequency-dependent.  As we shall see, this frequency dependence leads to some very interesting phenomena.  

Using the relationship [Eq.~(\ref{eq13})] among ${\rm {\bf D}}$, ${\rm {\bf E}}$, and ${\rm {\bf P}}$ we have
\begin{equation}
\label{eq65}
{\rm {\bf \tilde {D}}}( {\rm {\bf r}} ) = \varepsilon _0 \,{\rm {\bf \tilde {E}}}( {\rm {\bf r}} ) + {\rm {\bf \tilde {P}}}( {\rm {\bf r}} ),
\end{equation}

\noindent which results in
\begin{equation}
\label{eq66}
{\rm {\bf \tilde {D}}}( {\rm {\bf r}} ) = \varepsilon _0 \left( {1 + {\frac{N_c \,q^2}{\varepsilon_0 m^* } \, \frac{1}{ {\omega_0^2 - \omega ^2} }} } \right)  {\rm {\bf \tilde {E}}}( {\rm {\bf r}} ).
\end{equation}

\noindent Comparing this equation with Eq.~(\ref{eq52}) we immediately identify the dielectric function $\varepsilon(\omega)$ associated with excitation of a single normal mode as 
\begin{equation}
\label{eq67}
\varepsilon ( \omega ) =  {1 +  {\frac{\omega_{p}^2}{ {\omega_0^2 - \omega^2} }} },
\end{equation}

\noindent where we have defined the \textbf{plasma frequency} $\omega_{p}$ via
\begin{equation}
\label{eq68}
\omega_{p}^2 = \frac{N_c \,q^2}{\varepsilon _0 m^*}.
\end{equation}

\begin{figure}
\includegraphics[scale=0.70]{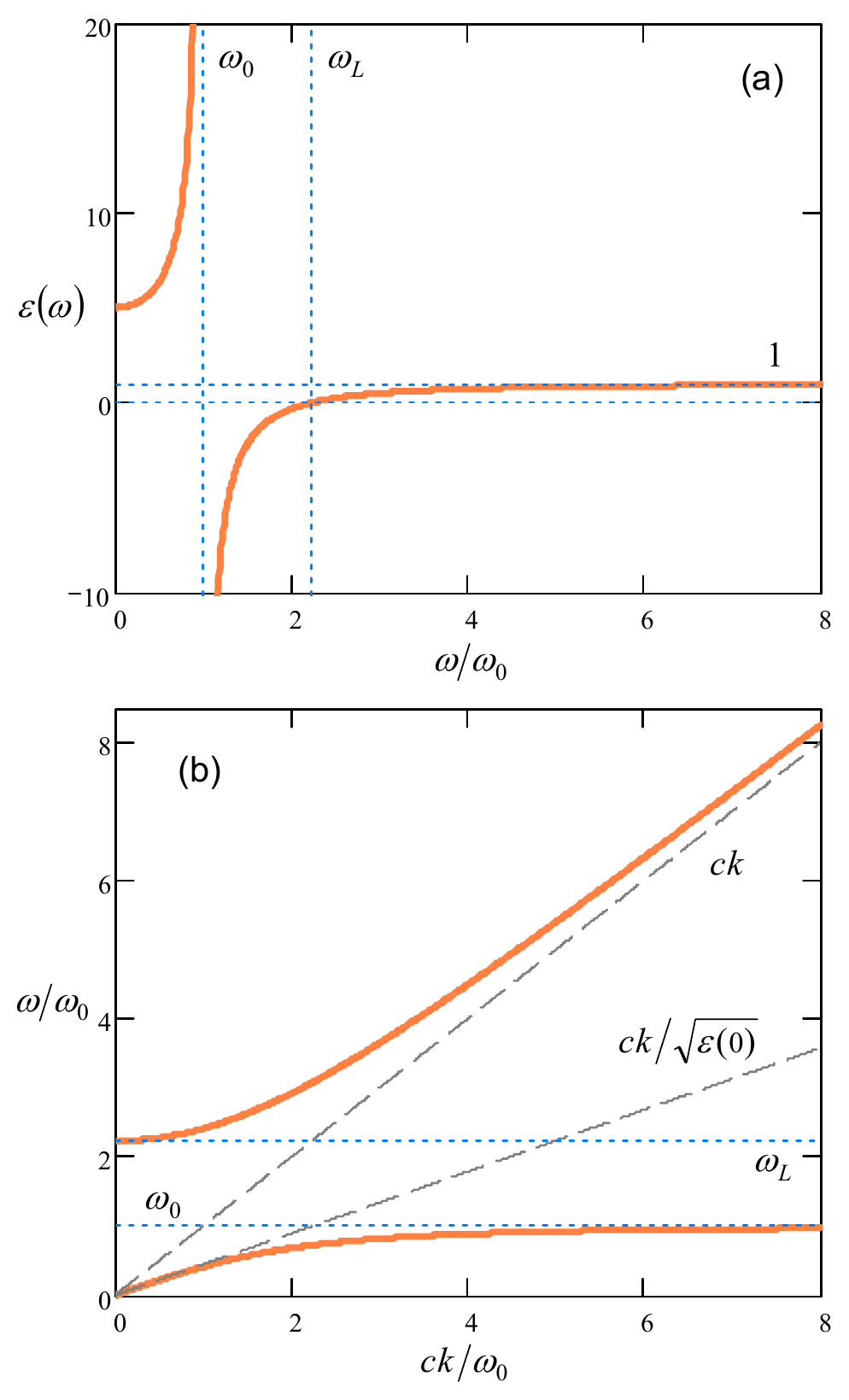}
\caption{\label{fig1}Dielectric function (a) and dispersion curves (b) for $\varepsilon(\omega)$ is given by Eq.~(\ref{eq67}).  As indicated, the high-frequency and low-frequency asymptotes (dashed lines) corresponds to $\omega = ck$ and $\omega = ck / \sqrt{\varepsilon(0)}$, respectively.  Here we have set $\omega_{p}$/$\omega_0$ = 2.}
\end{figure}

Although any solid will have more than one normal mode that couples to a particular EM wave, let's presently investigate the consequences of this one-mode model by studying Eq.~(\ref{eq67}) in some detail.  Later we shall expand the dielectric function to include the possibility of multiple modes interacting with the field.

The low and high-frequency limits of Eq.~(\ref{eq67}) are quite revealing.  First, for $\omega \to \infty $ we obtain
\begin{equation}
\label{eq69}
\varepsilon ( \infty ) =  1,
\end{equation}

\noindent the vacuum result for the dielectric function. This occurs because at very high frequencies the charge cannot respond to the rapidly varying electric field, and so the EM wave propagates without any notice of the material.  In the opposing limit ($\omega \to 0$) we have
\begin{equation}
\label{eq70}
\varepsilon (0) =  {1 +  {\frac{\omega_{p}^2}{\omega_0^2 }} }.
\end{equation}

\noindent That is, the dielectric constant is greater than the vacuum value. This is the result of a dc electric field being able to statically polarize the bound charge.

Let's now consider what happens when $\omega $ is in the vicinity of the natural frequency $\omega_0$.  In (a) of Fig.~\ref{fig1} we plot $\varepsilon(\omega)$ as a function of $\omega$.  Due to the denominator containing $\omega_0^2 - \omega^2$, the function diverges at $\omega = \omega_0$, as illustrated.  This is an unphysical result of our neglect of any damping associated with the oscillator.\footnote{We shall rectify this a bit later.}  For frequencies such that $\omega < \omega_0$ or $\omega > \omega_L$ (where $\omega_L^2 = \omega_{p}^2 +  \omega_0^2$) the dielectric function is positive.  In these two frequency regions EM waves freely propagate through the material.  However, for $\omega_0 <\omega < \omega_L$ the dielectric function is negative.  In this frequency region EM waves do not propagate, but are exponentially damped.  To see why this is the case we can appeal to the dispersion relation, Eq.~(\ref{eq57}).  As this equation shows, if $\varepsilon(\omega)$ is negative (and $\mu$ is positive), then the wave vector $k$ must be imaginary, and so the spatial part of the wave ($\sim e^{ikz}$) exponentially decays.

We can gain more insight into the behavior of EM waves coupled to normal modes of the solid by further considering the dispersion relation 
\begin{equation}
\label{eq70b}
\omega (k) = \frac{ck}{\sqrt{\varepsilon(\omega)}}, 
\end{equation}

\noindent where here (and here on out) we set $\mu = 1$.  Substituting the expression in  Eq.~(\ref{eq67}) for $\varepsilon(\omega)$ into this general form of the dispersion relation and solving explicitly for $\omega (k)$ yields \marginpar{\footnotesize{$\mathbb{EX} \,$\ref{E5}}}
\begin{eqnarray}
\label{eq72a}
\omega^2(k) &=& \frac{1}{2} \bigg\{  \omega_{p}^2 + \omega_0^2 + c^2k^2  \pm \big[  (\omega_{p}^2 +\omega_0^2)^2 \nonumber \\
&+& (2\omega_{p}^2 -2\omega_0^2 +c^2k^2) \, c^2k^2\big]^{\frac{1}{2}}\bigg\}.
\end{eqnarray}

In spite of this rather complicated formula, the result is fairly simple (although quite interesting), as shown in Fig.~\ref{fig1}(b).  As illustrated, there are two branches to the dispersion relation.  The lower branch, which approaches linearity at long wavelengths (small $k$), corresponds to EM waves traveling through the material with an index of refraction  $n \approx \sqrt{\varepsilon(0)}$.  The response of the system serves to reduce the speed of the propagating EM waves, but because the frequencies are well away from the natural frequency $\omega_0$ of the system's normal modes, the material is only weakly involved in the EM fields.  At high frequencies this branch flattens out at the natural frequency $\omega_0$ of the normal modes.  Indeed, this part of this dispersion curve corresponds to the normal modes the system essentially \textit{uncoupled} from the EM fields.  The behavior of the upper branch is quite different from that of the lower branch.  This branch approaches linear dispersion at high frequencies; this part of the branch corresponds to EM waves essentially uncoupled from the normal modes of the system, and so the EM waves travel at nearly $c$.  However, at long wavelengths this branch flattens out to $\omega \to$ $\omega_L$ as $k \to 0$.  Obviously, there is a strong coupling of the EM waves and the material's normal modes.  We note for later that at $\omega$ = $\omega_L$ the dielectric function $\varepsilon$ vanishes, as illustrated in Fig.~\ref{fig1}(a).  The parameter $\omega_L$ is known as the \textbf{longitudinal frequency} of the system.

Figure~\ref{fig1} shows there are no solutions for $\omega$ (for real values of $k$) in the gap between $\omega_0$ and $\omega_L$.  In this frequency region there are solutions, but all of these solutions have wave vector $k$ values that are purely imaginary, indicating that propagating waves are not allowed in this range of frequencies.  This is analogous to the absence of extended electronic states between the valence and conduction bands of a semiconductor.

\subsection{Multiple Modes}

We now expand our simple dielectric function to include the possibility that a given EM wave couples to more that one polarization mode of the solid.  In this case the dipole moment \textbf{p} will have contributions from all modes involved in the response of the system.  The upshot of this is that the dielectric function [as written in Eq.~(\ref{eq67})] expands to include a sum over all of the  involved modes,
\begin{equation}
\label{eq73}
\varepsilon ( \omega ) =  {1 + \sum_{n=0}^N {\frac{\omega_{pn}^2}{ {\omega_n^2 - \omega^2} }} }.
\end{equation}

\noindent Here $\omega_n$ is the natural frequency associated with the normal mode (set) labeled by $n$, and 
\begin{equation}
\label{eq74}
\omega_{pn}^2 = \frac{N_c \,q_n^2}{\varepsilon _0 m_n^*}.
\end{equation}

\noindent is the plasma frequency associated with these modes.  

A canonical application of this model dielectric function is the response of a diatomic ionic crystal, such as NaCl.  Such a crystal has one cation and one anion (Na$^+$ and Cl$^-$, respectively, e.g.) per unit cell.  This structure results one phonon mode (for each value of \textbf{k}) that couples to the EM fields.  Typically the dispersion of these \textbf{optic phonon} modes is quite flat.  At frequencies well above the optic-phonon response are electronic interband excitations (typically in the UV).  These excitations cover a range of frequencies and a given EM wave may couple to more than one.  Thus, in a typical ionic crystal there is one low frequency mode well separated from a band of much higher frequency electronic modes.

Let's now focus on the dielectric function at frequencies well below the resonant frequencies of the interband excitations.  In this case the response of the electrons can be approximated by their zero-frequency limit $\sum_{n=1}^N \omega_{pn}^2 / \omega_n^2$,\footnote{Fairly obviously, the $n$ = 0 term represents the optic-phonon response.} which simplifies the dielectric function to \marginpar{\footnotesize{$\mathbb{EX} \,$\ref{E7},$\,$\ref{E8}}}
\begin{equation}
\label{eq75}
\varepsilon ( \omega ) =  {\varepsilon_{\infty} +  {\frac{\omega_{p0}^2}{ {\omega_0^2 - \omega^2} }} },
\end{equation}

\noindent where $\varepsilon_{\infty}$ is defined to be
\begin{equation}
\label{eq75b}
\varepsilon_{\infty} =  1 +  \sum_{n=1}^N \omega_{pn}^2 / \omega_n^2.
\end{equation}

\begin{figure}
\includegraphics[scale=0.70]{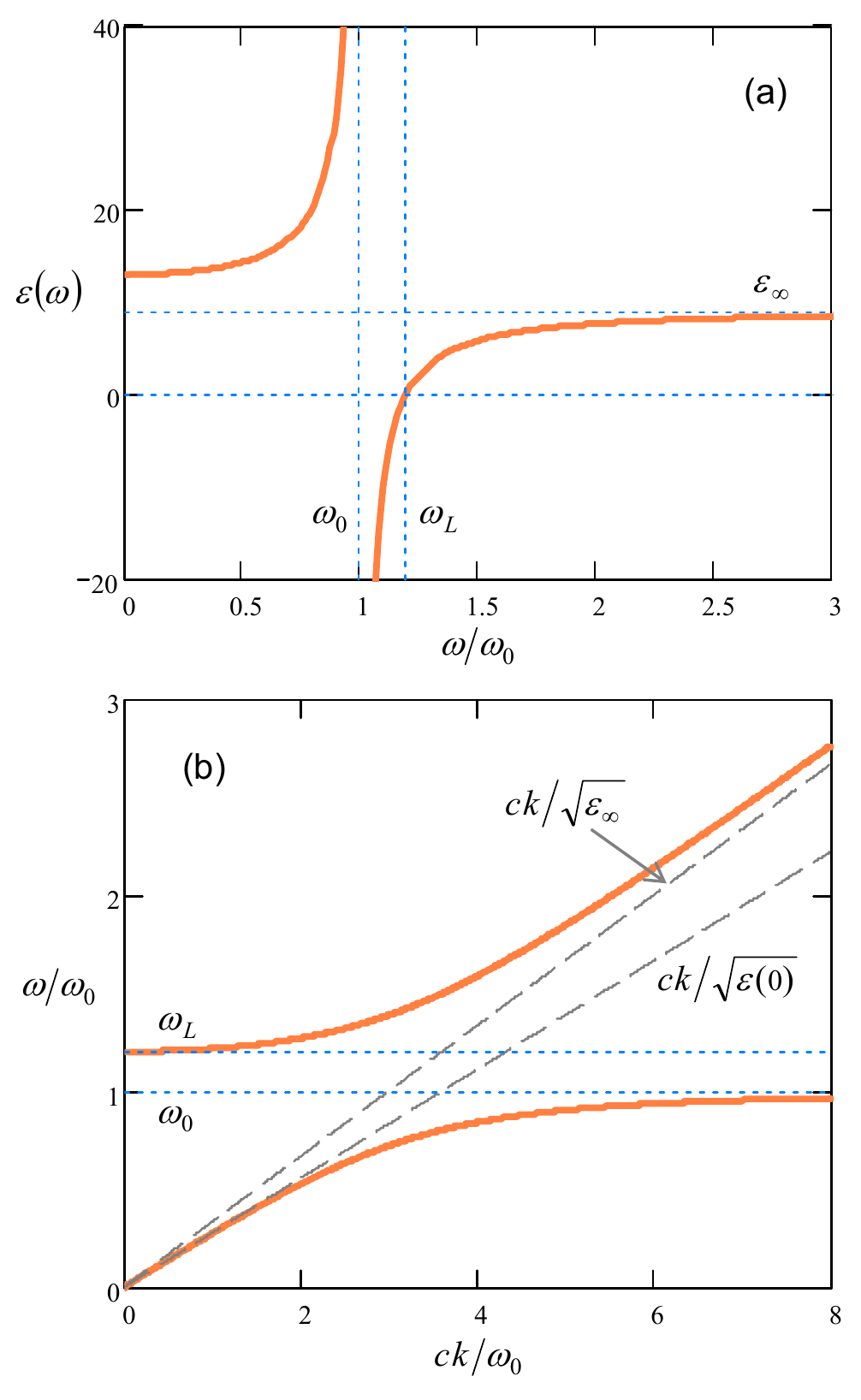}
\caption{\label{fig2}Dielectric function (a) and dispersion curves (b) for $\varepsilon(\omega)$ as given by Eq.~(\ref{eq75}).  As indicated, the high-frequency and low-frequency asymptotes (dashed lines) corresponds to $\omega = ck / \sqrt{\varepsilon_{\infty}}$ and $\omega = ck / \sqrt{\varepsilon(0)}$, respectively.  As in Fig.~\ref{fig1} $\omega_{p}$/$\omega_0$ = 2, while (consistent with Fig.~\ref{fig3} below) $\varepsilon_{\infty} = 9$.}
\end{figure}

Comparing the dielectric function of Eq.~(\ref{eq75}) with that of the single oscillator [Eq.~(\ref{eq67})], we see that they are the same except that the constant 1 on the right hand side of Eq.~(\ref{eq67}) has been replaced by $\varepsilon_{\infty}$.  Figure \ref{fig2}(a) plots $\varepsilon(\omega)$ vs $\omega$ as given by Eq.~(\ref{eq75}).  As shown there, $\varepsilon_{\infty}$ is the high-frequency limit of Eq.~(\ref{eq75}).\footnote{Hence the use of the symbol $\infty$}

This factor of $\varepsilon_{\infty}$ has several important consequences, all of which are manifest in the dispersion relation $\omega(k)$, plotted in Fig.~\ref{fig2}(b).\marginpar{\footnotesize{$\mathbb{EX} \,$\ref{E6}}}  First, the longitudinal frequency $\omega_L$ where the dielectric function vanishes is now given by  
\begin{equation}
\label{eq75bb}
\omega_L^2 = \frac{\omega_{p0}^2}{ \varepsilon_{\infty} } +  \omega_0^2, 
\end{equation}

\noindent making $\omega_L$ relatively closer to $\omega_0$.  Second, the large $k$ asymptote of the upper branch is modified to be $\omega = ck/\sqrt{\varepsilon_{\infty}}$.  Third, while the small $k$ asymptote of the lower branch is still given by $\omega = ck / \sqrt{\varepsilon(0)}$, this dispersion now has a contribution from the high-frequency electronic response.  This is because $\varepsilon(0)$ = $\varepsilon_{\infty} + \omega_{p0}^2 / \omega_0^2$.

\begin{figure}
\includegraphics[scale=0.75]{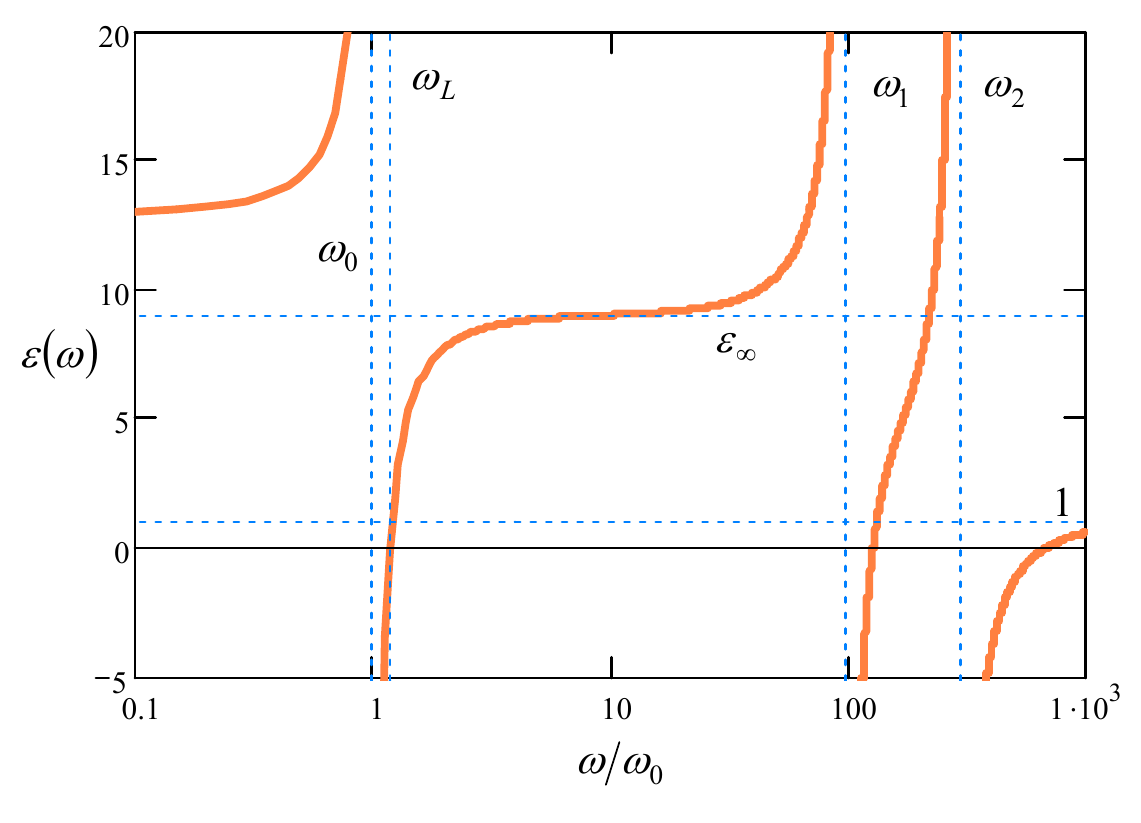}
\caption{\label{fig3}Dielectric function given by Eq.~(\ref{eq73}) for three modes.  Here $\omega_{p1} / \omega_{p0} = 100$, $\omega_{p2} / \omega_{p0} = 300$, and $\omega_{pn} / \omega_n = 2$ for all three modes.  }
\end{figure}

To examine the multiple-mode dielectric function [Eq.~(\ref{eq73})] over all relevant frequencies, in Fig.~\ref{fig3} we have plotted $\varepsilon(\omega)$ assuming a total of three coupled modes.  The $n$ = 0 mode (representing the optic phonons) is well separated from the other two (electronic interband) modes.  There are two features worth mentioning.  First, consistent with the dielectric function  illustrated in Fig~\ref{fig2}(a), in the region between the phonon and electron responses the dielectric function is quite close to $\varepsilon_{\infty}$.  Furthermore, here the slope of $\varepsilon(\omega)$ is positive.  As this region typically encompasses the near IR and visible, we can infer that the index of refraction $n = \sqrt{\varepsilon(\omega)}$ increases with $\omega$, consistent with the behavior of transparent dielectrics (consider light transmission through a prism).  Second, at the largest frequencies shown in Fig.~\ref{fig3}, we observe $\varepsilon(\omega)$ to be positive, but less than 1.  Hence, the phase velocity $v_p = c / \sqrt{\varepsilon}$ is \textit{greater} than $c$~!\marginpar{\footnotesize{$\mathbb{EX} \,$\ref{E9}}}  Light in this region can thus undergo total \textit{external} reflection from a material.  This phenomenon is utilized to make surface sensitive x-ray diffraction measurements.

\subsection{Damping Included}

A feature that is acutely missing from the equation of motion for \textbf{p} [Eq.~(\ref{eq61})] is any description of the forces that serve to dissipate any induced moment.  The simplest way to account for any such damping is to add in a term that opposes the dipole-moment velocity $d\mathbf{p} / dt$.\footnote{From a classical physics point of view this is a drag force that is proportional to the velocities of the oscillating charges.   From a quantum point of view this dissipation corresponds to an excitation decaying into other degrees of freedom of the solid.  For example, an optic phonon typically decays into two or three acoustic phonons.}  Doing so, we obtain
\begin{equation}
\label{eq75c}
 \frac{d^2 \mathbf{p(r},t)}{dt^2} + \gamma \, \frac{d\mathbf{p(r},t)}{dt} + \omega_0^2 \, \mathbf{p(r},t)   = \frac{q^2}{m^*} \mathbf{\tilde E(r)} \,e^{- i  \omega \, t },
\end{equation}

\noindent where $\gamma$ is known as the \textbf{damping parameter}.  

Let's now derive the dielectric function for multiple oscillators coupled to an EM wave.  If we carry out the the same steps as above for one oscillator, we first obtain the polarization amplitude \marginpar{\footnotesize{$\mathbb{EX} \,$\ref{E9b}}}
\begin{equation}
\label{eq76}
{\rm {\bf \tilde {P}}}( {\rm {\bf r}} ) =  {\frac{N_c \,q^2}{m^* \left( {\omega_0^2 - \omega ^2} - i \gamma \omega \right)}\;} {\rm {\bf \tilde {E}}}( {\rm {\bf r}}),
\end{equation}

\noindent which leads to the dielectric function
\begin{equation}
\label{eq77}
\varepsilon ( \omega ) =  1 +  \frac{\omega_{p}^2}{ {\omega_0^2 - \omega^2} - i \gamma \omega } .
\end{equation}

\begin{figure}
\includegraphics[scale=0.70]{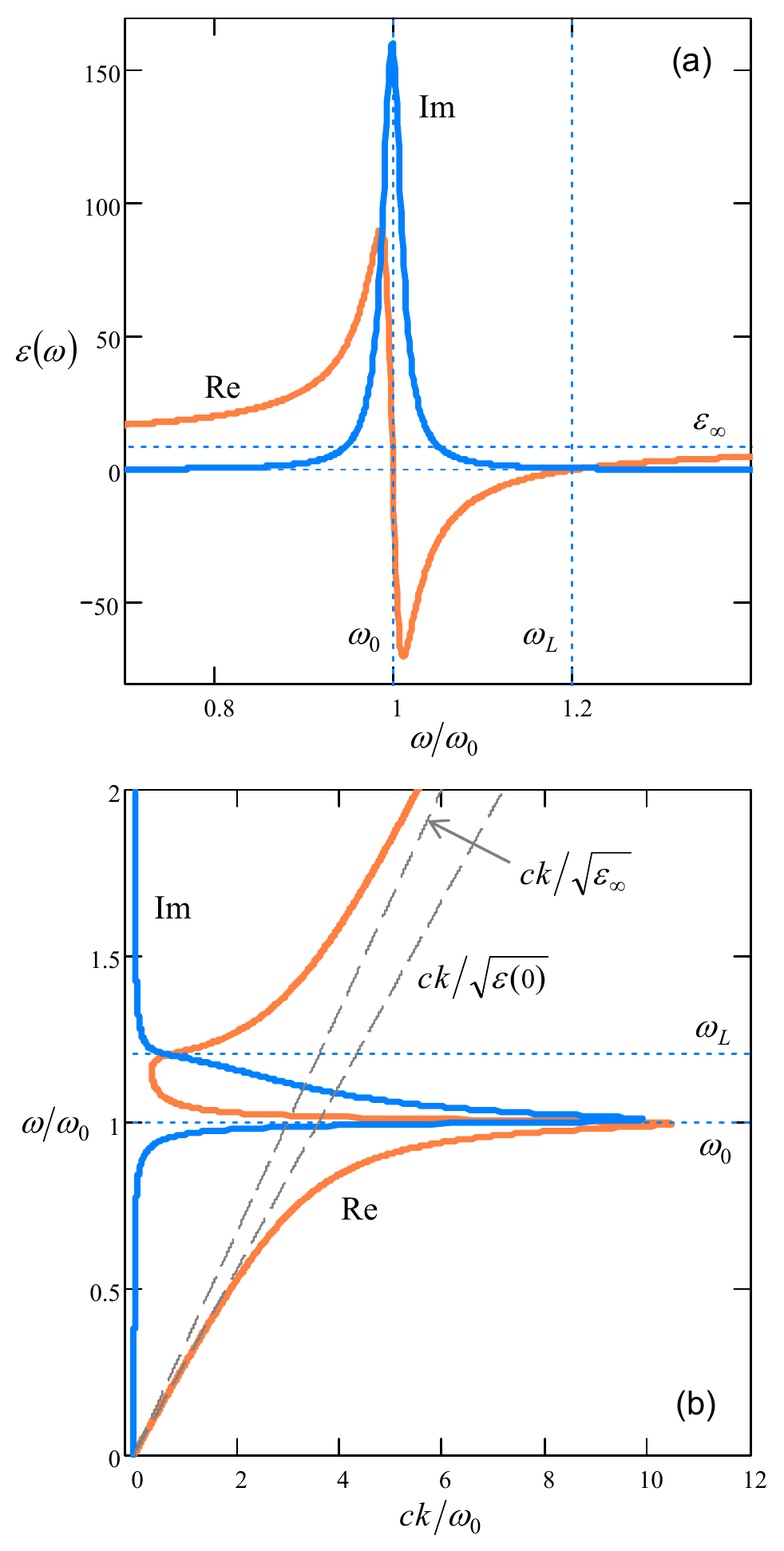}
\caption{\label{fig4}Dielectric function (a) and dispersion curves (b) for $\varepsilon(\omega)$ as given by Eq.~(\ref{eq80}).  In (a) and (b) the Re and Im parts of $\varepsilon$ and $k$ are plotted, respectively.  Parameters for $\varepsilon(\omega)$ are the same as for Fig.~\ref{fig2} with the addition of $\gamma_0 = 0.025 \, \omega_0$.}
\end{figure}

\noindent The obvious new feature associated with $\varepsilon(\omega)$ is that it is now a \textit{complex} function.  As we shall see, this results in \textit{any} electromagnetic wave being damped as it propagates through a material.\footnote{Of course, in some circumstances the damping may be so small that it may be neglected.  Consider visible light traveling through a thin piece of glass, for example.}  Under the general condition that multiple modes interact with the EM fields these last two equations become
\begin{equation}
\label{eq78}
{\rm {\bf \tilde {P}}}( {\rm {\bf r}} ) = \sum_{n=0}^N {\frac{N_c \,q_n^2}{m_n^* \left( {\omega_n^2 - \omega ^2} - i \gamma_n \omega \right)}\;} {\rm {\bf \tilde {E}}}( {\rm {\bf r}})
\end{equation}

\noindent and
\begin{equation}
\label{eq79}
\varepsilon ( \omega ) =  1 +  \sum_{n=0}^N  \, \frac{\omega_{pn}^2}{ {\omega_n^2 - \omega^2} - i \gamma_n \omega }.
\end{equation}

\noindent 

As above, let's now focus on the situation of having one coupled phonon mode far removed (in frequency) from any electronic interband excitations.  In this case the expression for the dielectric function in Eq.~(\ref{eq79}) is well represented by
\begin{equation}
\label{eq80}
\varepsilon ( \omega ) =  {\varepsilon_{\infty} +  {\frac{\omega_{p0}^2}{ {\omega_0^2 - \omega^2} - i \gamma_0 \omega}} }.
\end{equation}
 
\noindent  This dielectric function and its consequential dispersion relation are plotted in Fig.~\ref{fig4}.  For this figure the parameters are the same as in Fig.~\ref{fig2} with the addition of $\gamma_0 = 0.025 \, \omega_0$.  

Focusing specifically on the dielectric function, plotted in Fig.~\ref{fig4}(a), we first note that Re$(\varepsilon)$ no longer diverges at the natural frequency $\omega_0$.  Second, we see that Im$(\varepsilon)$ peaks very close to $\omega_0$, but away from this frequency it becomes negligible.  Those familiar with the response of a driven, damped harmonic oscillator will recognize this (underdamped) response and know that this peak narrows (broadens) with decreasing (increasing) $\gamma_0$.  Third, Re$(\varepsilon)$ still vanishes close to $\omega_L$.  This is a result of $\gamma_0$ being relatively small in this example.  Fourth, the behavior of Re$(\varepsilon)$ is unchanged at the frequency extremes.  

The dielectric constant having a non-zero imaginary part has consequences for the dispersion relation $\omega(k) = ck / \varepsilon(\omega)$.\marginpar{\footnotesize{$\mathbb{EX} \,$\ref{E10}}}  For a given (real) $\omega$, $k$ now has both real and imaginary parts at all frequencies,  as illustrated in Fig.~\ref{fig4}(b).  When Im$(k)  \ll$ Re$(k)$, EM waves still freely propagate, but with a decaying amplitude as they travel through the material.  This happens when $\omega$ is well below $\omega_0$ or well above $\omega_L$.  However, when Im$(k)  \agt $ Re$(k)$, as is the case when $\omega_0 < \omega < \omega_L$, the waves are strongly damped within at least a few wavelengths.  We note that this nonpropagating nature of the solutions for $\omega_0 < \omega < \omega_L$ was already apparent before the inclusion of damping, and so is not a consequence of its inclusion.

\subsection{Free Carriers}

We now turn to describing the response of charge carriers that are free to move throughout a crystal.\footnote{We remind the reader of our present convention that $\rho_{\rm p}$ comprises both free charge and bound charge in the solid.}  Such \textbf{free carriers} are present whenever an electronic band is partially filled.  The response that we are interested in describing is due to transitions between states within the same electronic band, and are thus know as \textbf{intraband transitions}.  The most common types of materials with significant numbers of free carriers are metals, semimetals, and doped semiconductors.  The charge carriers can have either negative charge (electrons) or positive charge (holes).

The model dielectric functions discussed so far are appropriate for charge carriers that have a well defined equilibrium positions, and so are inappropriate for charge that is free to move about a crystal.  However, with slight modification of our equation of motion [Eq.~(\ref{eq75c})], we can describe the response of the free carriers.  The required change is simple:  all we need to do is remove the restoring force $\omega_n^2 \, \mathbf{p(r},t)$ from the equation of motion for each polarization term corresponding to a particular type of free carrier.  Doing this we have
\begin{equation}
\label{eq80b}
 \frac{d^2 \mathbf{p(r},t)}{dt^2} + \gamma \, \frac{d\mathbf{p(r},t)}{dt} = \frac{q^2}{m^*} \mathbf{\tilde E(r)} \,e^{- i  \omega \, t }.
\end{equation}

\noindent In order to study the simplest case, let's make the following assumptions: (i) there is only one set of free carriers,\footnote{For example, we might be modeling a semiconductor that has electrons in the conduction band but no holes in the valence band.} (ii) there are no excitable optic phonon modes, and (iii) all other (interband) electronic modes have natural frequencies much greater than the plasma frequency $\omega_{p}$ associated with the free carriers. 

With these assumptions we straightforwardly obtain \marginpar{\footnotesize{$\mathbb{EX} \,$\ref{E10b}}}  
\begin{equation}
\label{eq81}
\varepsilon ( \omega ) =  \varepsilon_{\infty} -   \frac{\omega_p^2}{\omega^2} \, \frac{1}{ 1 + i / (\omega \tau)} 
\end{equation}

\noindent  for frequencies far below the other electronic excitations.  This dielectric function is canonically known as the \textbf{Drude} dielectric function.  Here $\omega_p$ is associated with the zero-frequency free-carrier mode.  The parameters that appear in the expression for $\omega_p$ [Eq.~(\ref{eq68})] now have the following meanings:  (i) $N_c$ is the free-carrier density, (ii)  $q$ is the electron charge magnitude $e$, and (iii) $m^*$ is the effective mass of the free carriers.\footnote{In a metal $m^*$ is often (but not always) close to the free-electron mass $m_e$.  In some semiconductor bands, $m^*$ is significantly different from $m_e$.  For example, in GaAs carriers in the conduction, light-hole, and heavy-hole bands have effective masses $m^* / m_e$ = 0.067, 0.082, and 0.45, respectively.}  We shall see later that $\tau = 1 / \gamma$ is the \textbf{momentum relaxation time} associated with the free carriers.  As before, $\varepsilon_{\infty}$ is the low-frequency contribution from interband modes.

\begin{figure*}
\includegraphics[scale=0.90]{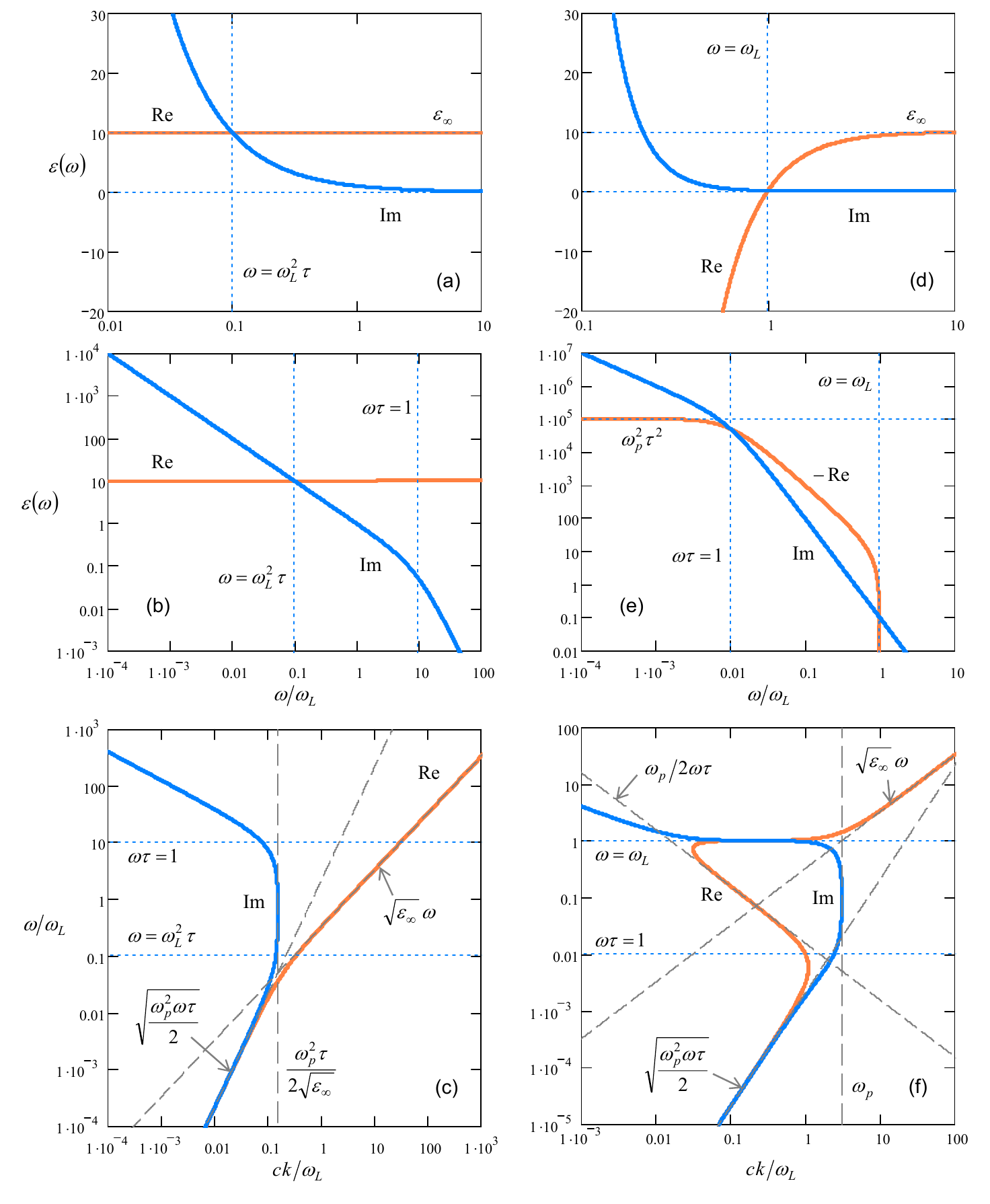}
\caption{\label{fig5}Drude dielectric function $\varepsilon(\omega)$ [Eq.~(\ref{eq82})] and resulting dispersion curves $\omega$ vs $ck$.  A poor [good] conductor is illustrated in (a), (b), and (c) [(d), (e), and (f)].  In (a), (b), (d), and (e) the Re and Im parts of $\varepsilon$ are plotted.  In (c) and (f) the Re and Im parts of $ck$ are shown.  In contrast to previous figures, $\omega$ and $ck$ are scaled by $\omega_L$.  The dashed lines in the dispersion-curve plots are approximate expressions for $ck$ in the appropriate regions.  For the poor (good) conductor $\omega_L \tau$ = 0.1 (100).  For both conductors $\varepsilon_{\infty} = 10$.}
\end{figure*} 

\begin{figure*}
\includegraphics[scale=0.70]{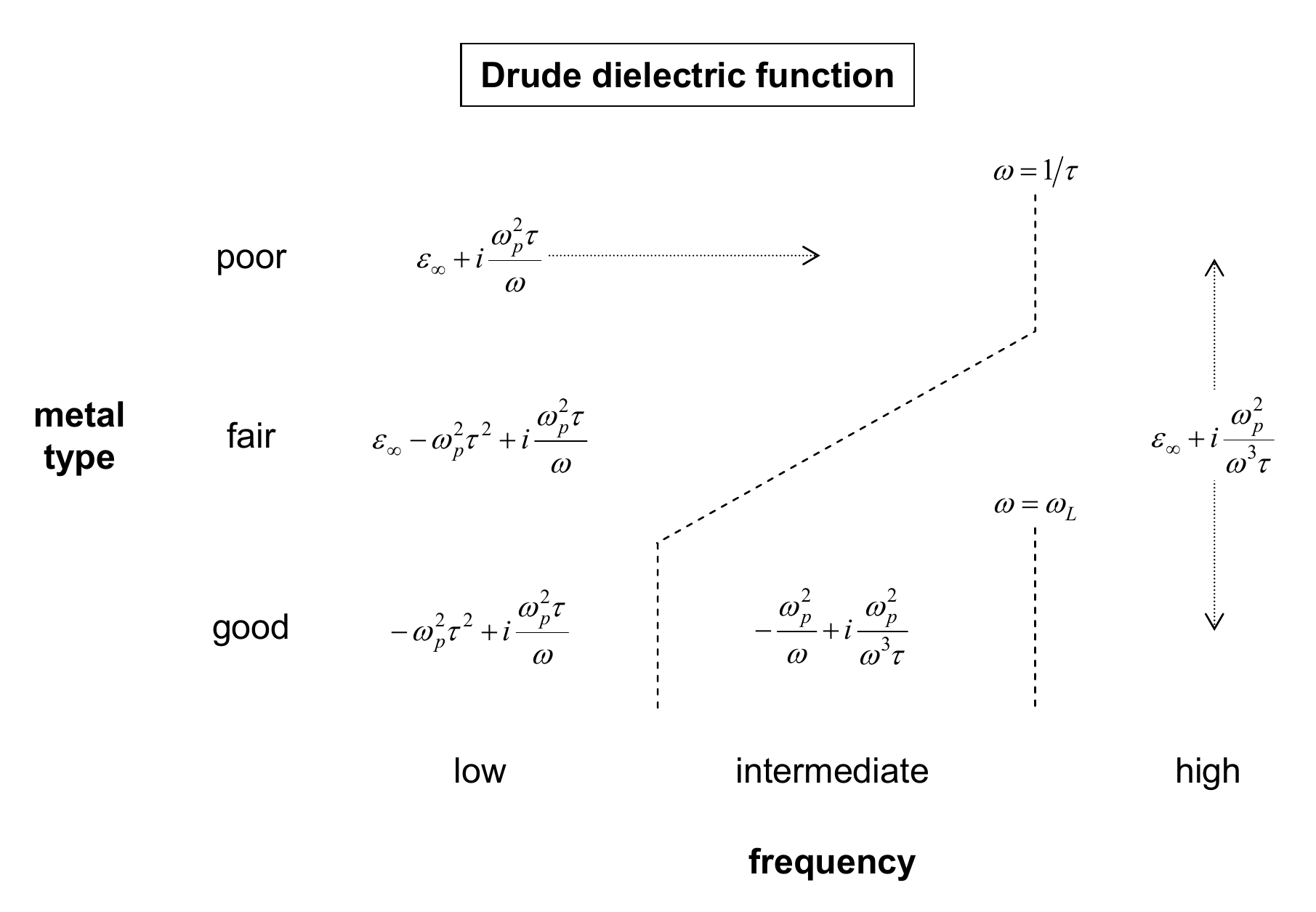}
\caption{\label{figDE}Approximate expressions for Drude dielectric function [Eq.~(\ref{eq82})] for poor ($\omega_L \tau \ll 1$), fair ($\omega_L \tau \sim 1$), and good ($\omega_L \tau \gg 1$) conductors in low, intermediate, and high frequency regions.}
\end{figure*}

\begin{figure*}
\includegraphics[scale=0.70]{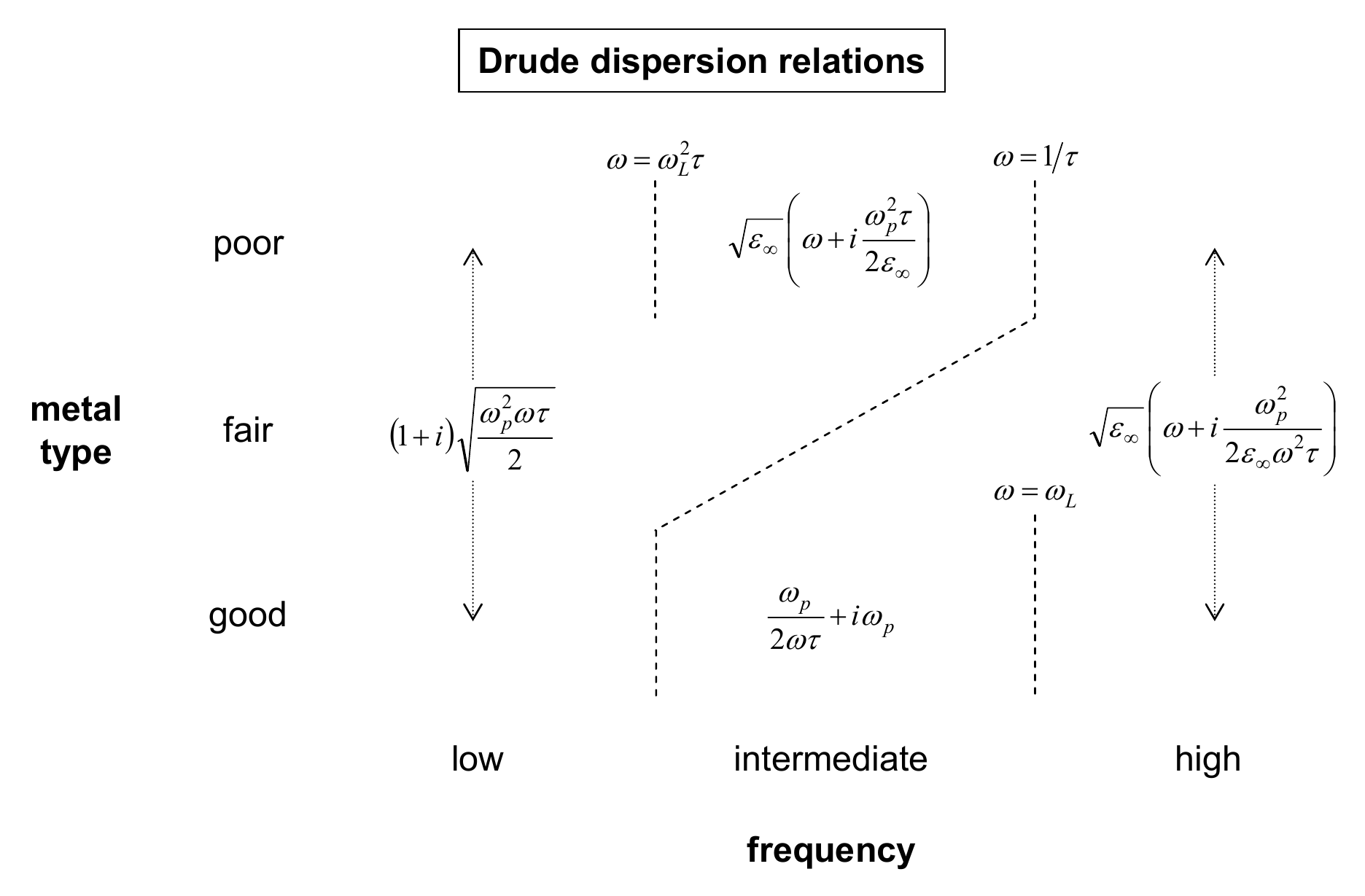}
\caption{\label{figDD}Approximate expressions $ck$ resulting from Drude dispersion relation [Eq.~(\ref{eq82})] for poor ($\omega_L \tau \ll 1$), fair ($\omega_L \tau \sim 1$), and good ($\omega_L \tau \gg 1$) conductors in low, intermediate, and high frequency regions.}
\end{figure*}

An unsuspecting richness lies within the Drude response given by Eq.~(\ref{eq81}).  In order to facilitate an analysis of this response, we first rewrite the right-hand-side of Eq.~(\ref{eq81}) to make the real and imaginary parts more obvious, \marginpar{\footnotesize{$\mathbb{EX} \,$\ref{E10c}}}
\begin{equation}
\label{eq82}
\varepsilon ( \omega ) =  \varepsilon_{\infty} \left[ 1 - \frac{\omega_L^2}{\omega^2} \, \frac{1 - i  / (\omega \tau)}{ 1 + 1 / (\omega^2 \tau^2)} \right].
\end{equation}

\noindent Because for free carriers the natural frequency of oscillation $\omega_0$ is zero, the longitudinal frequency is simply $\omega_L = \omega_p / \sqrt{\varepsilon_{\infty}}$.   As one might surmise from inspection of this equation, the parameters $\tau$ and $\omega_L$ are the keys to the frequency dependence of $\varepsilon$.  In fact, the combination $\omega_L \tau$ allows us to characterize three classes of conductors:  $\omega_L \tau \gg 1$ defines a \textbf{good conductor}, $\omega_L \tau  \sim 1$ a \textbf{fair conductor}, and $\omega_L \tau  \ll 1$ a \textbf{poor conductor}.  Examples of good conductors are abundant; most elemental metals qualify as good conductors.  Below we discuss Pb and Au in this context.  Doped semiconductors can provide examples of poor conductors, but this is by no means universal.  An interesting case is $n$-type GaAs.  From dc-resistivity and dielectric-constant data \cite{Sze1981}, one can infer that $\omega_L \tau$ = 0.14, 1.6, and 7.2 at carrier densities of 10$^{14}$, 10$^{16}$, and 10$^{18}$ cm$^{-3}$, respectively.\footnote{Below we discuss conductivity, the inverse of resistivity.}  Thus, GaAs can be a poor, fair, or good conductor, depending upon the doping level.

Let's consider the frequency dependence of Im$(\varepsilon)$.  As Im$(\varepsilon)$ depends upon $\tau$ (but not $\omega_L$), its frequency dependence is the same for all three types of conductors.  For $\omega \ll 1 / \tau$ we have
\begin{equation}
\label{eq83}
{\rm Im}(\varepsilon) \approx  \frac{\omega_p^2 \tau}{\omega},
\end{equation}  

\noindent while for $\omega \gg 1 / \tau$
\begin{equation}
\label{eq84}
{\rm Im}(\varepsilon) \approx \frac{\omega_p^2}{\omega^3 \tau}.
\end{equation}

\noindent  Thus, $\omega \tau =1$ demarcates $1 / \omega$ behavior at lower frequencies from $1 / \omega^3$ behavior at higher frequencies.  This response is apparent in parts (b) and (e) of Fig.~\ref{fig5}, which plot $\varepsilon(\omega)$ on log-log plots for examples of good and poor conductors, respectively.    

The behavior of Re$(\varepsilon)$ is substantially more complicated, as the parameter $\omega_L$ is also involved.  For both good and poor conductors there are three frequency regions with different behaviors.  For good conductors the regions are separated by $\omega = 1 / \tau$ and $\omega = \omega_L$ (lower and higher frequency boundaries, respectively), while for poor conductors the regions are separated by $\omega = \omega_L^2 \tau$ and $\omega = 1 / \tau$.  Because $\omega_L  \sim 1 / \tau$ for a fair conductor, the middle-frequency region collapses, leaving only two regions separated by $\omega \sim 1 / \tau \sim \omega_L$.

We now discuss how Re$(\varepsilon)$ varies with $\omega$.  For all three types of conductors the real part of the dielectric response at high frequencies is simply
\begin{equation}
\label{eq85a}
{\rm Re}(\varepsilon) \approx  \varepsilon_{\infty},
\end{equation}

\noindent which shows that the response of the bound electrons makes the major contribution to Re$(\varepsilon)$.  This is illustrated in (a) and (d) of Fig.~\ref{fig5}.  For poor conductors Eq.~(\ref{eq85a}) is also valid in the intermediate- and low-frequency regimes. Thus, a poor conductor is one where Re$(\varepsilon)$ is always dominated by the bound electrons, as evident in Fig.~\ref{fig5}(a) and (b). For fair conductors the low-frequency expression for Re$(\varepsilon)$ is 
\begin{equation}
\label{eq85b}
{\rm Re}(\varepsilon) \approx  \varepsilon_{\infty} - \omega_p^2 \tau^2.
\end{equation}

\noindent Both terms on the right hand side appear because $\omega_L \tau \sim 1$ is equivalent to $\epsilon_{\infty} \sim \omega_p^2 \tau^2$.  Thus, in a fair conductor both the free and bound electron contributions to Re$(\varepsilon)$ are significant.  For good conductors the corresponding relation is  
\begin{equation}
\label{eq85c}
{\rm Re}(\varepsilon) \approx   - \omega_p^2 \tau^2,
\end{equation}

\noindent  which shows (perhaps not unexpectedly) that at low frequencies the response of a good conductor is primarily due to the free carriers.  This low-frequency limit is indicated in Fig.~\ref{fig5}(e).  The only region not yet addressed is the intermediate-frequency region of good conductors.  Here the approximate equation is
\begin{equation}
\label{eq85d}
{\rm Re}(\varepsilon) \approx   - \frac{\omega_p^2}{\omega^2}.
\end{equation}

\noindent  As at low frequencies, the response of the bound electrons ($\varepsilon_{\infty}$) does not appear.  The intermediate-frequency response of a good conductor is also illustrated in Fig.~\ref{fig5}(e).  All of the approximate results for $\varepsilon(\omega)$ are summarized in Fig.~\ref{figDE}.  \marginpar{\footnotesize{$\mathbb{EX} \,$\ref{E11}}}

We point out a useful expression for $\varepsilon(\omega)$ for a good conductor.  Frequencies within or below the intermediate-frequency region are defined by $\omega^2 \ll \omega_L^2$.  With solely this condition, in Eq.~(\ref{eq81}) the bound electron response $\varepsilon_{\infty}$ of Eq.~(\ref{eq81}) can be ignored compared to the free-carrier response, and so $\varepsilon(\omega)$ simplifies to 
\begin{equation}
\label{eq85e}
\varepsilon ( \omega ) \approx   -   \frac{\omega_p^2}{\omega^2} \, \frac{1}{ 1 + i / (\omega \tau)}.
\end{equation}
  
\noindent It is easily seen that this reduces to the good-conductor expressions at low and intermediate frequencies in Fig.~\ref{figDE} in the appropriate limits of $\omega \tau$. 

Approximate dispersion relations analogous to the $\varepsilon(\omega)$ expressions in Fig.~\ref{figDE} are presented in Fig.~\ref{figDD}.\marginpar{\footnotesize{$\mathbb{EX} \,$\ref{E11}--\ref{E13}}}  Specifically, this figure shows approximate expressions for the wave vector $ck$ as a function of frequency $\omega$. 

As Fig.~\ref{figDD} indicates, at the lowest and highest frequencies the conductor types are not distinguishable via the dispersion relations.  In the low-frequency regime this is due to  Im$(\varepsilon) = \omega_p^2 \tau / \omega$ being much larger than Re$(\varepsilon)$ for all three conductor types.  Consequently, at these smallest values of $\omega$
\begin{equation}
\label{eq86}
c k \approx (1 + i) \sqrt{\frac{\omega_p^2 \omega \tau}{2}}.
\end{equation}

\noindent  Notice Re$(k) =$ Im$(k)$. This asymptotic limit is indicated in Fig.~\ref{fig5}(c) and (f).  Similarly, EM waves in all three conductor types have the same approximate dispersion at high frequencies,
\begin{equation}
\label{eq87}
c k \approx \sqrt{\varepsilon_{\infty}} \left(  \omega + i \frac{\omega_p^2}{2 \varepsilon_{\infty} \omega^2 \tau}  \right).
\end{equation}

\noindent It is worth noting that Im$(k)$ falls off as $1 / \omega^2$, and so in this region with increasing $\omega$ a Drude conductor becomes more transparent.  The real part of this dispersion relation is also indicated in Fig.~\ref{fig5}.

It is the intermediate frequency regimes that distinguish a good conductor from a poor conductor, as (c) and (f) of Fig.~\ref{fig5} vividly illustrate.  As shown, in a poor conductor Re$(ck) \approx  \sqrt{\varepsilon_{\infty}} \omega$ has already taken on its high-frequency behavior while 
\begin{equation}
\label{eq88}
{\rm Im} (c k) \approx   \frac{\omega_p^2 \tau}{2 \varepsilon_{\infty}}
\end{equation}

\noindent is frequency independent and smaller than Re$(ck)$.  Conversely, in a good conductor 
\begin{equation}
\label{eq89}
{\rm Im} (c k) \approx   \omega_p,
\end{equation}

\noindent which is also frequency independent, dominates
\begin{equation}
\label{eq89b}
{\rm Re} (c k) \approx  \frac{\omega_p^2}{2 \omega \tau}.
\end{equation}

\noindent Although neither type of metal is transparent in this region, these difference do impact the frequency dependence of the reflectivity (not discussed here).

\subsection{Debye Polarization Response}

Let's now consider the response of a set of dipoles that have a natural frequency of oscillation $\omega_0$ about some equilibrium (as in the above case of optic phonons), but instead of relatively small damping, let's assume the damping to be highly viscous.  As it turns out, if the damping is large enough then the acceleration of the dipoles can be neglected without much error, and the equation of motion for the dipoles [Eq.~(\ref{eq75c})] reduces to
\begin{equation}
\label{eq90}
\gamma \, \frac{d\mathbf{p(r},t)}{dt} + \omega_0^2 \, \mathbf{p(r},t)   = \frac{q^2}{m^*} \mathbf{\tilde E(r)} \,e^{- i  \omega \, t }.
\end{equation}

\noindent This equation of motion leads to the dielectric function

\begin{equation}
\label{eq91}
\varepsilon ( \omega ) =  \varepsilon_{\infty} +  {\frac{\omega_{p}^2}{ { \omega_0^2} - i \gamma \omega}}.
\end{equation}

\noindent  Following tradition we rewrite this last equation as \marginpar{\footnotesize{$\mathbb{EX} \,$\ref{E18}}}
\begin{equation}
\label{eq92}
\varepsilon ( \omega ) =  \varepsilon_{\infty} +  {\frac{\varepsilon(0) - \varepsilon_{\infty}}{ 1 - i \omega \tau}}.
\end{equation}

\noindent where $\varepsilon(0) - \varepsilon_{\infty} = \omega_p^2 / \omega_0^2$, and $\tau = \gamma / \omega_0^2$ is the relaxation time associated with decay of this highly damped polarization.  As before, $\varepsilon_{\infty}$ represents the response of any other polarization modes at frequencies much higher than the highly-damped one that is the focus of our attention. The dielectric function given by either of these last two equations is know as the \textbf{Debye} dielectric function.\footnote{Often this dielectric function is written with $+i \omega \tau$ rather than $-i \omega \tau$ in the denominator.  This comes about if one assumes an harmonic time dependence $e^{i \omega t}$ rather than $e^{-i \omega t}$ (as we have assumed).  In order to match up the two conventions one must change the sign on the imaginary part of $\varepsilon(\omega)$ associated with one of these conventions.}

This response function was first introduced by Debye in order to describe the (low-frequency) dielectric response of polar liquids \cite{Debye1929}.  Specifically, Debye was interested in describing the response of the permanent dipoles associated with the molecules that make up a liquid.  An applied electric field will serve to align the moments, while thermal agitation will relax any induced dipole orientation back to some equilibrium value.  One can thus imagine how such response would map onto an overdamped harmonic oscillator.  Unfortunately, interactions between dipoles often cause the response to be more complicated than the simple model of Debye \cite{Cole1941,Kalmykov2004}.  However, it can be an excellent description when the dipoles are sufficiently dilute so that the interactions between them are negligible \cite{Kalmykov2004}.

\begin{figure}
\includegraphics[scale=0.70]{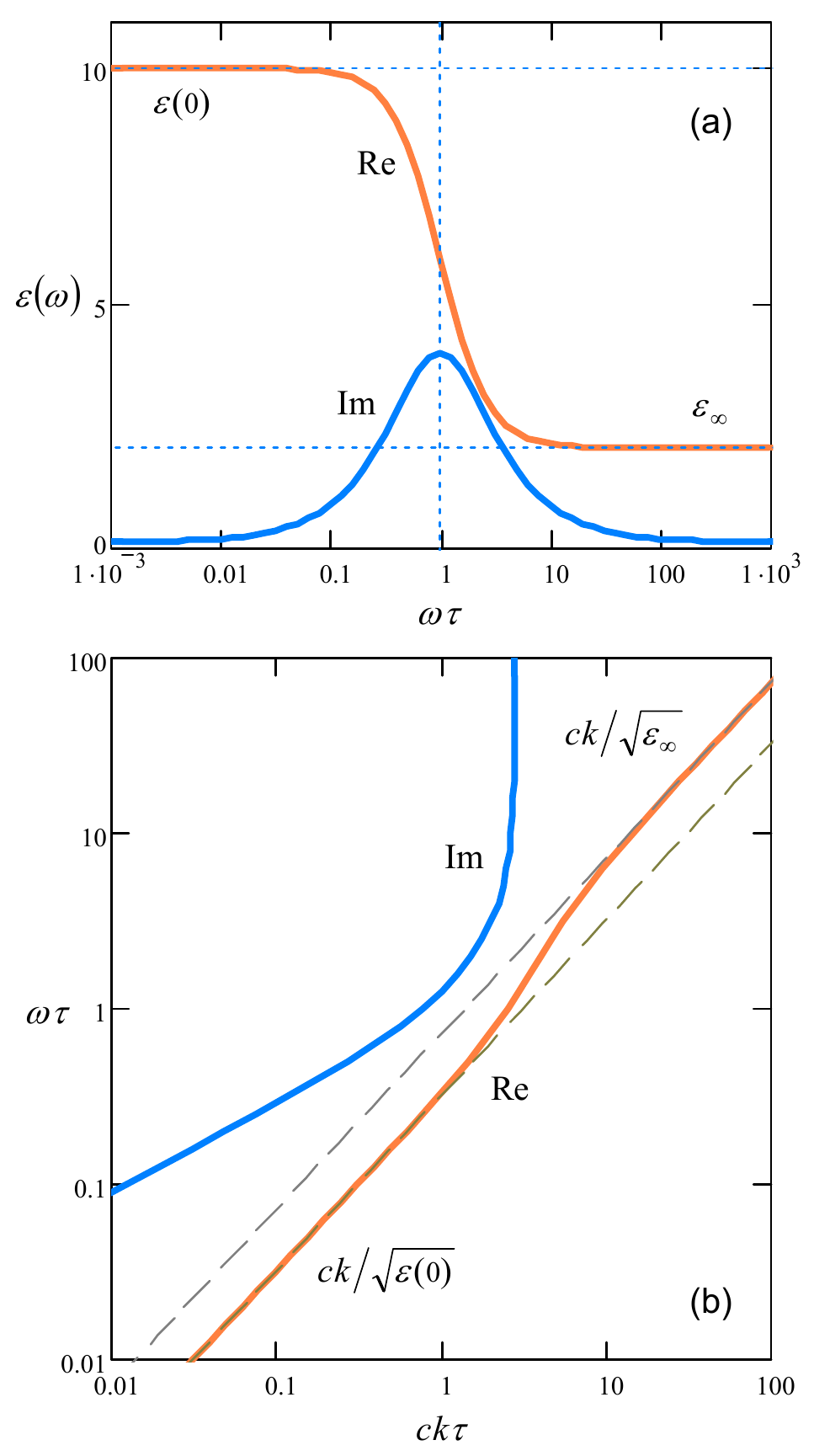}
\caption{\label{fig6}Dielectric function (a) and dispersion curves (b) for $\varepsilon(\omega)$ as given by Eq.~(\ref{eq92}).  In (a) and (b) the Re and Im parts of $\varepsilon$ and $k$ are plotted, respectively.  In contrast to previous figures, $\omega$ and $ck$ are scaled by $1 / \tau$.  Here we have arbitrarily chosen $\varepsilon_{\infty} = 2$ and $\varepsilon (0)$ = 10.}
\end{figure}

As with the other dielectric functions it is instructive to look at plots of $\varepsilon(\omega)$ and the dispersion curves.  As shown in Fig.~\ref{fig6}(a), with increasing $\omega$ the real part of $\varepsilon(\omega)$ smoothly falls off from $\varepsilon(0)$ to $\varepsilon_{\infty}$, reaching the midway point at $\omega \tau =1$.  Associated with this falloff is a broad\footnote{Notice the log scale on the $\omega \tau$ axis.} peak in the imaginary part of $\varepsilon(\omega)$.

In comparison with the Drude model, the dispersion curves [shown in part (b) of Fig.~\ref{fig6}] are relatively simple. With increasing $\omega$ the real part of $k$ indicates a smooth transition from waves propagating at $v_p = c / \sqrt{\varepsilon(0)}$ to waves with $v_p = c / \sqrt{\varepsilon_{\infty}}$.  Concurrently, the imaginary part of $k$ is relatively small at the frequency extremes [where Re$(\varepsilon) \gg$ Im$(\varepsilon)$], but for $\omega \tau \sim 1$ there is substantial damping of the EM waves.

At the frequency extremes the dielectric function simplifies considerably.  For $\omega^2 \tau^2 \ll 1$ one obtains \marginpar{\footnotesize{$\mathbb{EX} \,$\ref{E19}}}
\begin{equation}
\label{eq93}
{\rm Re}(\varepsilon) \approx  \varepsilon(0)
\end{equation}      

\noindent and
 \begin{equation}
\label{eq94}
{\rm Im}(\varepsilon) \approx   [\varepsilon(0) - \varepsilon_{\infty}] \,  \omega \tau.
\end{equation} 

\noindent  In this limit the dispersion relation is well represented by
\begin{equation}
\label{eq95}
c k \approx  \omega \sqrt{\varepsilon(0)} + i \frac{\varepsilon(0) - \varepsilon_{\infty}}{\sqrt{\varepsilon(0)}} \frac{\omega^2 \tau}{2}.
\end{equation}

\noindent For $\omega^2 \tau^2 \gg 1$ the dielectric function reduces to
\begin{equation}
\label{eq95}
{\rm Re}(\varepsilon) \approx  \varepsilon_{\infty}
\end{equation}      

\noindent and
 \begin{equation}
\label{eq96}
{\rm Im}(\varepsilon) \approx   [\varepsilon(0) - \varepsilon_{\infty}] \,  \frac{1}{\omega \tau},
\end{equation} 

\noindent while the dispersion relation becomes
\begin{equation}
\label{eq97}
c k \approx  \omega \sqrt{\varepsilon_{\infty}} + i \frac{\varepsilon(0) - \varepsilon_{\infty}}{\sqrt{\varepsilon_{\infty}}} \frac{\tau}{2}.
\end{equation}

\noindent Notice in this limit that Im$(k)$ is independent of $\omega$, as can be observed in Fig.~\ref{fig6}(b).

\subsection{Canonical Dielectric Functions}

To summarize so far, we have discussed in detail examples of the dielectric response associated with three different types of polarization that can be induced in a solid:  polarization due to (i) optic-phonon excitation, (ii) free carriers, and (iii) orientable permanent dipoles.  As a point of reference in moving forward, here we collect the canonical dielectric functions associated with each of these types of material response.  For optic-phonons we have the damped harmonic-oscillator dielectric function
\begin{equation}
\label{eqHO}
\varepsilon_{h} ( \omega ) =  {\varepsilon_{\infty} +  {\frac{\omega_{p0}^2}{ {\omega_0^2 - \omega^2} - i \gamma_0 \omega}} },
\end{equation}

\noindent for free carriers the Drude dielectric function
\begin{equation}
\label{eqDrude}
\varepsilon_f ( \omega ) =  \varepsilon_{\infty} -   \frac{\omega_p^2}{\omega^2} \, \frac{1}{ 1 + i / (\omega \tau)},
\end{equation}

\noindent and for dipoles the Debye dielectric function
\begin{equation}
\label{eqDebye}
\varepsilon_{\rm D} ( \omega ) =  \varepsilon_{\infty} +  {\frac{\varepsilon(0) - \varepsilon_{\infty}}{ 1 - i \omega \tau}}.
\end{equation}

\noindent Note that each of these dielectric functions includes the response of interband excitations as the constant term $\varepsilon_{\infty}$.  In the (strictly theoretical) limit that such excitations are nonexistent, $\varepsilon_{\infty}$ can be replaced by 1.

\subsection{Experimental Examples}

Here we discuss the measured dielectric response of three different types of materials as (at least approximate) examples of the dielectric functions in Eqs.~(\ref{eqHO}) -- (\ref{eqDebye}).  We start with sodium chloride (NaCl), which provides a beautiful example of the optic phonon response in an ionic crystal.  We next discuss lead (Pb) and gold (Au) as examples of a good Drude metal.  Lastly, we illustrate Debye polarization response with data from erbium (Er) doped calcium fluoride (CaF$_2$).

\begin{figure}
\includegraphics[scale=0.62]{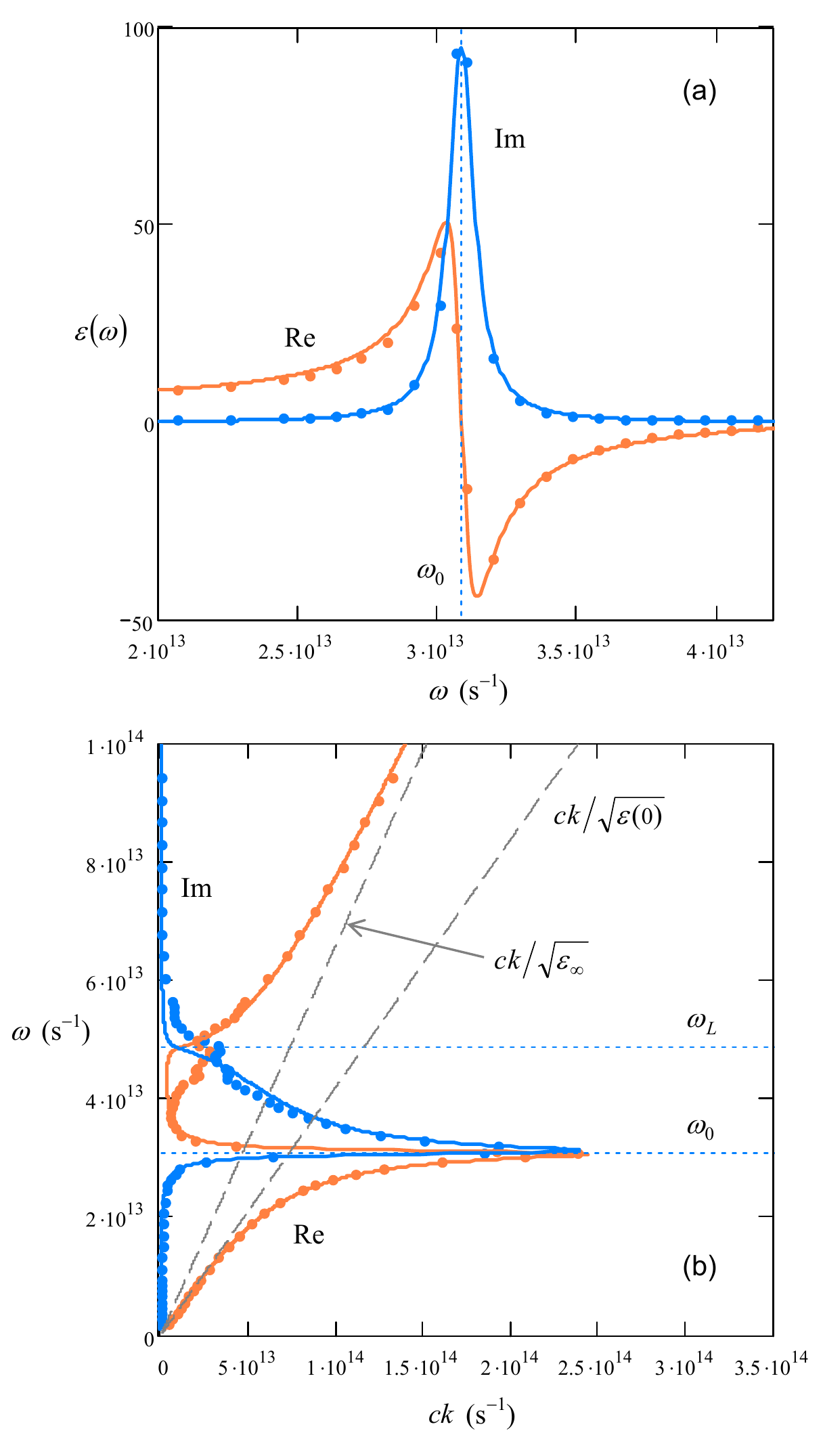}
\caption{\label{fig7}Dielectric function (a) and dispersion curves (b) for sodium chloride (NaCl).  In (a) and (b) the Re and Im parts of $\varepsilon$ and $k$ are plotted, respectively.  Experimental data (filled circles) along with fit (solid lines) using damped harmonic oscillator model are shown; see text for details.}
\end{figure}

\begin{figure*}
\includegraphics[scale=0.77]{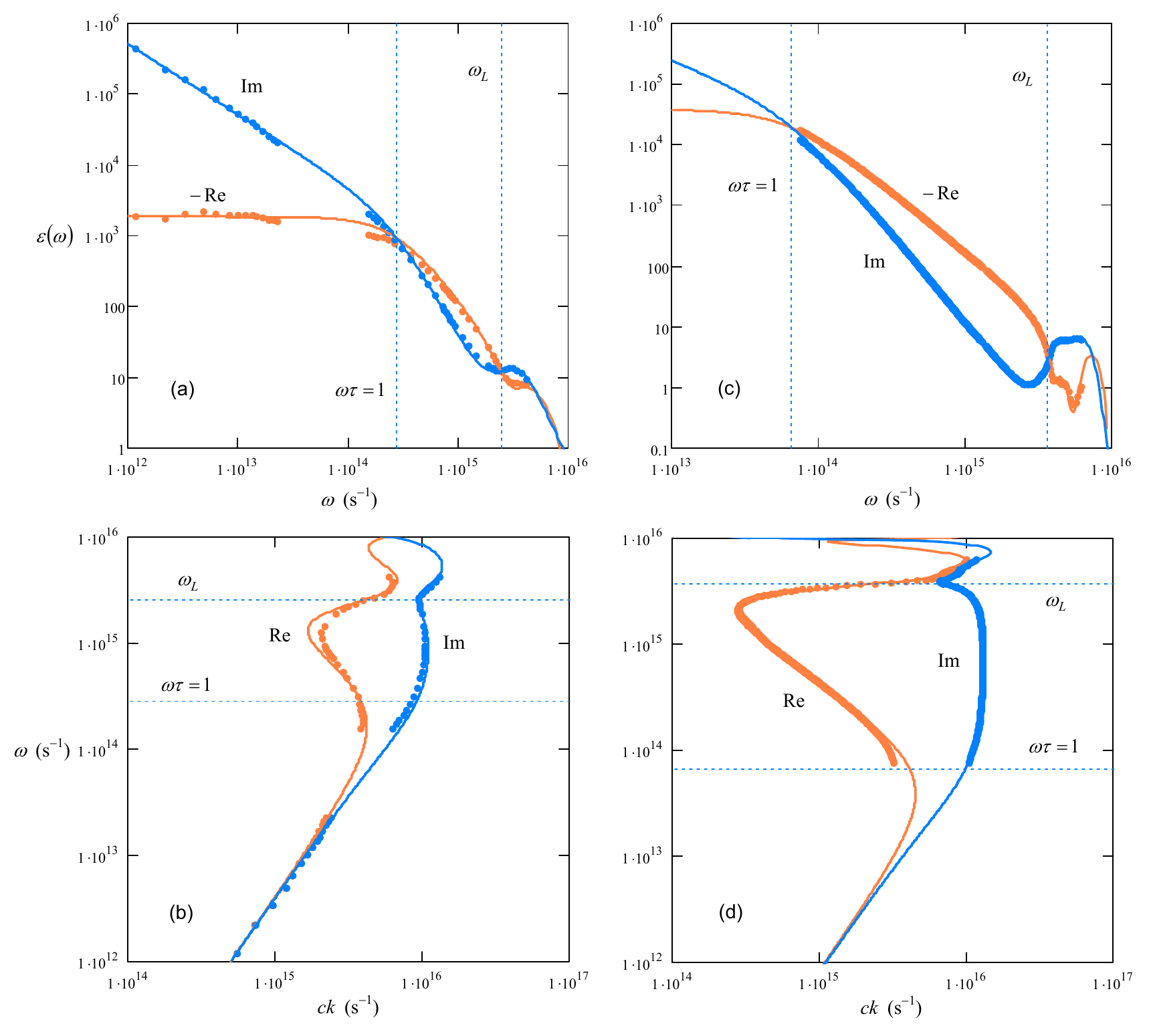}
\caption{\label{fig8}Dielectric function and dispersion curves for Pb (a,b) and Au (c,d).  In (a,c) and (b,d) the Re and Im parts of $\varepsilon$ and $k$ are plotted, respectively.  Experimental data (filled circles) along with fits (solid lines) using Drude model (plus harmonic-oscillator components at high frequencies) are shown; see text for details.}
\end{figure*}

\begin{figure}
\includegraphics[scale=0.60]{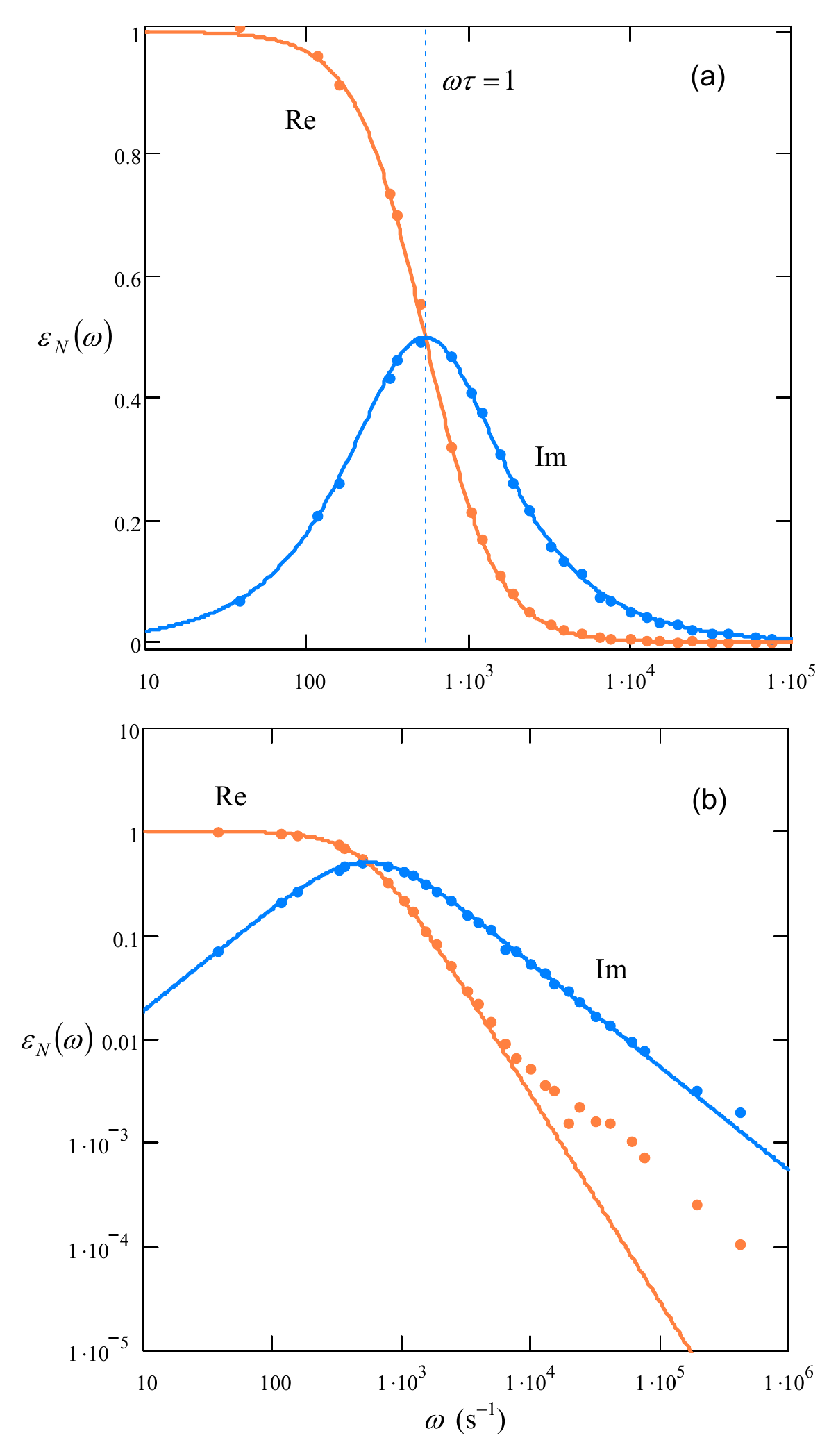}
\caption{\label{fig9}Normalized dielectric function $\varepsilon_N(\omega)$ = $ (\varepsilon(\omega) - \varepsilon_{\infty}) / (\varepsilon(0) - \varepsilon_{\infty})$ of erbium (Er) doped calcium fluoride (CaF$_2$) at very low frequencies.  Experimental data (filled circles) along with fit (solid lines) using Debye mode are shown; see text for details}
\end{figure}

\subsubsection{NaCl}

Figure \ref{fig7} plots both the dielectric function and dispersion curves of NaCl in the vicinity of the optic phonon.  The experimental data are the solid circles \cite{Palik1985}.  These data are compared with theoretical curves that come from the damped harmonic-oscillator dielectric function given by Eq.~(\ref{eqHO}).  The parameters in the model are $\varepsilon_{\infty}$ = 2.32, $\omega_{p0}$ = $5.7 \times 10^{13}$ s$^{-1}$, $\omega_{0}$ = $3.09 \times 10^{13}$ s$^{-1}$  [which together imply $\omega_L$ = $4.85 \times 10^{13}$ s$^{-1}$; see Eq.~(\ref{eq75bb})], and $\gamma_0$ = $11.1 \times 10^{11}$ s$^{-1}$.  As the figure shows, this simple model describes the overall measured response quite well.  The exception occurs in the vicinity of $\omega_L$, where there is appreciable deviation in the dispersion curves, evident in part (b) of the figure.  This deviation is due to the damping parameter $\gamma_0$ having a strong dependence on $\omega$ in this region \cite{Eldridge1977}, which is certainly not accounted for in the model.

\subsubsection{Pb and Au}

Nice experimental examples of the Drude model as applied to a good metal are provided by the elements Pb and Au, illustrated in Fig.~\ref{fig8}.  Again, the solid circles are experimental data [from \cite{Brandli1972,Golavashkin1968} (Pb) and \cite{Olmon2012} (Au)].  The solid lines are model fits, described in detail below.

Before discussing the low-frequency Drude-like behavior of these two metals, we must first address the high-frequency response, which has significant (nonconstant) contributions from interband transitions.  In Pb and Au (and essentially all other metals), some of these transitions have resonant frequencies in the vicinity of what would otherwise be the crossover to transparency at $\omega_L$.  Owing to the presence of these  excitations, the lower end of the transparency region is pushed to higher frequencies.  We can, however, define a phenomenological $\omega_L$ as the frequency where $-$Re($\varepsilon$) = Im($\varepsilon$), as is illustrated in Fig.~\ref{fig8}.  This gives a good estimation of the frequency below which the free carriers dominate the optical response of the metal.  For Pb and Au this phenomenological $\omega_L$ is 2.5 $\times$ 10$^{15}$ Hz and 3.8 $\times$ 10$^{15}$ Hz, respectively.  Both of these frequencies are in the visible part of the spectrum.

For $\omega \lesssim \omega_L$ the data from both metals are well described by the Drude dielectric function [Eq.~(\ref{eqDrude})] with $\omega_p = 1.20 \times 10^{16}$ s$^{-1}$ ($1.29 \times 10^{16}$ s$^{-1}$) and $\tau = 3.58 \times 10^{-15}$ s  ($15.0 \times 10^{-15}$ s) for Pb (Au).  We note that $\omega_L \tau$ equals 8.9 for Pb and 57 for Au, putting Pb just within and Au well within the good-metal category defined by $\omega_L \tau \gg 1$.  The solid lines in Fig.~\ref{fig8} are calculated using the Drude dielectric function plus some number of harmonic-oscillator modes to describe the interband transitions.  Similar to Pb and Au, many other elemental metals display infrared dielectric functions characteristic of a good metal \cite{Rakic1998}.

\subsubsection{Er doped CaF$_2$}

For our last example, we consider the very low frequency response of Er doped CaF$_2$ \cite{Jonscher1980}.  The impurity Er atoms create dipolar complexes that response in a very Debye-like manner to an electric field.  The data shown in Fig.~\ref{fig9} are for a doping level of 0.01\%, which results in an average spacing between Er atoms of 22 lattice spacing.  Interactions between the complexes should thus be minimal, a necessary requirement for applicability of the Debye model.  As Fig.~\ref{fig9} illustrates, except at the very highest frequencies shown the experimental data (solid circles) are described quite well by the Debye dielectric function [Eq.~(\ref{eqDebye})] (solid lines).  Notice that the relevant frequencies are quite low; the peak in Im$(\varepsilon)$ is at $\omega =  550$ s$^{-1}$.

\section{Conductivity}

\subsection{Relationship to the Dielectric Function}

One of the most important properties of a material is the conductivity $\sigma$.  In general, this response function is the quantity that connects the current density \textbf{j} in the material to the electric field \textbf{E}.  The linear-response ansatz that $\sigma$ is a simple proportionality constant,
\begin{equation}
\label{eq98}
{\rm \bf j (r},t) = \sigma \, {\rm \bf E(r},t),
\end{equation}

\noindent is known as \textbf{Ohm's law}.  This relationship is akin to the simple notion that $\varepsilon$ and $\mu$ are both constants for a given material.  However, as we have seen with regards to the dielectric function $\varepsilon$, the conductivity $\sigma$ generally has some frequency dependence.\footnote{Indeed, in general $\mu$ is also frequency dependent.  However, we are only considering materials with negligible response to magnetic fields.}

In fact, the conductivity $\sigma(\omega)$ is intimately related to the dielectric function $\varepsilon(\omega)$.  To see this we start by defining $\sigma(\omega)$ via
\begin{equation}
\label{eq99}
{\rm \bf \ \tilde j (r)} = \sigma(\omega) \, {\rm \bf \tilde E(r)}.
\end{equation} 

\noindent This equation is entirely analogous to the expression [Eq.~(\ref{eq57})] that defines $\varepsilon(\omega)$.  As has been the case in all of our discussion so far, we treat $\rho_{\rm P}$ as comprising all of the charge in the material.  With this viewpoint, Eq.~(\ref{eq11}), which relates the current density to the polarization, transforms into
\begin{equation}
\label{eq100}
{\rm \bf \ \tilde j (r)} = -i \omega \, {\rm \bf \tilde P(r)}.
\end{equation} 

\noindent If we combine Eqs.~(\ref{eq57}) and  (\ref{eq65}) we can write

\begin{equation}
\label{eq101}
{\rm \bf \ \tilde P(r)} = \varepsilon_0 \, \big[\varepsilon(\omega) - 1 \big] \, {\rm \bf \tilde E(r)}.
\end{equation}

\noindent This expression allow us to eliminate the polarization ${\rm \bf \tilde P(r)}$ from Eq.~(\ref{eq100}), yielding
\begin{equation}
\label{eq102}
{\rm \bf \ \tilde j (r)} = -i \omega \, \varepsilon_0 \, \big[\varepsilon(\omega) - 1 \big] \, {\rm \bf \tilde E(r)}.
\end{equation}

\noindent  Comparing this equation with Eq.~(\ref{eq99}) reveals the \textit{general} relationship between the conductivity and dielectric function,
\begin{equation}
\label{eq109}
\sigma(\omega) = -i \omega \, \varepsilon_0 \, \big[\varepsilon(\omega) - 1 \big].
\end{equation}

\noindent This result is important because it shows that the conductivity and dielectric function are not independent quantities.  Rather, they are just two different ways of expressing the response of the charge in a material to an electric field.

\begin{figure*}
\includegraphics[scale=0.90]{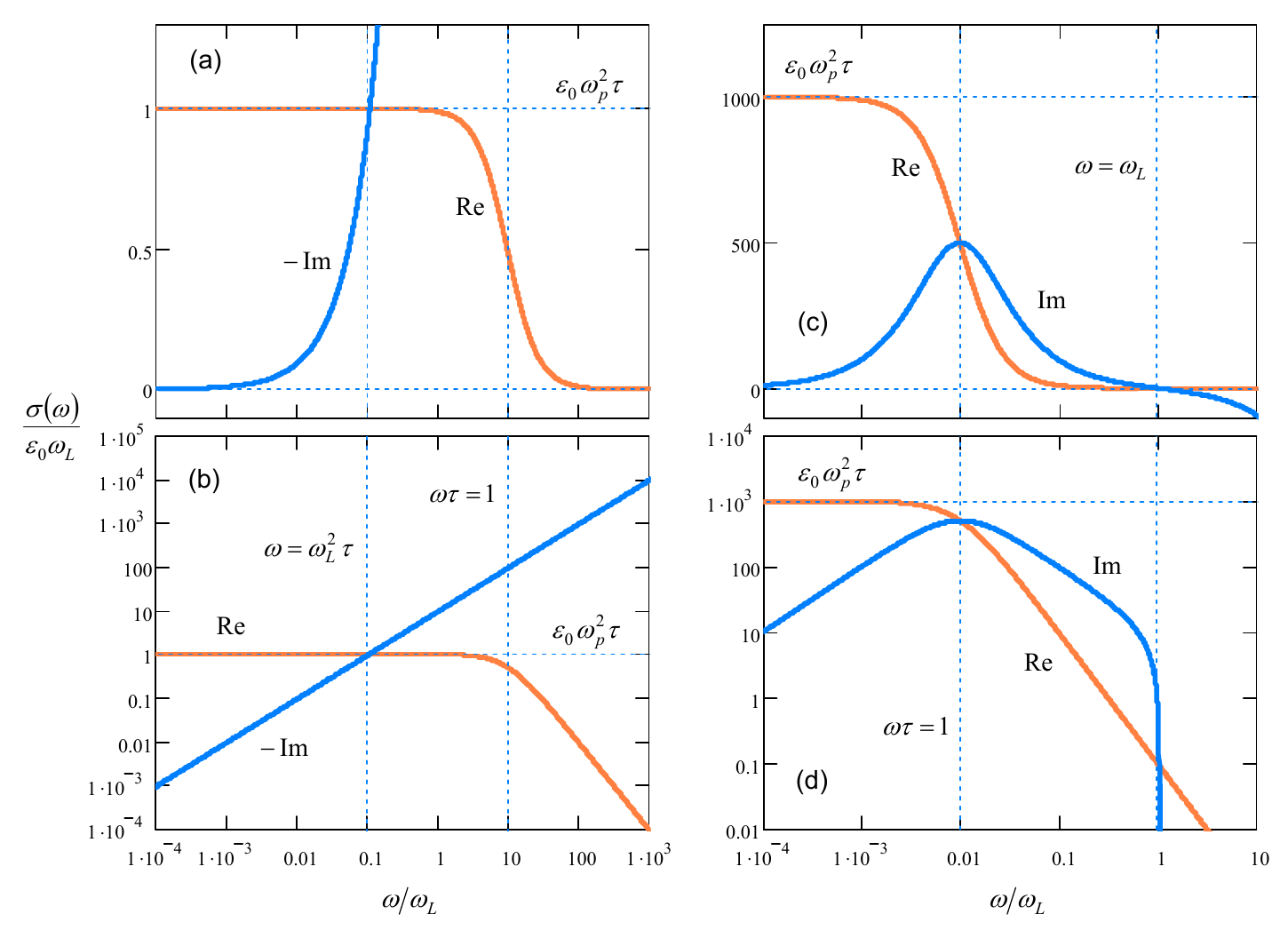}
\caption{\label{fig12} Normalized total conductivity $\sigma(\omega) / (\varepsilon_0 \, \omega_L)$ versus frequency $\omega$.  A poor [good] conductor is illustrated in (a) and (b) [(c) and (d)].  As in Fig.~\ref{fig5}, for the poor (good) conductor $\omega_L \tau$ = 0.1 (100) and $\varepsilon_{\infty} = 10$.}
\end{figure*}

\begin{figure*}
\includegraphics[scale=0.68]{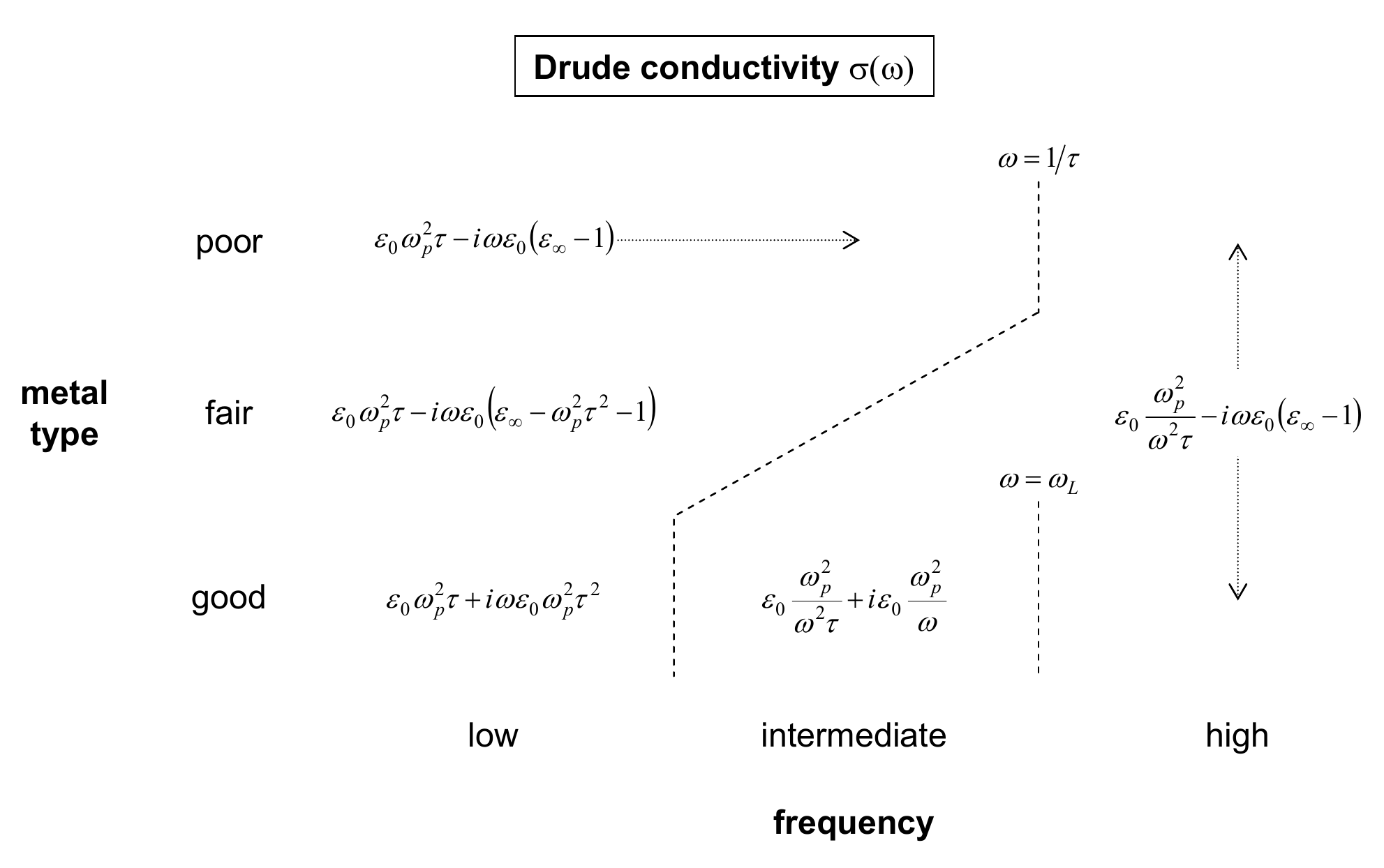}
\caption{\label{fig13}Approximate expressions for total conductivity $\sigma(\omega)$ of poor ($\omega_L \tau \ll 1$), fair ($\omega_L \tau \sim 1$), and good ($\omega_L \tau \gg 1$) conductors in low, intermediate, and high frequency regions.}
\end{figure*}

Let's see what this implies for a Drude conductor, which has the dielectric function given by Eq.~(\ref{eqDrude}).  Analogous to Fig.~\ref{fig5} where we plot $\varepsilon(\omega)$, in Fig.~\ref{fig12} we plot the conductivity $\sigma(\omega)$ for examples of both poor and good conductors.  Furthermore, in Fig.~\ref{fig13} we present approximate expressions for the conductivity $\sigma(\omega)$ that are analogous to those in Fig.~\ref{figDE} for the dielectric function. \marginpar{\footnotesize{$\mathbb{EX} \,$\ref{E20}}}

The first thing to note is that in the low frequency regions ($\omega \ll \omega_L^2 \tau$ for a poor metal,  $\omega \ll 1 / \tau$ for a good metal) the real part of the conductivity dominates the imaginary part, with the result that to good approximation the conductivity is simply given by its dc limit   
\begin{equation}
\label{eq110}
\sigma_f(0) = \varepsilon_0 \, \omega_p^2 \, \tau.
\end{equation}

\noindent The subscript $f$ has been added to $\sigma$ to indicate that this response is solely due to the free carriers.  If we now use the relationship $\omega_p^2 = (N_c \, e^2) / (\varepsilon_0 \, m^*)$, we obtain the classic result for the \textbf{Drude dc conductivity}
\begin{equation}
\label{eq111}
\sigma_f(0) = \frac{N_c \, e^2 \, \tau}{m^*}.
\end{equation}

\noindent From dc transport theory the relaxation time $\tau$ can be identified as the momentum relaxation time associated with the free carriers \cite{Ashcroft1976}.

The conductivity is a bit more interesting in the intermediate- and high-frequency regions. For poor conductors the imaginary part of the conductivity
\begin{equation}
\label{eq112}
{\rm Im}(\sigma) \approx -i \omega \varepsilon_0 ( \varepsilon_{\infty} -1 )
\end{equation}

\noindent dominates the real part.  Notice it is negative and due only to the bound charge characterized by $\varepsilon_{\infty}$.  This is also the dominant contribution to Im$(\sigma)$ for fair and good conductors at high frequencies.  For a good conductor at intermediate frequencies the imaginary part also dominates the real part, but here Im$(\sigma) > 0$ and is due to the free carriers.  Notice that Re$(\sigma)$ is always $>$ 0.

Analogous to Eq.~(\ref{eq85e}) for the dielectric function of a good conductor in the intermediate- and low-frequencies regimes, there is an analogous expression for the conductivity, \marginpar{\footnotesize{$\mathbb{EX} \,$\ref{E21}}}
\begin{equation}
\label{eq113}
\sigma_f(\omega) = \varepsilon_0 \, \omega_p^2 \tau \, \frac{1}{1 - i \omega \tau}.
\end{equation}

\noindent This is often referred to as the \textbf{Drude ac conductivity}.  Again, the subscript $f$ denotes that this conductivity is entirely from the free carriers.

\begin{figure}
\includegraphics[scale=0.635]{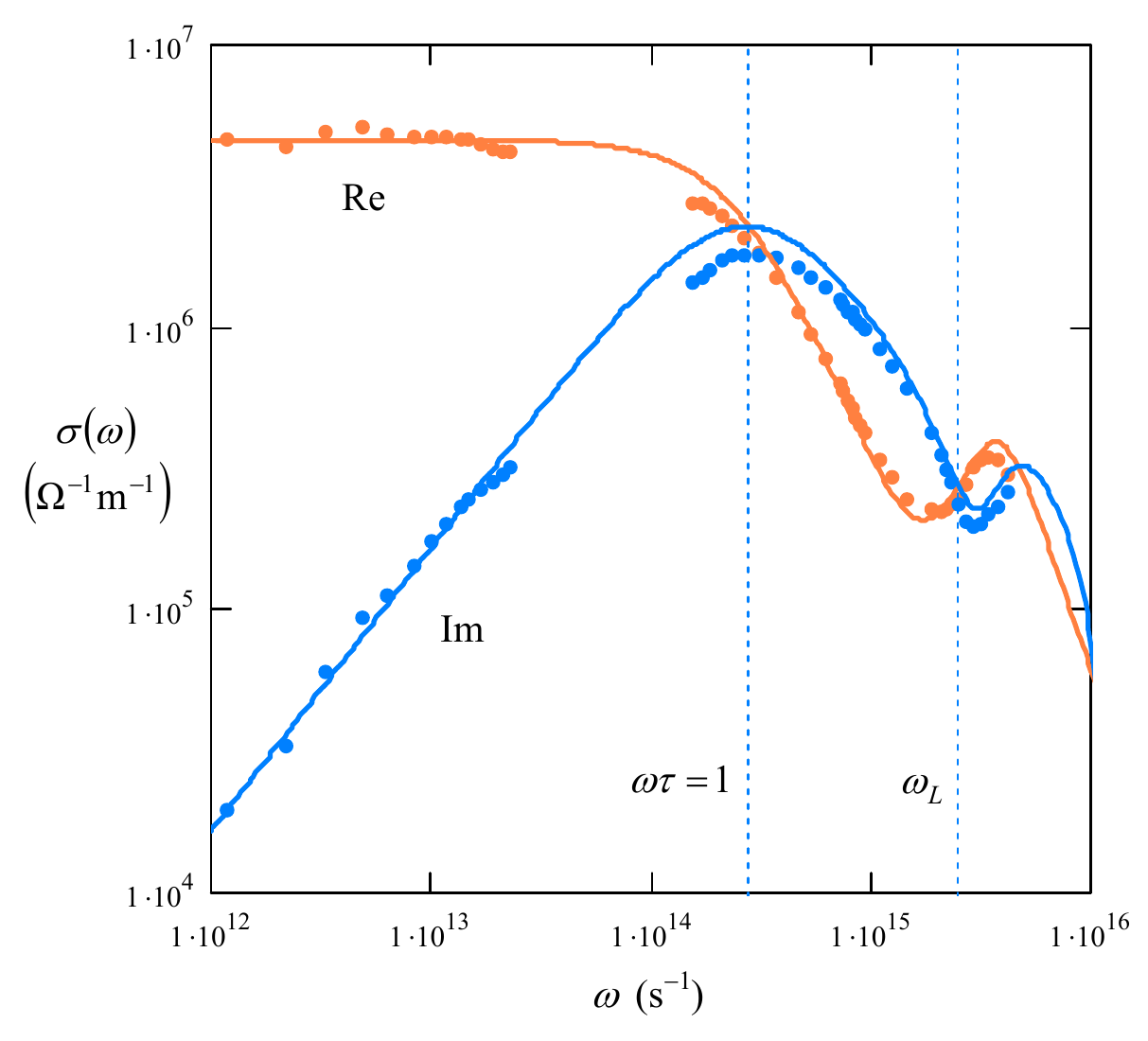}
\caption{\label{fig10}Conductivity $\sigma(\omega)$ of lead (Pb).  Experimental data (filled circles) along with fit (solid lines) using same Drude/harmonic-oscillator model as in Fig.~\ref{fig8}; see text for details.}
\end{figure}

To see what $\sigma(\omega)$ looks like for an actual metal, in Fig.~\ref{fig10} we plot the conductivity of Pb obtained from the dielectric function data (solid circles) and model (solid lines) shown in Fig.~\ref{fig8}.  Notice for $\omega \tau \ll 1$ (in this case frequencies up to $\sim$10$^{14}$ s$^{-1}$) that, as expected, the real part of the conductivity is essentially constant and dominates the imaginary part.  Indeed, in this frequency region Re$(\sigma)$ = 4.56 $\times 10^{6}$ $\Omega^{-1}$m$^{-1}$ corresponds to a resistivity $\rho \, (= 1 / \sigma)$ of 21.9 $\times$ 10$^{-8}$ $\Omega$ m.  This is quite close to the experimental dc resistivity of 20.6 $\times$ 10$^{-8}$ $\Omega$ m \cite{Emsley1993}. 

By comparison, if we start with the dielectric function for a pure, crystalline insulator [Eq.~(\ref{eqHO})] or a collection of Debye dipole moments [Eq.~(\ref{eqDebye})], we end up with $\varepsilon(0)$ equal to a finite constant.  With this result Eq.~(\ref{eq109}) tells us that $\sigma(0) = 0$, as expected for a material with no free charge.

\subsection{Free and Bound Charge Response}    

Our approach of lumping all material charge into $\rho_{\rm p}$ is rather canonical, at least within the research literature.   However, another convention is to keep the responses of the free and bound charges separate from each other, using the dielectric function to account for the bound charge and the conductivity to account only for the free carriers.  This approach can be found in some undergraduate textbooks; see for example \cite{Griffiths2013}.  In this scheme $\rho_{\rm p}$ and \textbf{j}$_{\rm p}$ refer only to bound charge while $\rho_{other}$ and \textbf{j}$_{other}$ are set equal to the free-charge density $\rho_f$ and current density \textbf{j}$_{f}$, respectively.  Although this separation of charge may be impossible to distinguish experimentally, it can certainly be maintained theoretically.  In this section we explore this alternative point of view, in part to show how it connects to our approach above where these two types of charge are formally undifferentiated.

We start by first identifying the response functions.  Because the dielectric function is to be associated solely with the bound charge in the material, we use $\varepsilon_b(\omega)$ to represent this response.  The displacement field is thus defined via
\begin{equation}
\label{eq114}
{\rm \bf \tilde D} = \varepsilon_0 \varepsilon_b(\omega)\, {\rm \bf \tilde E}.
\end{equation}

\noindent It is important to note that because the free charge is not included in the dielectric function, the displacement field  in this formulation is \textit{not} the same as our previous field ${\rm \bf \tilde D}$.  This is apparent in Gauss' law below.  Likewise, we represent the free-electron conductivity with $\sigma_f(\omega)$, which is defined via 
\begin{equation}
\label{eq115}
{\rm \bf \tilde j}_f = \sigma_f(\omega) \, {\rm \bf \tilde E}.
\end{equation}

We now derive the Helmholtz equation for the electric field.  With distinct free and bound charge, Maxwell's equations with harmonic time dependence can be written in terms of the fundamental fields as \marginpar{\footnotesize{$\mathbb{EX} \,$\ref{E22}}}
\begin{equation}
\label{eq116}
\varepsilon_0 \, \varepsilon_b(\omega) \nabla \cdot {\rm {\bf \tilde E}} = {\tilde \rho}_f,
\end{equation}

\begin{equation}
\label{eq117}
\nabla \cdot {\rm {\bf \tilde B}} = 0,
\end{equation}

\begin{equation}
\label{eq118}
\nabla \times {\rm {\bf \tilde E}} = i \omega {\rm {\bf \tilde B}},
\end{equation}

\noindent and
\begin{equation}
\label{eq119}
\nabla \times {\rm {\bf \tilde B}} =  \mu_0  \big[ \sigma_f(\omega) - i \omega \varepsilon_0  \varepsilon_b(\omega) \big] {\rm {\bf \tilde E}}.
\end{equation}

\noindent With manipulations similar to those previously carried out, the last two equations can be combined to yield

\begin{equation}
\label{eq120}
\nabla ( \nabla \cdot {\rm {\bf \tilde {E}}} ) - \nabla ^2{\rm {\bf \tilde {E}}} = \mu _0  \big[ \omega^2 \varepsilon_0 \varepsilon_b(\omega) + i \omega \, \sigma_f(\omega) \big] {\bf \tilde {E}}.
\end{equation}

\noindent  We previously showed for oscillations at a real frequency $\omega$ that $\nabla \cdot {\rm \bf \tilde E} =0$;\footnote{Notice $\nabla \cdot {\rm \bf \tilde E} =0$ and Eq.~(\ref{eq116}) also imply ${\tilde\rho}_f = 0$.} we can thus simplify this last equation to
\begin{equation}
\label{eq121}
\nabla ^2{\rm {\bf \tilde {E}}} = -\mu_0 \varepsilon_0 \, \bigg[  \varepsilon_b(\omega) + i \, \frac{\sigma_f(\omega)}{\varepsilon_0 \, \omega} \bigg] \, \omega^2 \, {\bf \tilde {E}},
\end{equation} 

\noindent an alternative form of the Helmholtz equation for ${\bf \tilde {E}}$.  Comparing this with our original version of the wave equation [Eq.~(\ref{eq57})], we can identify the (complete) dielectric function as
\begin{equation}
\label{eq122}
\varepsilon(\omega) = \varepsilon_b(\omega) + i \, \frac{\sigma_f(\omega)}{\varepsilon_0 \, \omega}.
\end{equation}   

\noindent As the response functions for both free and bound electrons are generally complex, this last equation is  not always particularly enlightening.

However, it does have its appeal if the frequency $\omega$ is low enough that the bound-charge response is simply the constant $\varepsilon_{\infty}$, simplifying Eq.~(\ref{eq122}) to
\begin{equation}
\label{eq123}
\varepsilon(\omega) = \varepsilon_{\infty} + i \, \frac{\sigma_f(\omega)}{\varepsilon_0 \, \omega}.
\end{equation}

\noindent As is obvious, all of the frequency response is now contained in the free-carrier conductivity $\sigma_f(\omega)$.  Indeed, if $\sigma_f(\omega)$ is taken to be the Drude ac conductivity [Eq.~(\ref{eq113})], then $\varepsilon(\omega)$ given by this last equation becomes the Drude dielectric function given by Eq.~(\ref{eqDrude}).  As we shall see below when discussing the decay of charge fluctuations, describing {\it any} interesting frequency-dependent response with a conductivity while leaving the high-frequency response in the term $\varepsilon_{\infty}$ can be quite useful at times.

\section{Wave Propagation -- Further Considerations}

Now that we have examined several dielectric functions, it is appropriate that we go back to EM wave propagation and investigate this topic in more depth.  We first concentrate on the spatial nature of an EM wave inside a material.  Certainly, we have touched on this topic in our discussion of the dispersion relation $\omega(k)$; here we extend this discussion by introducing the complex index of refraction $N = n + i \kappa$.   As we shall see, $N$ is proportional to the wave vector $k$, and so $n$ (the real part of $N$) controls the wavelength while $\kappa$ (the imaginary part) controls the decrease of the wave's amplitude along the direction of propagation.  Next, we introduce the optical impedance $Z$.  This property controls the ratio of the electric field to the magnetic field in an EM wave.  Although not discussed here, the optical impedance controls the reflectivity of EM radiation from the surface of a solid.    

Before delving into these subjects, a few background remarks are in order regarding the exact situation being considered here.   First, we assume that the material of interest [described by $\varepsilon(\omega)$ (and $\mu = 1$)] occupies the half space defined by $z \geq 0$, while $z < 0$ is filled with vacuum. Consequently, the material-vacuum interface is  infinitely sharp.\footnote{Any realistic interface consists of a finite-sized transition region between the two materials of interest.  We shall not consider the consequences of such a transition region.} Second, we assume that an EM wave of frequency $\omega$ traveling in the $+ \hat{z}$ direction is (normally) incident on the ($z=0$) surface of the material.\footnote{Just to be clear, $\hat{z}$ is the unit vector pointing in the $+z$ direction.}  This \textbf{incident wave} generates both a \textbf{reflected wave} (propagating through the vacuum in the $- \hat{z}$ direction) and a \textbf{transmitted wave} (propagating into the material in the $+ \hat{z}$ direction).  Our present interest is in this transmitted wave.

Furthermore, for simplicity we assume that (i) the transmitted wave is linearly polarized and (ii) the electric field points along the $\hat{x}$ direction.  We can then write the electric-field part of the EM wave as 
\begin{equation}
\label{eqWP1}
{\bf E}(z,t) = \hat{x} \, E_0 \,  e^{i(kz-\omega t)},
\end{equation}

\noindent where $k$ and $\omega$ are related via the dispersion relation
\begin{equation}
\label{eqWP2}
k = \frac{\omega}{c} \, \sqrt{\varepsilon(\omega)}.
\end{equation}

\subsection{Index of Refraction}

The \textbf{complex index of refraction} $N = n + i\kappa$ is defined in terms of the wave vector $k$ via

\begin{equation}
\label{eqWP3}
k = \frac{\omega}{c} \, N.
\end{equation}

\noindent A comparison of this equation with Eq.~(\ref{eqWP2}) immediately reveals that (i) the index of refraction is frequecy dependent (no surprise) and (ii) simply related to the dielectric constant;
\begin{equation}
\label{eqWP3b}
N \! (\omega) = \sqrt{\varepsilon(\omega)}.
\end{equation}

Let's see how $N$ determines the spatial behavior of an EM wave in the solid.  Because $\omega / c = 2 \pi / \lambda_0$, where $\lambda_0$ is the vacuum wavelength of an EM wave of frequency $\omega$, we can write
\begin{equation}
\label{eqWP4}
k = \frac{2 \pi}{\lambda_0} \, (n + i\kappa),
\end{equation}

\noindent which allows us to re-express Eq.~(\ref{eqWP1}) as
\begin{equation}
\label{eqWP5}
{\bf E}(z,t) = \hat{x} \, E_0 \, e^{-(2 \pi / \lambda_0) \kappa \,  z} \, e^{i (2 \pi / \lambda_0) n \, z} \, e^{-i\omega t}.
\end{equation}

\noindent Inspection of this equation reveals that $n$ controls the wavelength $\lambda$ (inside the material) and $\kappa$ controls the exponential decay of the electric field as the wave propagates into the solid.  Indeed, writing Eq.~(\ref{eqWP5}) in the more generic form
\begin{equation}
\label{eqWP6}
{\bf E}(z,t) = \hat{x} \, E_0 \, e^{-z / \delta_{\! f}} \, e^{i (2 \pi / \lambda)\, z} \, e^{-i\omega t},
\end{equation}

\noindent we see that the specific relationships are
\begin{equation}
\label{eqWP7}
\frac{\lambda}{\lambda_0} = \frac{1}{n},
\end{equation}

\noindent and
\begin{equation}
\label{eqWP8}
\frac{\delta_{\! f}}{\lambda_0} = \frac{1}{2 \pi \kappa},
\end{equation}

\noindent The parameter $\delta_{\! f}$ is known as the \textbf{field skin depth} of the material.\footnote{The field skin depth is twice as large as the intensity skin depth $\delta_{I}$.  As there is no clear convention as to whether the term  ``skin depth'' refers to the fields or intensity, it is best to be explicit about which quantity is being discussed.  Hence, we use the subscripts $f$ and $I$ to differentiate these two quantities.}  The relative importance of spatial decay vs the wavelength in the material is obtained by taking the ratio of these last two equations,
\begin{equation}
\label{eqWP9}
\frac{\delta_{\! f}}{\lambda} = \frac{n}{2 \pi \kappa}.
\end{equation} 

\noindent  Notice that a material with negligible $\kappa$ compared to $n$ exhibits a decay length that is much longer that a wavelength.  This approximation applies, for example, to a typical glass in the visible region of the spectrum.  The other extreme occurs when $\kappa \gg n$.  As we see below, this applies, for example, to a good conductor in the intermediate frequency region.    

\begin{figure*}
\includegraphics[scale=0.90]{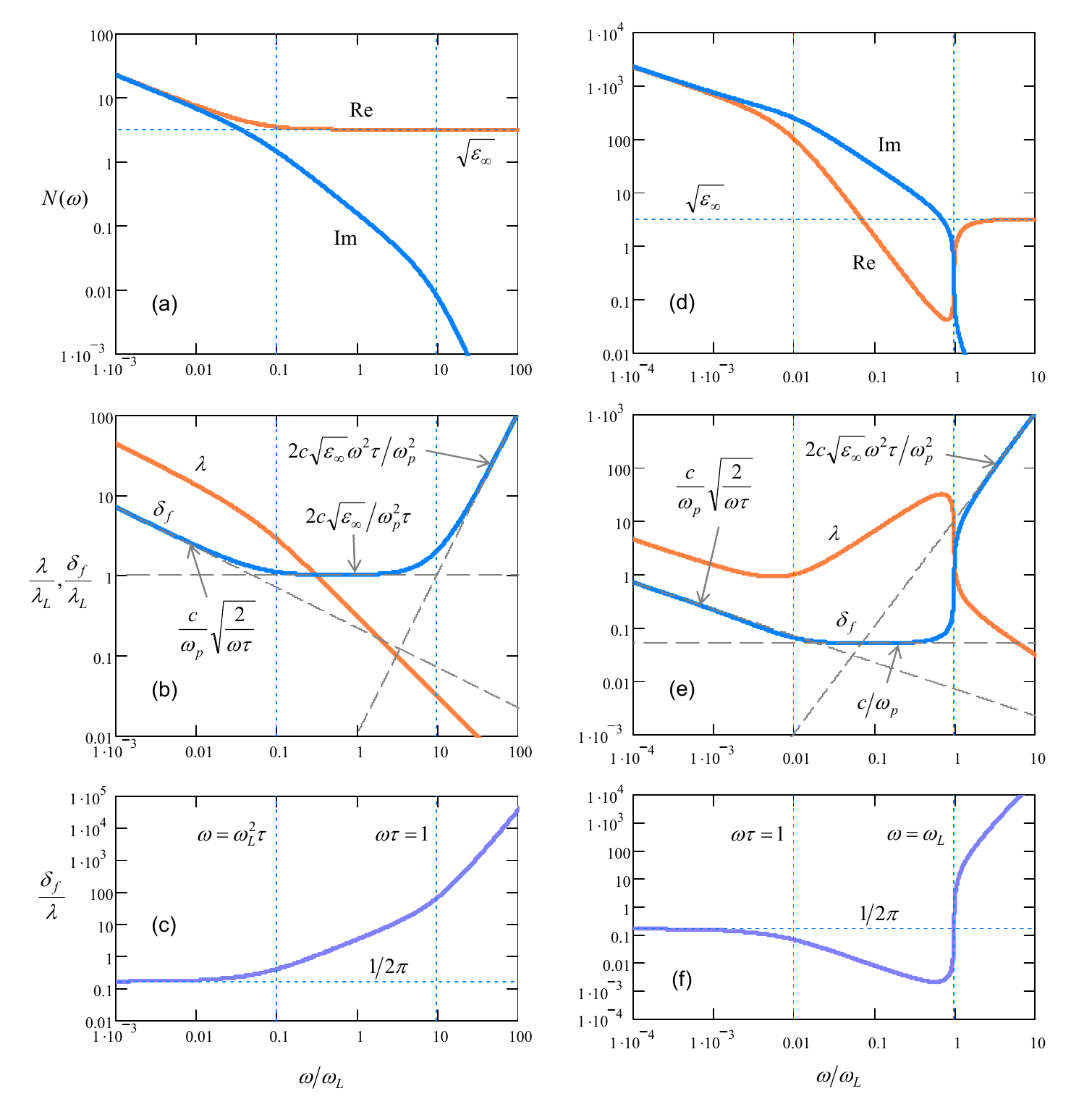}
\caption{\label{fig15}Drude index of refraction $N(\omega) = \sqrt{\varepsilon(\omega)}$ and related quantities.  A poor [good] conductor is illustrated in (a), (b), and (c) [(d), (e), and (f)].  In (a) and (d) the Re and Im parts of $N$ are plotted.  In (b) and (e) the wavelength $\lambda$ and field skin depth $\delta_{\! f}$ are plotted, normalized by $\lambda_L = 2 \pi c / \omega_L$.  In (c) and (f) the ratio $\delta_{\! f} / \lambda$ is plotted.  The dashed lines in (b) and (e) are labeled by approximate expressions for $\delta_{\! f}$ appropriate for each frequency region.  As in Figs.~\ref{fig5} and \ref{fig12}, for the poor (good) conductor $\omega_L \tau$ = 0.1 (100) and $\varepsilon_{\infty} = 10$.}
\end{figure*}

\begin{figure*}
\includegraphics[scale=0.67]{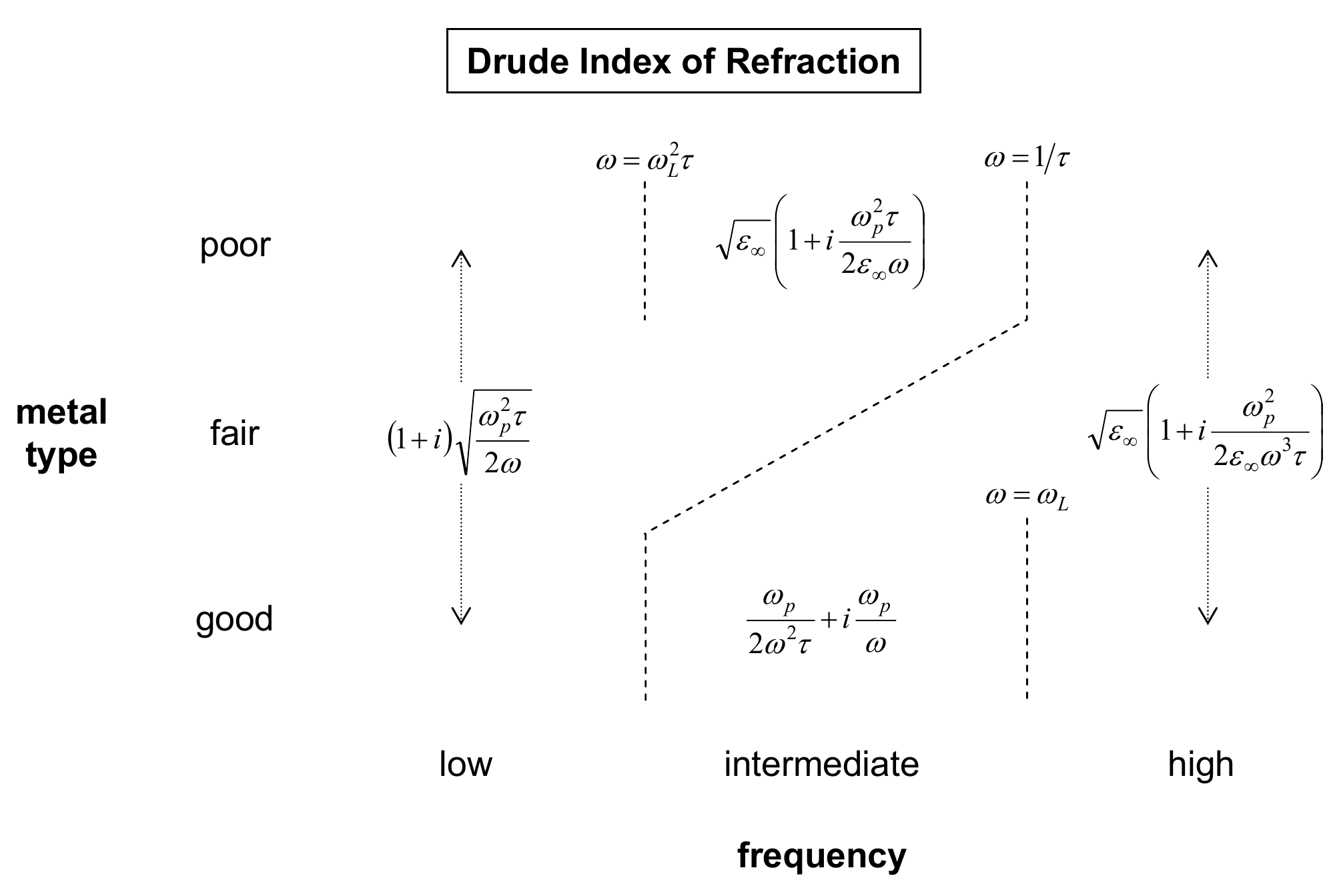}
\caption{\label{fig16}Approximate expressions for index of refraction $N(\omega)$ of poor ($\omega_L \tau \ll 1$), fair ($\omega_L \tau \sim 1$), and good ($\omega_L \tau \gg 1$) conductors in low, intermediate, and high frequency regions.}
\end{figure*}

We now look at $N$ and related quantities for both poor and good conductors.  Similar to our previous figures for the dielectric function (Fig.~\ref{fig5}) and conductivity (Fig.~\ref{fig12}), in Fig.~\ref{fig15} we plot the frequency dependent index of refraction $N \! (\omega) = \sqrt{\varepsilon_{\rm D}(\omega)}$, where the Drude dielectric function $\varepsilon_{\rm D}(\omega)$ is given by Eq.~(\ref{eqDrude}).  Also in this figure the wavelength $\lambda$ and skin depth $\delta_{\! f}$ are plotted, where both are normalized by $\lambda_L = 2 \pi c / \omega_L$, the vacuum wavelength of EM radiation at the longitudinal frequency $\omega_L$.  We also include graphs of $\delta / \lambda$, which gives some indication of the relative transparency of the material.  As we have also previous done for several quantities, in Fig.~(\ref{fig16}) we indicate approximate expressions for $N$ in the low, intermediate, and high frequency regions of poor, fair, and good conductors.  As $N = ck / \omega$, these expressions are trivially obtained from those for $ck$ in Fig.~\ref{figDD}.           

\begin{figure}
\includegraphics[scale=0.635]{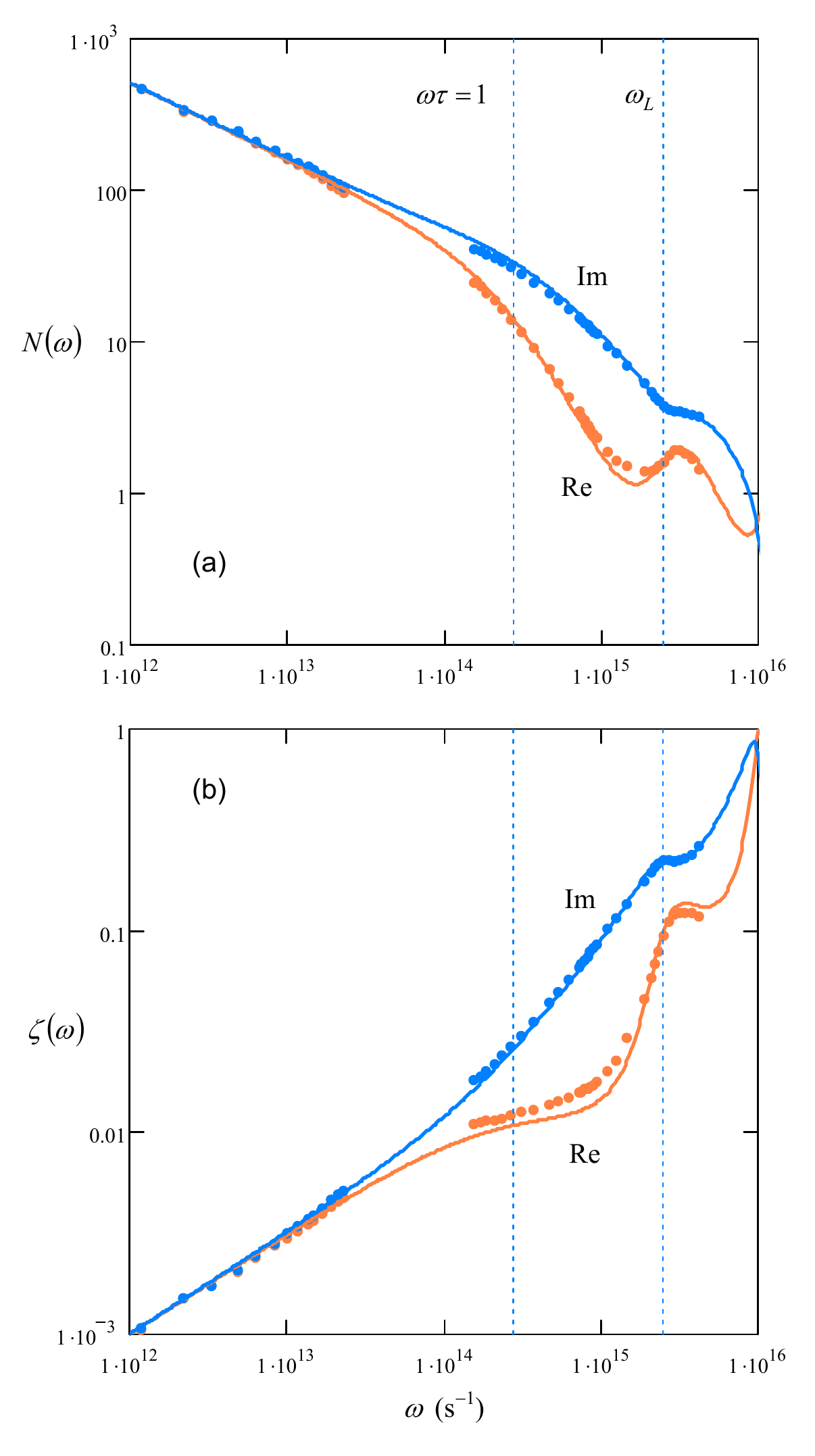}
\caption{\label{fig20}Index of refraction $N \! (\omega)$ (a) and normalized optical impedance $\zeta(\omega)$ (b) of lead (Pb).  Experimental data (filled circles) along with fit (solid lines) using same Drude/harmonic-oscillator model as in Fig.~\ref{fig8}; see text for details.}
\end{figure}

There are several noteworthy observations.  First, at low and high frequencies poor, fair, and good conductors are indistinguishable.  At low frequencies  $n \approx \kappa$ so that $\delta_{\! f} \approx \lambda / (2 \pi)$, resulting in strong attenuation of the wave on a length scale somewhat smaller than one wavelength.  At high frequencies $n \approx \sqrt{\varepsilon_{\infty}}$ while $\kappa \sim 1 / \omega^3$; this makes $\delta_{\! f} \sim \omega^2$, and so with increasing frequency the conductor becomes more transparent.  It is only at intermediate frequencies that differences in behavior are found.  For both poor and good conductors the skin depth is constant in this region, and so the absolute transparency of the material is constant.  However, the ratio of $\delta_{\! f}$ to $\lambda$ is quite different for these two types of conductors:  the ratio increases as $\omega$ in a poor conductor but decreases as $1 / \omega$ for a good conductor.  Thus, relative to the wavelength $\lambda$, with increasing frequency a poor conductor becomes more transparent and a good conductor less transparent.  In a good conductor this behavior produces an abrupt transparency edge at $\omega = \omega_L$.  In a poor conductor the transition to transparency is more gradual.  These last observations are perhaps most apparent in the plots of $\delta_{\! f} / \lambda$ shown in (c) and (f) of Fig.~\ref{fig15}.  

In Fig.~\ref{fig20}(a) we plot the index of refraction for Pb.  Given our previous discussion of Pb, it is not surprising that for frequencies below $\omega_L$ the response is quite well described by a good Drude metal.  

\begin{figure}
\includegraphics[scale=0.70]{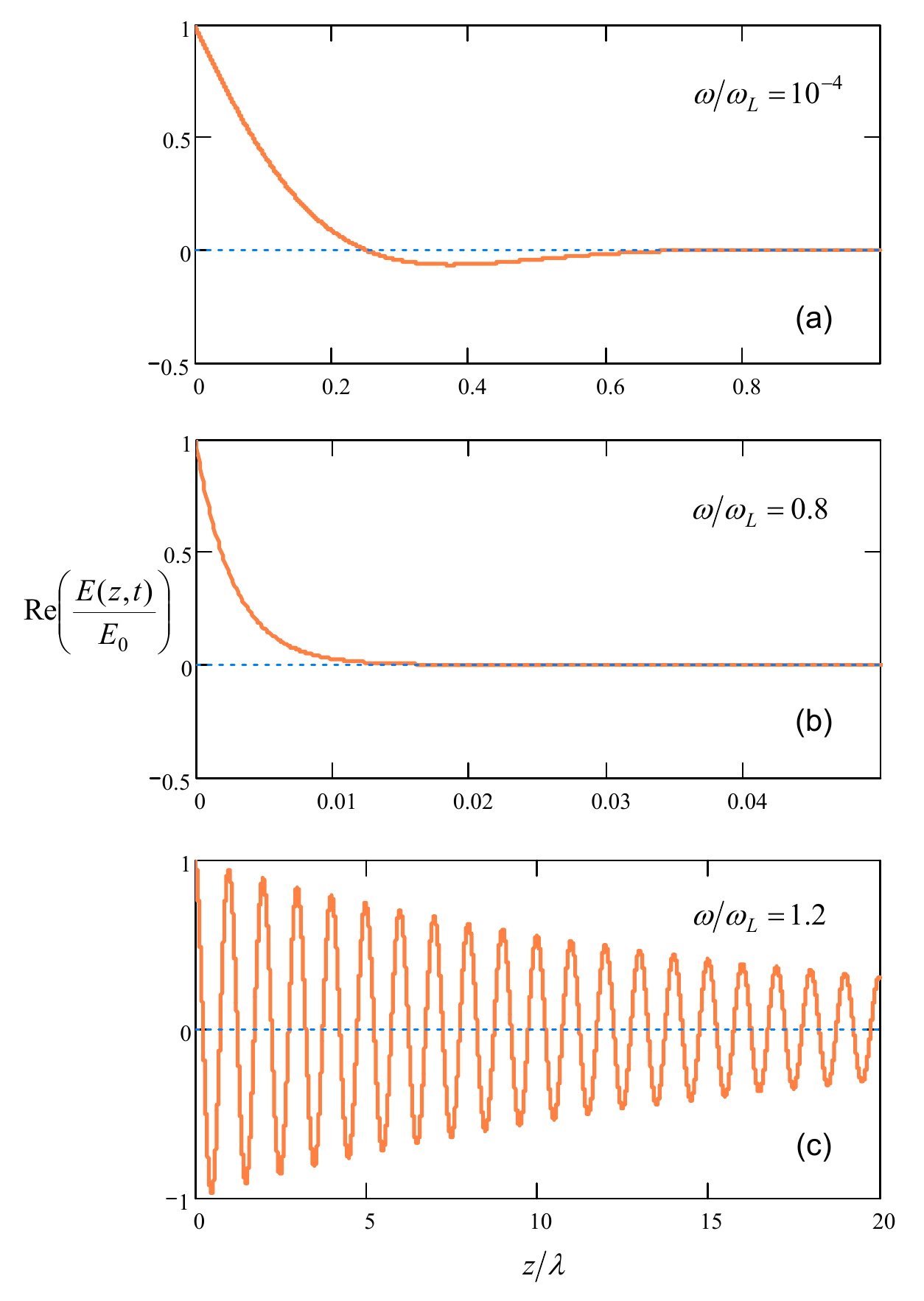}
\caption{\label{fig17}Normalized electric field Re$(E(z,t) / E_0)$ vs normalized distance $z / \lambda$ (at $t=0$) for a good conductor (same parameters as in Fig.~\ref{fig15}) in three different frequency regions:  (a) low frequencies ($\omega / \omega_L = 10^{-4}$), (b) intermediate frequencies ($\omega / \omega_L = 0.8$), (c) high frequencies ($\omega / \omega_L = 1.2$).  Notice different scales for $z / \lambda$.}
\end{figure}

To see how the electric field ${\bf E}(z,t)$ as given by Eq.~(\ref{eqWP5}) varies with distance $z$ into the material, in Fig.~\ref{fig17} we plot the real part of the normalized electric field $E(z,t) / E_0$ as a function of the normalized distance $z / \lambda$ (with $t=0$) for a good conductor.  Examples from the three frequency regimes are illustrated.  In the low frequency region (where $n \approx \kappa$) the field barely oscillates before the amplitude of the field becomes negligible.  In the intermediate region (where $n \ll \kappa$) the electric field is even more strongly damped; here no oscillations are observed.  In contrast to these two regions, at high frequencies (where $n \gg \kappa$) many oscillations in the field are observed within the field skin depth $\delta_{\! f}$.  For the examples illustrated in Fig.~\ref{fig17}, $\delta_{\! f} / \lambda =$ 0.16, 0.0028, and 16 at frequencies given by $\omega / \omega_L =$ 10$^{-4}$, 0.8, and 1.2, respectively.  At $\omega / \omega_L = 1.2$, the damping is still relatively important for all but the thinnest materials.  However, with increasing frequency the relative transparency becomes much stronger; for example, $\delta_{\! f} / \lambda = 3.2 \times 10^{4}$ at $\omega / \omega_L = 10$.  

\begin{figure*}
\includegraphics[scale=0.81]{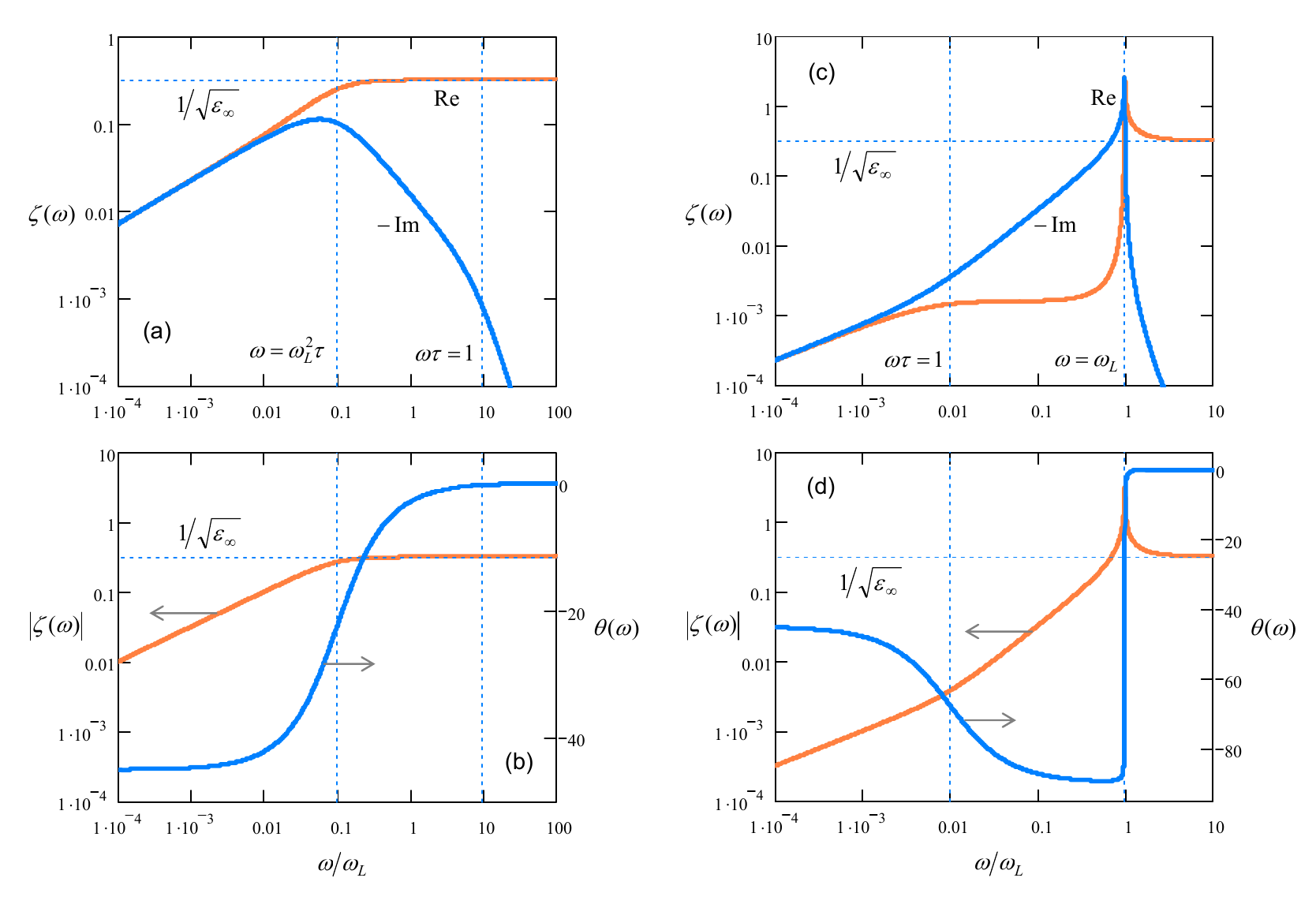}
\caption{\label{fig18}Optical impedance $\zeta(\omega) = 1 / \sqrt{\varepsilon(\omega)}$ associated with Drude dielectric function.  A poor [good] conductor is illustrated in (a) and (b) [(c) and (d)].  In (a) and (c) the Re and Im parts of $\zeta$ are plotted.  In (b) and (d) the modulus $|\zeta(\omega)|$ and phase $\theta(\omega)$ are plotted; see text for details.  As in previous figures, for the poor (good) conductor $\omega_L \tau$ = 0.1 (100) and $\varepsilon_{\infty} = 10$}.
\end{figure*} 

\begin{figure*}
\includegraphics[scale=0.66]{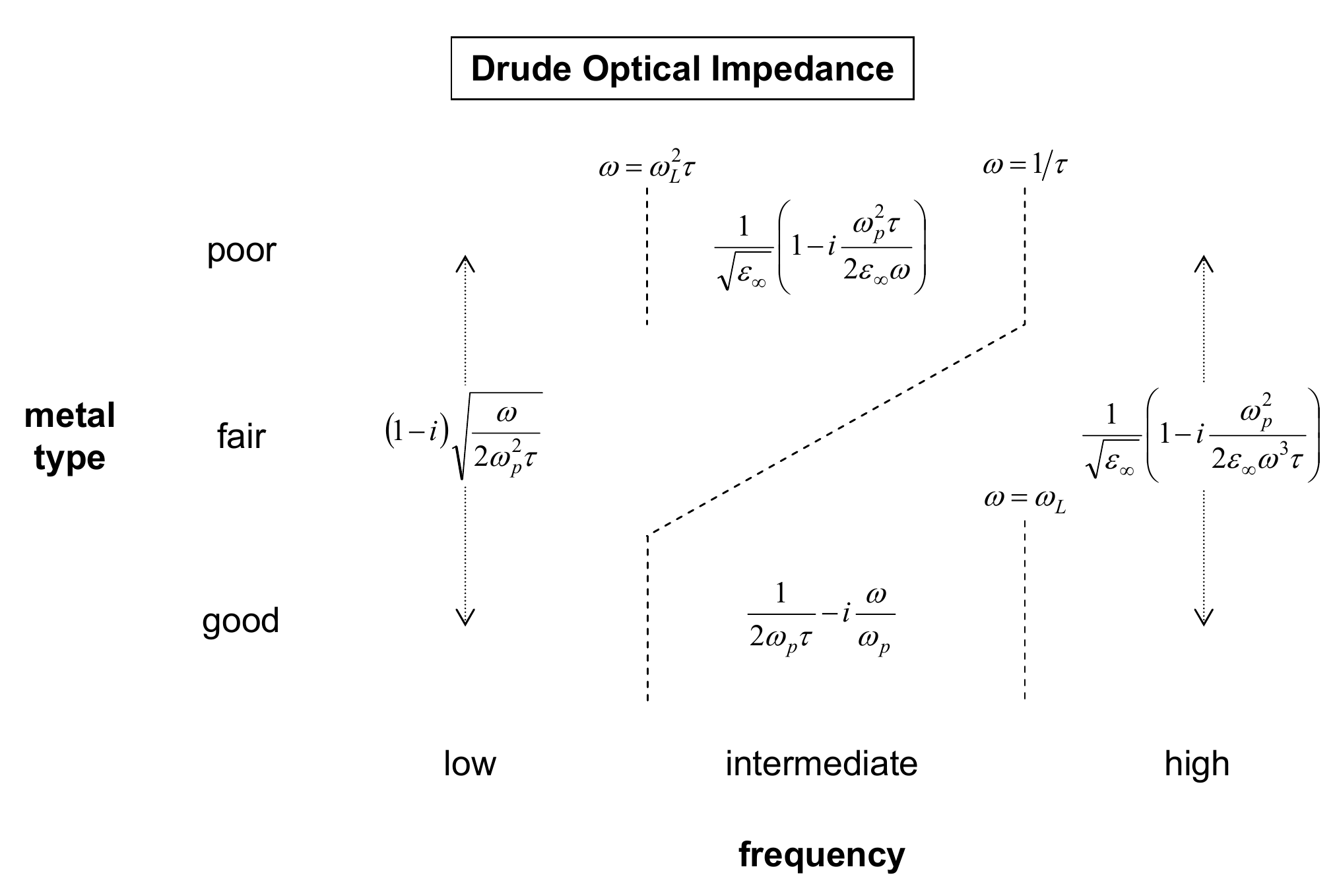}
\caption{\label{fig19}Approximate expressions for optical impedance $\zeta(\omega)$ of poor ($\omega_L \tau \ll 1$), fair ($\omega_L \tau \sim 1$), and good ($\omega_L \tau \gg 1$) conductors in low, intermediate, and high frequency regions.}
\end{figure*}

\subsection{Optical Impedance}

In any EM wave there is not only a propagating electric field, but also an accompanying magnetic field;\footnote{Hence the term -- \textit{electromagnetic} wave.} in this section we consider this magnetic field and its relationship to the electric field.  Following tradition we use the \textbf{H} field rather than the \textbf{B} field, although for nonmagnetic materials that are our present interest, we recall that these fields are related via ${\bf B} = \mu_0 {\bf H}$.  Starting with the electric field given by Eq.~(\ref{eqWP1}), it is not hard to show (using Maxwell's equations) that the \textbf{H} field can be written a
\begin{equation}
\label{eqWP10}
{\bf H}(z,t) = H_0 \, \hat{y} \, e^{i(kz-\omega t)},
\end{equation}  

\noindent where the \textbf{H}-field and \textbf{E}-field amplitudes are related via
\begin{equation}
\label{eqWP11}
\frac{E_0}{H_0} = \sqrt{\frac{\mu_0}{\varepsilon_0}} \, \frac{1}{\sqrt{\varepsilon(\omega)}}.
\end{equation} 

Interestingly, the ratio $Z = E_0 / H_0$ has units of resistance.  For a wave traveling in vacuum $\varepsilon(\omega) = 1$, and so $Z_0 = \sqrt{\mu_0 / \varepsilon_0} \approx 377 \, \Omega$ is known as the \textbf{impedance of free space}.  By extension, the right side of Eq.~(\ref{eqWP4}) is known as the (optical) \textbf{impedance of the material}.  That is, 
\begin{equation}
\label{eqWP12}
Z(\omega) = \sqrt{\frac{\mu_0}{\varepsilon_0}} \, \frac{1}{\sqrt{\varepsilon(\omega)}}.
\end{equation}

 The \textbf{normalized impedance} $\zeta(\omega)$ is defined as the impedance $Z(\omega)$ divided by the free-space impedance $Z_0$, which leads to the simple relations
 \begin{equation}
\label{eqWP13}
\zeta(\omega) = \frac{1}{\sqrt{\varepsilon(\omega)}}
\end{equation}

\noindent and
\begin{equation}
\label{eqWP14}
\frac{E_0}{Z_0 H_0} =  \zeta(\omega).
\end{equation}

In Fig.~\ref{fig19} we plot the normalized optical impedance $\zeta(\omega)$ for both a poor and good conductor.  Perhaps the most interesting features of these curves are associated with the phase $\theta(\omega)$ of the impedance $\zeta(\omega)$, defined via
\begin{equation}
\label{eqWP14}
\zeta(\omega) = |\zeta(\omega)| e^{i \, \theta}.
\end{equation} 

\noindent For both good and poor conductors $\theta \approx - \pi / 4$ at low frequencies and  $\theta \approx 0$ at high frequencies. The difference between conductor types occurs at intermediate frequencies.  As illustrated, in a poor conductor $\theta(\omega)$ smoothly transitions between its two extremes, while in a good conductor $\theta(\omega)$ first approaches $ - \pi / 2$ before abruptly transitioning to values close to zero.

Similar to previous figures for the other response functions, in Fig~\ref{fig19} we provide approximate expressions for the optical impedance of poor, fair, and good Drude metals.  As is the case with the index of refraction $ N(\omega)$ (see Fig.~\ref{fig16}), differences in the expressions only differ in the intermediate frequency regime.   

We conclude this section by noting the graph of optical impedance for Pb in Fig.~\ref{fig20}(b).  A comparison with the analogous graphs in Fig.~\ref{fig18} again illustrates that Pb is a fine example of a good Drude metal.   

\section{Relaxation of Charge Fluctuations}

In our discussions above we have noted several times that for harmonic fields $\nabla \cdot {\tilde {\bf E}} = 0$, implying the important result ${\tilde \rho} =0$.  In some respects this can be viewed as an extension of the well known result $\rho_f =0$ in a conductor at equilibrium.  However, it might certainly be the case that at some point in time an external perturbation induces a nonequilibrium deviation in the charge density $\rho$ within the solid.

Here we consider the relaxation of such nonequilibrium fluctuations in $\rho$.  After building up the requisite mathematical machinery, we first look at the relaxation associated with the free-carrier charge density $\rho_f$ in a conductor.  We might expect a perturbation in $\rho_f$ to die away with some characteristic relaxation time $\tau_R$.  Indeed, in poor conductors this is exactly what happens.  Perhaps surprisingly, in a good conductor oscillations in the density accompany the decay.  We then extend this discussion to disturbances in the charge density associated with a set of dipoles described by the Debye model.  As in the case of a poor conductor, these fluctuations also exponentially die away.  Lastly, we look at charge-density fluctuations associated with optic-phonon modes in a diatomic insulator such as NaCl.  Similar to the free-charge density decay in a good conductor, oscillations also accompany the relaxation to equilibrium.

To facilitate our study of these charge-density fluctuations, we divide the charge as follows:  $\rho_{other}$ comprises the charge density $\rho_i$ of interest ($i$ = $f$, D, or $h$, as appropriate; see below), while $\rho_{\rm p}$ comprises the (remaining) charge density associated with high-frequency electronic response.  Naturally following from this division are a conductivity $\sigma_i(\omega)$ associated with $\rho_i$ and the dielectric constant $\varepsilon_{\infty}$ to describe the high-frequency interband response.  That is, analogous to Eq.~(\ref{eq123}), we think of the overall dielectric responses given in Eqs.~(\ref{eqHO}) -- (\ref{eqDebye}) as \marginpar{\footnotesize{$\mathbb{EX} \,$\ref{E23}}}
\begin{equation}
\label{eq126}
\varepsilon(\omega) = \varepsilon_{\infty} + i \, \frac{\sigma_i(\omega)}{\varepsilon_0 \, \omega},
\end{equation}        

\noindent where $\sigma_i$ is either the free-carrier Drude ac conductivity
\begin{equation}
\label{eq127}
\sigma_f(\omega) = \varepsilon_0 \, \omega_p^2 \tau \, \frac{1}{1 - i \omega \tau},
\end{equation}

\noindent the overdamped Debye-dipole conductivity
\begin{equation}
\label{eq128}
\sigma_{\rm D}(\omega) = -i \varepsilon_0 \, (\varepsilon(0) - \varepsilon_{\infty} ) \, \frac{\omega}{1 - i \omega \tau},
\end{equation}

\noindent or the damped harmonic-oscillator conductivity 
\begin{equation}
\label{eq129}
\sigma_h(\omega) = -i \, \varepsilon_0 \, \omega_{p0}^2  \, \frac{\omega}{\omega_0^2 -\omega^2 -i \gamma_0 \omega}.
\end{equation}

\noindent  In what follows we first construct an equation of motion for the density $\rho_i(t)$.  We then solve that equation for each of the specific conductivities in Eqs.~(\ref{eq127}) -- (\ref{eq129}).

Our equation of motion comes directly from the continuity equation for the charge density $\rho_i$,
\begin{equation}
\label{eq131}
\frac{\partial \rho_i }{\partial t} + \nabla \cdot {\rm \bf j}_i = 0.
\end{equation}

\noindent  Starting here, we shall (i) relate {\bf j}$_i$ to the electric field {\bf E} and (ii) use Gauss' law to relate {\bf E} to $\rho_i$, which will then produce our desired equation for $\rho_i$.

\subsection{Local Response}

Before dealing with the specifics that arise from the particular conductivities listed above, we discuss a rather  generic model that has been presented in many textbooks on electrodynamics \cite{Saslow1971,Ashby1975}.  Although this model is only correct in one limit (free carriers in a poor conductor), its pervasiveness warrants a brief diversion from a more appropriate description. 

Let's first consider Gauss' law, as this is the simpler of the two relationships required to transform the continuity equation into an equation of motion for $\rho_i$.  With our division of charge densities we have only the high-frequency-excitation bound charge associated with the polarization {\bf P}.  As we are describing the response of this charge with the constant $\varepsilon_{\infty}$, the displacement is given by {\bf D} $= \varepsilon_0 \varepsilon_{\infty}${\bf E}, which leads to Gauss' law [Eq.~(\ref{eq31})] in the form 
\begin{equation}
\label{eq132}
\varepsilon_0 \, \varepsilon_{\infty} \nabla \cdot {\rm {\bf E}} = \rho_i.
\end{equation}

\noindent A tacit assumption underlies the simple equation {\bf D} $= \varepsilon_0 \varepsilon_{\infty}${\bf E}:  the polarization of the charge described by $\varepsilon_{\infty}$ is instantaneous with respect to the electric field.  As all charge carriers have inertia, {\bf D} $= \varepsilon_0 \varepsilon_{\infty}${\bf E} cannot be exact.  However, in what follows we assume that the resonant frequencies of the excitations that give rise to $\varepsilon_{\infty}$ are high enough that negligible error is introduced.  

We now consider {\bf j}$_i$ and {\bf E}.  Let's assume for the moment that  these two quantities are also related by a simple constant,
\begin{equation}
\label{eq133}
{\rm \bf j}_i = \sigma_i(0) \, {\rm \bf E}.
\end{equation} 

\noindent That is, let's assume that Ohm's law is valid.  If an instantaneous relationship is (at least approximately) valid, then the proportionality constant must be the zero-frequency conductivity $\sigma_i(0)$ (as indicated), as the relationship must hold in the limit of a dc electric field.  As we shall see below, this instantaneous relationship between {\bf j}$_i$ and {\bf E} must usually be abandoned.  Nonetheless, let's see where it leads.  

If we now use Eq.~(\ref{eq133}) to eliminate {\bf j}$_i$ in Eq.~(\ref{eq131}) and then use Eq.~(\ref{eq132}) to eliminate $\nabla \cdot {\bf E}$ we obtain
\begin{equation}
\label{eq134}
\frac{d \rho_i }{d t} = - \frac{\sigma_i(0)}{\varepsilon_0 \, \varepsilon_{\infty}} \rho_i,
\end{equation}

\noindent a simple first-order equation for $\rho_i$.  This has the solution
\begin{equation}
\label{eq135}
\rho_f(t) = \rho_f(0) e^{-t/\tau_R},
\end{equation}

\noindent where the relaxation time is given by
\begin{equation}
\label{eq136}
\tau_R = \frac{\varepsilon_0 \,\varepsilon_{\infty}}{\sigma_i(0)}.
\end{equation}

There are several problems with this result.  The first arises because for Debye dipoles or optic-phonon modes $\sigma_i(0) = 0$ [See Eqs.~(\ref{eq128}) and (\ref{eq129})].  We are thus forced to conclude that it would take infinitely long for charge fluctuations associated with these two responses to dissipate.  As we shall see below, a nonzero equilibrium solution is possible for these two cases, but charge-density fluctuations will at least decay to some extent in these systems.  There are also problems with Eq.~(\ref{eq136}) with regard to free carriers.  Using $\sigma_f(0) = \varepsilon_0 \omega_p^2 \tau$, this last equation becomes
\begin{equation}
\label{eq137}
\frac{\tau_R}{\tau} = \frac{1}{\omega_L^2 \tau^2}.
\end{equation}

\noindent  where $\omega_L^2 = \omega_p^2 / \varepsilon_{\infty}$.  For a poor conductor ($\omega_L \tau \ll 1$) we end up with $\tau_R \gg \tau$, a perfectly reasonable result which (as we show below) is also valid.  However, for a good conductor ($\omega_L \tau \gg 1$) this last result suggests that the charge density relaxes on a time scale much faster than the fundamental relaxation time $\tau$ of the charge carriers.  This is perfect nonsense.

\subsection{Nonlocal Response}

To more accurately describe the decay of a fluctuation in charge density $\rho_i$ we must recognize that on some (typically very short) time scale the electric field influences the motion of the charge carriers into the future.  As we shall see, this time scale is set by the underlying relaxation time in the system.  Consequently, the response function that connects the current density to the electric field is not local in time.\footnote{You might also wonder whether or not the response function should also be nonlocal in space.  Indeed, there are times when this is also the case.  One example is the anomalous skin effect in metals.}  A linear relationship between {\bf j}$_i$ and {\bf E} that is nonlocal in time can be mathematically expressed as
\begin{equation}
\label{eq138}
{\rm \bf j}_i({\rm \bf r},t) = \int_{-\infty}^{\infty} \breve{\sigma}_i(t-t') \, {\rm \bf E}({\rm \bf r},t') \, dt'.
\end{equation}

\noindent Causality dictates that $\breve{\sigma}_i(t) = 0$ for $t < 0$, and so when convenient we may alternatively write  
\begin{equation}
\label{eq139}
{\rm \bf j}_i({\rm \bf r},t) = \int_{-\infty}^{t} \breve{\sigma}_i(t-t') \, {\rm \bf E}({\rm \bf r},t') \, dt'.
\end{equation}

As we now show, the the \textbf{conductivity-operator kernal} $\breve{\sigma}_i(t)$ and the frequency-dependent conductivity $\sigma_i(\omega)$ are a Fourier-transform pair.  To see this we assume the current density and electric field in Eq.~(\ref{eq138}) to oscillate harmonically at some frequency $\omega$.  Then it is not hard to show that Eq.~(\ref{eq138}) becomes \marginpar{\footnotesize{$\mathbb{EX} \,$\ref{E24}}}
\begin{equation}
\label{eq140}
{\rm \bf \tilde j}_i ({\rm \bf r}) = \int_{-\infty}^{\infty} \breve{\sigma}_i(t) e^{i \omega t} \,  dt \, \, {\rm \bf \tilde E}({\rm \bf r}).
\end{equation}

\noindent  Comparing this equation with Eq.~(\ref{eq115}) (where $i = f$), we see 
\begin{equation}
\label{eq141}
\sigma_i (\omega) = \int_{-\infty}^{\infty} \breve{\sigma}_i(t) \, e^{i \omega t} \,  dt,
\end{equation}

\noindent which implies the inverse Fourier relation
\begin{equation}
\label{eq142}
\breve{\sigma}_i (t) = \frac{1}{2 \pi} \int_{-\infty}^{\infty} \sigma_i(\omega) \, e^{- i \omega t} \,  dt.
\end{equation}

To derive the equation of motion for $\rho_i$ we proceed as before, except this time we use Eq.~(\ref{eq139}) rather than Eq.~(\ref{eq133}).  This yields an integro-differential equation for $\rho_i$,
\begin{equation}
\label{eq143}
\frac{d \rho_i(t) }{d t} = - \frac{1}{\varepsilon_0 \varepsilon_{\infty}} \int_{-\infty}^{t} \breve{\sigma}_i(t-t') \, \rho_i(t') \, dt'.
\end{equation}

Equilibrium solutions to Eq.~(\ref{eq143}) lends some insight.  Assuming a constant solution $\rho_{eq}$, it is straightforward to reduce this equation to
\begin{equation}
\label{eq144}
0 = - \frac{1}{\varepsilon_0 \varepsilon_{\infty}} \sigma_i(0) \rho_{eq}.
\end{equation}         

\noindent If $\rho_{eq}$ describes free carriers, then $\sigma_i(0) \ne 0$.  Thus, $\rho_{eq}$ must be zero, as expected.  On the other hand, if $\rho_{eq}$ describes bound charge (such as that associated with Debye dipoles or optic-phonon modes), then $\sigma_i(0) = 0$ and a non-zero constant $\rho_{eq}$ is a valid solution.

We now imagine the following scenario.  Up until $t=0$ the charge density $\rho_i$ is zero.  At $t=0$ some external disturbance causes the charge density to take on some nonzero value $\rho_i(0)$.  Our goal to find out how this fluctuation dissipates.  With these conditions we can simplify Eq.~(\ref{eq143}) to
\begin{equation}
\label{eq145}
\frac{d \rho_i(t) }{d t} = - \frac{1}{\varepsilon_0 \varepsilon_{\infty}} \int_{0}^{t} \breve{\sigma}_i(t-t') \, \rho_i(t') \, dt'.
\end{equation}

\noindent Notice that setting $t=0$ in this expression implies $d\rho_i / dt (0) =0$, as long as the kernal $\breve{\sigma}_i(t)$ is a true function (rather than a distribution).  In the case where it has any distribution component (which it does for Debye dipoles and in the limit of free carriers in a very poor conductor), $d\rho_i / dt (0)$ is related to $\rho_i(0)$.  In either case, $d\rho_i / dt (0)$ is completely specified.

To gain further insight into Eq.~(\ref{eq145}) we take its Laplace transform ($s$ is the transform variable), which gives us \marginpar{\footnotesize{$\mathbb{EX} \,$\ref{E25}$\,$\ref{E23a}}}
\begin{equation}
\label{eq146}
\bar{\rho}_i(s) = \frac{\rho_i(0)}{s + \frac{1}{\varepsilon_0  \varepsilon_{\infty}} \, \bar{\sigma}_i(s)}.
\end{equation}

\noindent First, this expression explicitly demonstrates that the only required initial condition for the problem is $\rho_i(0)$.  Furthermore, Because
\begin{equation}
\label{eq147}
\bar{\sigma}_i(s) =  \int_0^{\infty} \breve{\sigma}_i(t) e^{-s t} dt,
\end{equation}

\noindent and $\breve{\sigma}_i(t) = 0$ for $t <0$, we have $\bar{\sigma}_i(s) = \sigma_i(i s)$ [see Eq.~(\ref{eq141})].  Hence, if one has an expression for $\sigma_i(\omega)$, then one need not calculate $\breve{\sigma}_i(t)$ when using the Laplace transform to find $\rho_i(t)$.

\subsection{Free Carriers}

We now solve Eq.~(\ref{eq145}) assuming the free carriers are described by the Drude ac conductivity $\sigma_f(\omega)$ given by Eq.~(\ref{eq127}).  Taking the Fourier transform of this particular $\sigma_i(\omega)$ yields the kernal \marginpar{\footnotesize{$\mathbb{EX} \,$\ref{E26},$\,$\ref{E27}}}
\begin{equation}
\label{eq148}
\breve{\sigma}_f(t) = \varepsilon_0 \, \omega_p^2 \, e^{-t/\tau} \, \Theta(t),
\end{equation}

\noindent where $\Theta(t)$ is the \textbf{Heaviside function}.  Satisfyingly, the Heaviside function ensures causality.  We now substitute this expression for $\breve{\sigma}_i(t)$ into Eq.~(\ref{eq145}), which gives us 
\begin{equation}
\label{eq149}
\frac{d \rho_f(t) }{d t} = - \omega_L^2 \int_{0}^{t} e^{-(t-t')/\tau} \, \rho_f(t') \, dt'.
\end{equation}

\noindent  The simplicity of the exponentially decaying kernal allows us to easily turn Eq.~(\ref{eq149}) into a familiar differential equation.  Taking a second derivative in time and then using Eq.~(\ref{eq149}) to eliminate the integral from the differentiated equation yields \marginpar{\footnotesize{$\mathbb{EX} \,$}\ref{E28}}
\begin{equation}
\label{eq150}
\frac{d^2 \! \rho_f }{d t^2} + \frac{1}{\tau} \frac{d \rho_f}{d t}+ \omega_L^2 \rho_f  = 0,
\end{equation}

\noindent the harmonic-oscillator equation of motion!  Notice that the damping is governed by the (momentum) relaxation time $\tau$ of the free carriers, while the natural frequency of oscillation is the longitudinal frequency $\omega_L$.    

\subsubsection{Good Conductors}

A good conductor is defined by $\omega_L \tau \gg 1$; hence, the solution to Eq.~(\ref{eq150}) is well into the underdamped regime ($\omega_L \tau > 1/2$) of a damped harmonic oscillator.  This results in the charge-density decay being most conveniently described by 
\begin{equation}
\label{eq151}
\rho_f(t) = e^{-t/(2\tau)} \, \Big( A_1 \, e^{i \sqrt{\omega_L^2 - (1/2\tau)^2 } \, t} + A_2 \, e^{-i \sqrt{\omega_L^2 - (1/2\tau)^2 } \, t} \Big),
\end{equation}

\noindent  where the constants $A_1$ and $A_2$ are determined by the initial conditions.  Because $\omega_L \tau \gg 1$, the square root in the exponents of Eq.~(\ref{eq151}) can simply be replaced by $\omega_L$, and so to excellent approximation  
\begin{equation}
\label{eq152}
\rho_f(t) = e^{-t/(2\tau)} \, \big( A_1 \, e^{i \omega_L t} + A_2 \, e^{-i \omega_L t} \big).
\end{equation}

\noindent From this result we see that any charge-density fluctuations decay with a characteristic time $2 \tau$ accompanied by oscillations at the frequency $\omega_L$.

The specific solution to our problem is obtained by imposing the initial condition of a given $\rho_f(0)$ and using  $d\rho / dt (0) = 0$ (discussed above).  With these conditions the solution can be written as \marginpar{\footnotesize{$\mathbb{EX} \,$}\ref{E29}}
\begin{equation}
\label{eq153}
\rho_f(t) = \rho_f(0) \, e^{-t/(2\tau)} \, \left[ {\rm cos}(\omega_L t) + \frac{1}{2 \omega_L \tau} {\rm sin}(\omega_L t) \right].
\end{equation}

\noindent  Because $\omega_L \tau \gg 1$, the cosine term dominates the sine term in this expression.

Let's apply this result to the ideal good conductor of Fig.~\ref{fig5}, for which $\omega_L \tau = 100.$\footnote{It would be more desirable to use a real metal (such as Pb or Au) as an example here.  However, the typical overlap of low lying interband transitions with what would be $\omega_L$ in most metals makes the situation significantly more complicated.}  Because the period of oscillation $T$ equals $2 \pi / \omega_L$, we have $2 \tau / T$ = 31.8, and so we have $\sim$32 oscillations within a decay time of the fluctuation.

\subsubsection{Poor Conductors}

In contrast, the charge-density decay for a poor conductor ($\omega_L \tau \ll 1$) is well into the overdamped regime.  In this case it is convenient to write the solution as
\begin{equation}
\label{eq154}
\rho_f(t) = e^{-t/(2\tau)} \, \Big( A_1 \, e^{\sqrt{ (1/2\tau)^2 - \omega_L^2 } \, t} + A_2 \, e^{-\sqrt{ (1/2\tau)^2 - \omega_L^2 } \, t } \Big).
\end{equation} 

\noindent  Because $\omega_L \tau \ll 1$, without much loss in accuracy this last expression simplifies to 
\begin{equation}
\label{eq155}
\rho_f(t) = A_1 \, e^{-t/\tau_R} + A_2 \, e^{-t/\tau},
\end{equation} 

\noindent where $\tau_R = 1 / (\omega_L^2 \tau)$.  Notice that this is the same relaxation time previously obtained with the instantaneous-response model discussed above [see Eq.~(\ref{eq137})].  Because $\omega_L \tau \ll 1$, $\tau_R \gg \tau$, and so the first term in Eq.~(\ref{eq155}) decays much more slowly than the second term.    

Imposing the initial conditions we obtain the solution \marginpar{\footnotesize{$\mathbb{EX} \,$}\ref{E30}}
\begin{equation}
\label{eq156}
\rho_f(t) = \rho_f(0) \frac{1}{1 - \tau / \tau_R}  \bigg( e^{-t/\tau_R} - \frac{\tau}{\tau_R} \, e^{-t/\tau} \bigg),
\end{equation} 

\noindent  This result demonstrates that the second (short lived) term has a much smaller amplitude than the first (long lived) term.  Indeed, if the ratio $\tau / \tau_R$ is neglected, we recover the local-response result, Eq.~(\ref{eq135}).  Notice for both poor and good conductors that $\rho_f$ approaches zero at long times.  

We can also obtain the local-response result via a slightly different route.  We first note $\breve{\sigma}(t)$ can be expressed as 
\begin{equation}
\label{eq157}
\breve{\sigma}_f(t) = \sigma_f(0)  \,   \frac{1}{\tau} \, e^{-t/\tau} \, \Theta(t).
\end{equation}

\noindent Because for a poor metal the relevant timescale $\tau_R$ is much longer than $1 / \tau$, then to good approximation
\begin{equation}
\label{eq158}
\breve{\sigma}_f(t) = \sigma_f(0) \, \delta(t), 
\end{equation}

\noindent where $\delta(t)$ is the \textbf{Dirac delta function}.  With this kernal Eq.~(\ref{eq145}) readily simplifies to Eq.~(\ref{eq134}), the local-response equation of motion, thus validating the Ohm's law approximation {\bf j}$_f = \sigma_f(0) ${\bf E} for a poor conductor.

It is instructive to consider this limit in the frequency domain.  First, we note that Ohm's law corresponds to $\sigma (\omega)$ being independent of frequency.  Second, the Drude ac conductivity $\sigma_f(\omega)$ [Eq.~(\ref{eq128})] is (essentially) constant for frequencies up to $\sim 1 / \tau$.  Therefore, if the local-model calculated timescale for the response of the system is longer than $\tau$, then the conductivity is indeed constant at all relevant frequencies, and the Ohm's law approximation is well founded.  That is, Ohm's law is valid as long as $\tau_R / \tau \gtrsim 1$.  As $\tau_R = 1 / (\omega_L^2 \tau)$ this is equivalent to $\omega_L \tau \lesssim 1$.  By definition, poor conductors easily meet this condition, while good conductors do not. 

As an example of a poor conductor, we consider $n$-type GaAs with a free-carrier density of 10$^{14}$ cm$^{-3}$.  For this material $\sigma_f(0) = 9.5$  $\Omega^{-1}$m$^{-1}$ and  $\varepsilon_{\infty} = 13.1$ \cite{Sze1981}, from which we obtain $\omega_L \tau = 0.14$ and $\tau = 2.3 \times 10^{-13}$ s.  These values imply $\tau_R / \tau = 54$ and $\tau_R = 1.2 \times 10^{-11}$ s.

\subsection{Debye Response}

To find the time dependence of the charge-density relaxation for a set of Debye dipoles, we proceed as above for free carriers.  We first calculate the Fourier transform of Eq.~(\ref{eq128}), which yields \marginpar{\footnotesize{$\mathbb{EX} \,$\ref{E31},$\,$\ref{E32}}}
\begin{equation}
\label{eq159}
\breve{\sigma}_{\rm D}(t) = \frac{\varepsilon_0 \, (\varepsilon(0) - \varepsilon_{\infty})}{\tau}  \, \bigg[ \delta(t) - \frac{1}{\tau} e^{-t/\tau} \, \Theta(t) \bigg].
\end{equation}

\noindent Interestingly, ${\breve \sigma}_{\rm D}(t)$ comprises both local and nonlocal terms.  Using this expression for ${\breve \sigma}_{\rm D}(t)$  $[= {\breve \sigma}_i(t)]$ in Eq.~(\ref{eq145}) gives us
\begin{equation}
\label{eq160}
\frac{d \rho_{\rm D}(t) }{d t} = -\frac{\varepsilon(0) - \varepsilon_{\infty}}{\varepsilon_{\infty}} \, \frac{1}{\tau} \, \bigg[ \rho_{\rm D}(t) - \int_{0}^{t} \frac{1}{\tau} e^{-(t-t')/\tau} \, \rho_{\rm D}(t') \, dt'  \bigg].
\end{equation}

\noindent Taking the second time derivative and eliminating the integral in the differentiated equation using Eq.~(\ref{eq160}) yields the simple differential equation\marginpar{\footnotesize{$\mathbb{EX} \,$}\ref{E32b}}
\begin{equation}
\label{eq161}
\frac{d^2 \! \rho_{\rm D} }{d t^2} =  - \frac{\varepsilon(0)}{\varepsilon_{\infty} \tau} \frac{d \rho_{\rm D}}{dt}.
\end{equation}

\noindent This equation can easily be integrated twice.  Doing so, and using the  $t=0$ relation
\begin{equation}
\label{eq162}
\frac{d \rho_{\rm D}(0) }{d t} = -\frac{\varepsilon(0) - \varepsilon_{\infty}}{\varepsilon_{\infty}} \, \frac{1}{\tau}  \, \rho_{\rm D}(0)
\end{equation}

\noindent [obtainable from Eq.~(\ref{eq160})] gives us \marginpar{\footnotesize{$\mathbb{EX} \,$}\ref{E33}}
\begin{equation}
\label{eq163}
\rho_{\rm D}(t)  = \rho_{\rm D}(0) \bigg[ \frac{\tau_R}{\tau}  + \bigg(   1 - \frac{\tau_R}{\tau} \bigg) e^{-t / \tau_R} \bigg],
\end{equation}

\noindent where the relaxation time $\tau_R$ is given by
\begin{equation}
\label{eq164}
\tau_R = \frac{\varepsilon_{\infty}}{\varepsilon(0)} \, \tau.
\end{equation} 

A few remarks are in order.  First, because $\varepsilon_{\infty} < \varepsilon(0)$, $\tau_R$ is smaller than the parameter $\tau$.  However, as $\tau$ is not a fundamental microscopic relaxation time [see the discussion just after Eq.~(\ref{eq92})], this result is not problematic in a manner similar to that for the local-model result applied to a good conductor.  Second, for $t \rightarrow \infty$ the charge density does not vanish, but rather decays to $[\varepsilon_{\infty} / \varepsilon(0)] \, \rho_D(0)$. Admittedly, this exact value is a curious result, but it is consistent with our previous observation that for a system with zero dc conductivity a constant nonzero charge density is a solution to Eq.~(\ref{eq143}).

\subsection{Optic Phonons}

For our third example of charge-density relaxation, we look at the response associated with optic phonons.  As in our previous two cases, we start with the Fourier transform of the appropriate frequency dependent conductivity [$\sigma_h(\omega)$, see Eq.~(\ref{eq129})], which is \marginpar{\footnotesize{$\mathbb{EX} \,$\ref{E34},$\,$\ref{E35}}}
\begin{equation}
\label{eq166}
\breve{\sigma}_h(t) = \varepsilon_0 \omega_{p0}^2 \, e^{-\gamma_0 t /2} \bigg[ {\rm cos}(\omega_1 t) - \frac{\gamma_0}{2 \omega_1} {\rm sin}(\omega_1 t) \bigg] \Theta(t).
\end{equation} 

\noindent Here $\omega_1 = \sqrt{\omega_0^2 - \gamma_0^2 / 4}$.\footnote{We remind the reader that optic phonon response is generally underdamped, so that $\omega_1$ is real.  If the damping is small enough, then $\omega_1 \approx \omega_0$, the phonon mode's natural frequency of oscillation.}  Substituting this expression for $\breve{\sigma}_h(t)$ into the general equation for the charge-density time dependence [Eq.~(\ref{eq145})] we obtain  
\begin{equation}
\label{eq167}
\frac{d \rho_h(t) }{d t} = - \frac{\omega_{p0}^2}{\varepsilon_{\infty}} \int_{0}^{t} e^{-\gamma_0 (t-t') /2} \bigg[ {\rm cos}(\omega_1 (t-t')) - \frac{\gamma_0}{2 \omega_1} {\rm sin}(\omega_1 (t-t')) \bigg] \, \rho_h(t') \, dt'.
\end{equation}

\noindent  Although its derivation is a bit more complicated than in the previous two examples, a differential equation equivalent to Eq.~(\ref{eq167}) can be obtained, this time by taking two time derivatives (and utilizing a bit of algebra), which results in\marginpar{\footnotesize{$\mathbb{EX} \,$}\ref{E37}}
\begin{equation}
\label{eq168}
\frac{d^3 \! \rho_h}{dt^3} + \gamma_0 \frac{d^2 \! \rho_h}{dt^2} + \omega_L^2 \frac{d\rho_h}{dt} = 0,
\end{equation}

\noindent yet another harmonic-oscillator equation of motion!  On this occasion it describes the motion of $d\rho_h / dt$.  Notice that, as in the case of free-charge-density relaxation in a conductor, the natural oscillation frequency is the longitudinal frequency $\omega_L = \sqrt{\omega_{p0}^2 / \varepsilon_{\infty} + \omega_0^2}$ [see Eq.~(\ref{eq75bb}].\footnote{For free carriers $\omega_0 = 0$.} Integrating Eq.~(\ref{eq168}) and using the initial conditions $d\rho_h / dt(0) =0$ and  $d^2 \! \rho_h / dt^2(0) =  - (\omega_p^2 / \varepsilon_{\infty} )\rho_h(0)$ [obtainable from Eq.~(\ref{eq167})] results in\marginpar{\footnotesize{$\mathbb{EX} \,$}\ref{E38}}
\begin{equation}
\label{eq169}
\rho_h(t) = \rho_h(0) \bigg\{ \frac{\omega_0^2}{\omega_L^2} + \bigg[ 1 - \frac{\omega_0^2}{\omega_L^2} \bigg] e^{-\gamma_0 t /2} \bigg[   {\rm cos}( \bar{\omega}_1 t) + \frac{\gamma_0}{2 \bar{\omega}_1} \, {\rm sin}( \bar{\omega}_1 t) \bigg] \bigg\},
\end{equation}

\noindent where $\bar{\omega}_1 = \sqrt{\omega_L^2 - \gamma_0^2 / 4}$.  Because in most cases we are interested in the significantly underdamped case, $\bar{\omega}_1 \approx \omega_L$.

As in the case of $\rho_{\rm D}(t)$ for Debye response  [Eq.~(\ref{eq163})], the solution for $\rho_h(t)$ does not decay to zero.  Instead, as evident from Eq.~(\ref{eq169}), the charge density decays to $(\omega_0 / \omega_L)^2 \rho_h(0)$.  Because $\omega_L / \omega_0 = \sqrt{\varepsilon(0) / \varepsilon_{\infty}}$ (see $\mathbb{EX}$~\ref{E7}), the long-time solution in both cases can be expressed as $[\varepsilon_{\infty} / \varepsilon(0) ] \rho_i(0)$, where $i$ = D or $h$, as appropriate.

\newpage

\section{Exercises}

\begin{enumerate}

\subsection{Basics of Maxwell's Equations}

\item Show that charge conservation [Eq.~(\ref{eq5})] is implied by the basic field equations for \textbf{E} and \textbf{B}.  \label{E1}

\item Show that Eqs.~(\ref{eq31}) - (\ref{eq34}) imply that the charge associated with $\rho_{other}$ is conserved.  \label{E1b}

\item Derive the wave equation for the magnetic field \textbf{B} that is analogous to Eq.~(\ref{eq44}) for the electric field \textbf{E}.  \label{E2}

\item Consider the plane wave given by Eq.~(\ref{eq47}).  Show that \textbf{E}$_0$ is orthogonal to the wave vector \textbf{k}.  \label{E3}

\item Assume that the electric field in a solid with simple linear response is described by the plane wave  ${\rm {\bf E}}( {\rm {\bf r}},t ) = {\rm {\bf E}}_0 \,e^{i\left( {{\rm {\bf k}} \cdot {\rm {\bf r}} - \omega \,t} \right)}$ [Eq.~(\ref{eq47})].  Now assume that the magnetic field the has the form ${\rm {\bf H}}( {\rm {\bf r}},t ) = {\rm {\bf H}}_0 \,e^{i\left( {{\rm {\bf k'}} \cdot {\rm {\bf r}} - \omega' \,t} \right)}$.  Show that Maxwell's equations imply (i) $\textbf{k}' = \textbf{k}$ and $\omega' = \omega$ [giving the field as expressed in Eq.~(\ref{eqZ1})] and (ii) the \textbf{H}-field amplitude is related to the \textbf{E}-field amplitude via Eqs.~(\ref{eqZ2}) and (\ref{eqZ3}).

\label{E4}

\item  Show that Eqs.~(\ref{eqM1}) -- (\ref{eqM4}) follow from the assumptions of (i) $\rho_{other} =0$ and \textbf{j}$_{other} =0$ and (ii) harmonic fields. \label{E4b}

\subsection{Model Dielectric Functions / Dispersion Relations}

\subsubsection{Harmonic Oscillator}

\item Starting with  the generic dispersion relation [Eq.~(\ref{eq70b})] and the dielectric function $\varepsilon(\omega)$ given by Eq.~(\ref{eq67}), derive the specific dispersion relation given by Eq.~(\ref{eq72a}).  \label{E5}

\item Beginning with Eq.~(\ref{eq75}), show that $\omega_L / \omega_0 = \sqrt{\varepsilon(0) / \varepsilon_{\infty}}$.  This is known as the \textbf{Lyndane-Sachs-Teller} (LST) relation.  \label{E7}

\item Show that the dielectric function in the vicinity of an (undamped) optic phonon [Eq.~(\ref{eq75})] can be written in the form
\begin{equation}
\label{eq1000}
\varepsilon ( \omega ) =  \varepsilon_{\infty} \, \frac{\omega_L^2 - \omega^2}{\omega_0^2 - \omega^2},
\end{equation}

This form clearly shows that $\omega_L$ is a zero and $\omega_0$ a pole of $\varepsilon(\omega)$.  \label{E8}

\item Show that the dielectric function given by Eq.~(\ref{eq75}) leads to a dispersion relation equivalent to that given by Eq.~(\ref{eq72a}), but with $(ck)^2 \to (ck)^2 / \varepsilon_{\infty}$ and $\omega_p^2 \to \omega_p^2 / \varepsilon_{\infty}$.  (Hint:  you need not derive the dispersion relation.)  \label{E6}

\item For the multiple-oscillator model [Eq.~(\ref{eq73})] the phase velocity $v_p = \omega / k$ is greater than $c$ at high frequencies.  Show that in this same frequency region the group velocity $v_g = d\omega / dk$ is less than $c$.  \label{E9}

\item Starting with the damped harmonic oscillator equation of motion for the dipole moment \textbf{p} [Eq.~(\ref{eq75c})], derive Eq.~(\ref{eq76}) for the polarization ${\tilde{ \bf P}}$ and Eq.~(\ref{eq77}) for the dielectric function $\varepsilon(\omega)$.  \label{E9b}

\item The dispersion curves for harmonic-oscillator response [Fig.~\ref{fig4}(b)] and a good conductor [Fig.~\ref{fig5}(d)] show for $\omega$ very close to $\omega_L$ that Re$(k) =$ Im$(k)$.  Find the conditions under which this occurs.  \label{E10}

\subsubsection{Drude Free Carriers}

\item Starting with the equation of motion for the polarization associated with free carriers [Eq.~(\ref{eq80b})], derive Eq.~(\ref{eq81}) for the dielectric function $\varepsilon(\omega)$.  \label{E10b}

\item  Show that the two expressions for the Drude dielectric function, Eqs.~(\ref{eq81}) and (\ref{eq82}), are equivalent.  \label{E10c}

\item  In the appropriate limits derive the simplified results for $\varepsilon(\omega)$ and $ck$ shown in Figs.~\ref{figDE} and \ref{figDD}.  \label{E11}

\item  The (complex) index of refraction $N$ is related to the dielectric function via $N = \sqrt{\varepsilon(\omega)}$.  Starting with the Drude dielectric function [Eq.~(\ref{eq82})], find approximate expressions for the real and imaginary part of $N$ in the regions illustrated in Fig.~\ref{figDD}.  \label{E12}

\item  The (complex) index of refraction $N$ is related to $k$ and $\omega$ via $ N = ck/\omega$.  Starting with this and the expressions in Fig.~\ref{figDD}, again find $\hat n$ in the regions illustrated in Fig.~\ref{figDD}, thus verifying your answers to the previous exercise.  \label{E13}

\subsubsection{Debye Response}

\item Starting with the equation of motion for the polarization associated with Debye dipoles [Eq.~(\ref{eq90})], derive Eq.~(\ref{eq92}) for the dielectric function $\varepsilon(\omega)$.  \label{E18}

\item In the appropriate limits, derive the approximate expressions [Eqs.~(\ref{eq93}) -- (\ref{eq96})] for the dielectric function and dispersion relation for Debye dipoles.  \label{E19}

\subsection{Conductivity}

\item  Starting with the Drude dielectric function [Eq.~(\ref{eqDrude})], derive all of the approximate expressions for the (total) conductivity found in Fig.~\ref{fig13}. \label{E20}

\item The Drude free-carrier conductivity $\sigma_f(\omega)$ can be obtained by starting with the following (averaged) equation of motion for a free carrier,
\begin{equation}
\label{eq1001}
\frac{d{\bf v}}{dt} + \frac{1}{\tau} {\bf v} = \frac{q}{m^*} \tilde{\bf E} e^{-i \omega t}.
\end{equation}

\noindent  (a) Clearly identify each symbol and the origin of each term in this equation.

\noindent  (b) Assuming the velocity \textbf{v} oscillates harmonically at at the same frequency $\omega$ as the electric field \textbf{E}, show that 
\begin{equation}
\label{eq1002}
\tilde{\bf v} = \frac{q \, \tau}{m^*} \frac{1}{1 - i \omega \tau} \tilde{\bf E}
\end{equation}

\noindent (c)  Given that the current density is related to the carrier velocity via $\tilde{\bf j} = N_c \, q \, \tilde{\bf v}$, shown that this leads to the Drude conductivity $\sigma_f(\omega)$ given by Eq.~(\ref{eq113}).  \label{E21}

\item  Show that the harmonic Maxwell's equations in the form of Eqs.~(\ref{eq116}) -- (\ref{eq119}) follow from the definitions given by Eqs.~(\ref{eq114}) and (\ref{eq115}).  \label{E22}

\item  Given Eq.~(\ref{eq126}), derive the conductivities in Eqs.~(\ref{eq127}) -- (\ref{eq129}) starting with the appropriate dielectric functions. \label{E23}

\subsection{Charge Relaxation}

\item  Show that Eq.~(\ref{eq140}) follows from Eq.~(\ref{eq138}) under the assumption of harmonic time dependence to the fields.  \label{E24}

\item  Laplace transform Eq.~(\ref{eq145}), and thence verify Eq.~(\ref{eq146}).  \label{E25}

\item Charge density decay $\rho_i(t)$ via Laplace transform.

In the text each integro-differential equation for $\rho_i(t)$ [$i = f$ (free carriers), D (Debye response), and $h$ (optic phonons)] is transformed into a familiar differential equation, and the solution for $\rho_i(t)$ is then determined.  Alternatively, $\rho_i(t)$ can be found using the Laplace transform $\bar{\rho}_i(s)$, the general form of which is given by Eq.~(\ref{eq146}).  In this exercise we find $\rho_i(t)$ for all three models using this approach.

(a)  For all three models first find $\bar{\sigma}_i(s)$.  This can be done quite simply starting with the appropriate expressions for $\sigma_i(\omega)$, given in Eqs.~(\ref{eq127}) -- (\ref{eq129}).

(b)  Next use each $\bar{\sigma}_i(s)$ to find appropriate expressions for $\bar{\rho}_i(s)$ for all three models.

(c)  Lastly, inverse transform each $\bar{\rho}_i(s)$ to find the corresponding $\rho_i(t)$, as expressed in Eqs.~(\ref{eq151}) and (\ref{eq154}) (good and poor conductors), Eq.~(\ref{eq163}) (Debye response), and Eq.~(\ref{eq169}) (optic phonons).  (Hint:  in each case $\bar{\rho}_i(s)$ can be written as a sum of terms of the form $ b / (s + a)$, which has the inverse transform $b \, e^{-at}$.)

\label{E23a}

\subsubsection{Free Carriers}

\item Starting with the Drude free-carrier conductivity $\sigma_f(\omega)$ [Eq.~(\ref{eq127})], calculate its inverse Fourier transform and thence obtain $\breve{\sigma}_f(t)$ as given by Eq.~(\ref{eq148}).  (Hint:  while there are various way to calculate the Fourier transform, perhaps the most straightforward is to continue the integral into the lower or upper complex plane (depending upon whether $t$ is positive or negative), and then use the residue theorem.) \label{E26}

\item  Time dependence of the free-carrier current density \textbf{j}$_f(t)$.

\noindent(a) If the Drude free-carrier conductivity kernal [Eq.~(\ref{eq148})] is used in the (nonlocal) expression for the current density \textbf{j}$_i(t)$ [Eq.~(\ref{eq139})], then we end up with
\begin{equation}
\label{eq1003}
{\rm \bf j}_f(t) =   \int_{-\infty}^{t}  \varepsilon_0\omega_p^2 \, e^{-(t-t') / \tau}  \, {\bf E}(t')  \, dt',
\end{equation}

\noindent  where we have suppressed the position variable \textbf{r}.  Show that this is equivalent to the first-order differential equation
\begin{equation}
\label{eq1004}
{\rm \bf j}_f + \tau \frac{d{\bf j}_f}{dt} =  \sigma_f(0) {\bf E}
\end{equation}

\noindent for ${\bf j}_f$.  Under what condition does this last equation reduce to Ohm's law?

\noindent (b) We now assume that (somehow) a constant electric field ${\bf E}_0$ can be turned on infinitely fast.  That is, we let ${\bf E}(t) = {\bf E}_0 \, \Theta(t)$.  Using this form for the electric field, calculate the integral on the right side of Eq.~(\ref{eq1003}) and show that the current density can be expressed as
\begin{equation}
\label{eq1005}
{\bf j}_f(t) = \sigma_f(0) \, {\bf E}_0 \, \big( 1 - e^{-t / \tau} \big) \, \Theta(t).
\end{equation} 

\noindent  Notice that under such idealized circumstances the relaxation time $\tau$ sets the time scale for establishing the stead-state current density $\sigma_f(0) \, {\bf E}_0$.

\noindent (c)  Show that Eq.~(\ref{eq1005}) is a solution to Eq.~(\ref{eq1004}) when ${\bf E}(t) = {\bf E}_0 \, \Theta(t)$. Pay particular attention to the point $t=0$.  

\label{E27}

\item Show how the integro-differential equation for the free-carrier charge density $\rho_f(t)$ [Eq.~(\ref{eq149})] leads to the harmonic-oscillator equation of motion for $\rho_f(t)$ [Eq.~(\ref{eq150})].  \label{E28}

\item Starting with the general solution [Eq.~(\ref{eq151})] to the harmonic-oscillator equation for the free-carrier charge density $\rho_f(t)$ [Eq.~(\ref{eq150})], derive the specific solution for $\rho_f(t)$ for good conductors given by   [Eq.~(\ref{eq153})].  \label{E29}

\item Starting with the general solution [Eq.~(\ref{eq154})] to the harmonic-oscillator equation for the free-carrier charge density $\rho_f(t)$ [Eq.~(\ref{eq150})], derive the specific solution for $\rho_f(t)$ for poor conductors given by   [Eq.~(\ref{eq156})].  \label{E30}

\subsubsection{Debye Response}

\item Starting with the Debye-dipole conductivity $\sigma_{\rm D}(\omega)$ [Eq.~(\ref{eq128})], calculate its inverse Fourier transform and thence obtain $\breve{\sigma}_{\rm D}(t)$ as given by Eq.~(\ref{eq159}).  (Hint 1:  see hint associated with $\mathbb{EX} \,$\ref{E26}.  Hint 2:  the Dirac Delta function is lurking about; make its presence explicit.) \label{E31}

\item Time dependence of the Debye-dipole current density ${\bf j}_{\rm D} (t)$.  

\noindent (a)  If the Debye conductivity kernal [Eq.~(\ref{eq159})] is used in the (nonlocal) expression for the current density [Eq.~(\ref{eq138})], then we end up with
\begin{equation}
\label{eq1007}
{\rm \bf j}_{\rm D}(t) =  \mathbb{C} \int_{-\infty}^{\infty}  \bigg[ \delta(t-t') - \frac{1}{\tau} \, e^{-(t-t') / \tau}  \, \Theta(t-t') \bigg] \, {\bf E}(t')  \, dt',
\end{equation}

\noindent  where we have suppressed the position variable \textbf{r}.  Identify the constant $\mathbb{C}$.

\noindent (b) Show that Eq.~(\ref{eq1007}) is equivalent to the first-order differential equation
\begin{equation}
\label{eq1008}
\frac{d{\bf j}_{\rm D}}{dt} + \frac{1}{\tau} \, {\rm \bf j}_{\rm D} = \mathbb{C} \, \frac{d{\bf E}}{dt}
\end{equation}

\noindent for ${\bf j}_{\rm D}$. 

(c) Assume that a constant electric field ${\bf E}_0$ can be turned on infinitely fast.  That is, assume ${\bf E}(t) = {\bf E}_0 \, \Theta(t)$.  (In contrast to the free-carrier response considered in $\mathbb{EX} \,$\ref{E27}, here this approximation is not so far fetched, as the response time of Debye dipoles is quite slow in some systems.)  Using this electric field calculate the integral on the right side of Eq.~(\ref{eq1007}) and show that the current density can be expressed as
\begin{equation}
\label{eq1009}
{\bf j}_{\rm D}(t) = \mathbb{C} \, {\bf E}_0 \, e^{-t / \tau} \Theta(t).
\end{equation} 

\noindent  It this current density continuous? When is ${\bf j}_{\rm D}(t)$ a maximum?  What is the steady state value of ${\bf j}_{\rm D}(t)$?

\noindent (d) Show that Eq.~(\ref{eq1009}) is a solution to Eq.~(\ref{eq1008}) when ${\bf E}(t) = {\bf E}_0 \, \Theta(t)$. Pay particular attention to the point $t=0$.  

\label{E32}

\item Show how the integro-differential equation for the Debye-response charge density $\rho_{\rm D}(t)$ [Eq.~(\ref{eq160})] leads to the differential equation of motion for $\rho_{\rm D}(t)$ [Eq.~(\ref{eq161})].  \label{E32b}

\item  Integrate Eq.~(\ref{eq161}) for the Debye-dipole density two times and then, using the initial condition expressed in  Eq.~(\ref{eq162}), obtain the solution for $\rho_{\rm D}(t)$ given by  Eq.~(\ref{eq163}).  \label{E33}

\subsubsection{Optic Phonons}

\item Starting with the optic-phonon conductivity $\sigma_h(\omega)$ [Eq.~(\ref{eq129})], calculate its inverse Fourier transform and thence obtain $\breve{\sigma}_h(t)$ as given by Eq.~(\ref{eq166}).  (Hint:  see hint associated with $\mathbb{EX} \,$\ref{E26}.) \label{E34}

\item Time dependence of the optic-phonon current density ${\bf j}_h(t)$.  

\noindent (a)  If the optic-phonon conductivity kernal [Eq.~(\ref{eq166})] is used in the (nonlocal) expression for the current density [Eq.~(\ref{eq139})], then we end up with
\begin{equation}
\label{eq1010}
{\rm \bf j}_h(t) =  \varepsilon_0 \omega_{p0}^2 \int_{-\infty}^{t} \! \! \! e^{-\gamma_0(t-t') / 2}  \bigg[ {\rm cos}(\omega_1 (t-t')) - \frac{\gamma_0}{2 \omega_1} {\rm sin}(\omega_1 (t-t')) \bigg] {\bf E}(t')  dt',
\end{equation}

\noindent  where we have suppressed the position variable \textbf{r}.

\noindent (b) Show that Eq.~(\ref{eq1010}) is equivalent to the first-order differential equation
\begin{equation}
\label{eq1011}
\frac{d^2{\bf j}_h}{dt^2}  +  \gamma_0 \, \frac{d{\bf j}_h}{dt} + \omega_0^2 \,\, {\rm \bf j}_{\rm D} = \varepsilon_0 \omega_{p0}^2 \, \frac{d{\bf E}}{dt}
\end{equation}

\noindent for ${\bf j}_h$. 

(c) Assume that a constant electric field ${\bf E}_0$ can be turned on infinitely fast.  That is, assume ${\bf E}(t) = {\bf E}_0 \, \Theta(t)$.  Using this electric field calculate the integral on the right side of Eq.~(\ref{eq1010}) and show that the current density can be expressed as
\begin{equation}
\label{eq1012}
{\bf j}_h(t) = \frac{\varepsilon_0 \omega_{p0}^2}{\omega_1} \, {\bf E}_0 \, e^{-\gamma_0 t / 2} \, {\rm sin}(\omega_1 t) \, \Theta(t).
\end{equation} 

\noindent  Is this current density continuous?  What is the steady state value of ${\bf j}_h(t)$?

\noindent (d) Show that Eq.~(\ref{eq1012}) is a solution to Eq.~(\ref{eq1011}) when ${\bf E}(t) = {\bf E}_0 \, \Theta(t)$. Pay particular attention to the point $t=0$.  

\label{E35}

\item Show how the integro-differential equation for the optic-phonon charge density $\rho_h(t)$ [Eq.~(\ref{eq167})] leads to the harmonic-oscillator equation of motion for $d\rho_h /dt$ [Eq.~(\ref{eq168})].  \label{E37}

\item  Find the general solution to Eq.~(\ref{eq168}) for the optic-phonon density and then, using the initial conditions given just prior to  Eq.~(\ref{eq169}), obtain the solution for $\rho_h(t)$ given by  Eq.~(\ref{eq169}).  \label{E38}

\end{enumerate}

\bibliography{QDs} 

\begin{thebibliography}{16}
\expandafter\ifx\csname natexlab\endcsname\relax\def\natexlab#1{#1}\fi
\expandafter\ifx\csname bibnamefont\endcsname\relax
  \def\bibnamefont#1{#1}\fi
\expandafter\ifx\csname bibfnamefont\endcsname\relax
  \def\bibfnamefont#1{#1}\fi
\expandafter\ifx\csname citenamefont\endcsname\relax
  \def\citenamefont#1{#1}\fi
\expandafter\ifx\csname url\endcsname\relax
  \def\url#1{\texttt{#1}}\fi
\expandafter\ifx\csname urlprefix\endcsname\relax\def\urlprefix{URL }\fi
\providecommand{\bibinfo}[2]{#2}
\providecommand{\eprint}[2][]{\url{#2}}

\bibitem[{\citenamefont{Ashby}(1975)}]{Ashby1975}
\bibinfo{author}{\bibnamefont{Ashby}, \bibfnamefont{N.}}, \bibinfo{year}{1975},
  \bibinfo{journal}{American Journal of Physics}
  \textbf{\bibinfo{volume}{43}}(\bibinfo{number}{6}), \bibinfo{pages}{553},
  \urlprefix\url{http://scitation.aip.org/content/aapt/journal/ajp/43/6/10.111%
9/1.9787}.

\bibitem[{\citenamefont{Ashcroft and Mermin}(1976)}]{Ashcroft1976}
\bibinfo{author}{\bibnamefont{Ashcroft}, \bibfnamefont{N.~W.}}, and
  \bibinfo{author}{\bibfnamefont{N.~D.} \bibnamefont{Mermin}},
  \bibinfo{year}{1976}, \emph{\bibinfo{title}{Solid State Physics}}
  (\bibinfo{publisher}{New York: Harcort Brace College Publishers}).

\bibitem[{\citenamefont{Br\"andli and Sievers}(1972)}]{Brandli1972}
\bibinfo{author}{\bibnamefont{Br\"andli}, \bibfnamefont{G.}}, and
  \bibinfo{author}{\bibfnamefont{A.~J.} \bibnamefont{Sievers}},
  \bibinfo{year}{1972}, \bibinfo{journal}{Phys. Rev. B}
  \textbf{\bibinfo{volume}{5}}, \bibinfo{pages}{3550},
  \urlprefix\url{http://link.aps.org/doi/10.1103/PhysRevB.5.3550}.

\bibitem[{\citenamefont{Cole and Cole}(1941)}]{Cole1941}
\bibinfo{author}{\bibnamefont{Cole}, \bibfnamefont{K.~S.}}, and
  \bibinfo{author}{\bibfnamefont{R.~H.} \bibnamefont{Cole}},
  \bibinfo{year}{1941}, \bibinfo{journal}{The Journal of Chemical Physics}
  \textbf{\bibinfo{volume}{9}}(\bibinfo{number}{4}), \bibinfo{pages}{341},
  \urlprefix\url{http://scitation.aip.org/content/aip/journal/jcp/9/4/10.1063/%
1.1750906}.

\bibitem[{\citenamefont{Debye}(1929)}]{Debye1929}
\bibinfo{author}{\bibnamefont{Debye}, \bibfnamefont{P.}}, \bibinfo{year}{1929},
  \emph{\bibinfo{title}{Polar Molecules}} (\bibinfo{publisher}{New York:
  Chemical Catalogue Company}).

\bibitem[{\citenamefont{Eldridge and Staal}(1977)}]{Eldridge1977}
\bibinfo{author}{\bibnamefont{Eldridge}, \bibfnamefont{J.~E.}}, and
  \bibinfo{author}{\bibfnamefont{P.~R.} \bibnamefont{Staal}},
  \bibinfo{year}{1977}, \bibinfo{journal}{Phys. Rev. B}
  \textbf{\bibinfo{volume}{16}}, \bibinfo{pages}{4608},
  \urlprefix\url{http://link.aps.org/doi/10.1103/PhysRevB.16.4608}.

\bibitem[{\citenamefont{Emsley}(1993)}]{Emsley1993}
\bibinfo{author}{\bibnamefont{Emsley}, \bibfnamefont{J.}},
  \bibinfo{year}{1993}, \emph{\bibinfo{title}{The Elements}}
  (\bibinfo{publisher}{Oxford: Clarendon Press}).

\bibitem[{\citenamefont{Golavashkin and Motulevich}(1968)}]{Golavashkin1968}
\bibinfo{author}{\bibnamefont{Golavashkin}, \bibfnamefont{A.~I.}}, and
  \bibinfo{author}{\bibfnamefont{G.~P.} \bibnamefont{Motulevich}},
  \bibinfo{year}{1968}, \bibinfo{journal}{Soviet Physics JETP}
  \textbf{\bibinfo{volume}{26}}(\bibinfo{number}{5}), \bibinfo{pages}{881},
  \urlprefix\url{http://www.jetp.ac.ru/cgi-bin/e/index/e/26/5/p881?a=list}.

\bibitem[{\citenamefont{Griffiths}(2013)}]{Griffiths2013}
\bibinfo{author}{\bibnamefont{Griffiths}, \bibfnamefont{D.~J.}},
  \bibinfo{year}{2013}, \emph{\bibinfo{title}{Introduction to Electrodynamics}}
  (\bibinfo{publisher}{New York: Pearson}).

\bibitem[{\citenamefont{Jonscher}(1980)}]{Jonscher1980}
\bibinfo{author}{\bibnamefont{Jonscher}, \bibfnamefont{A.~K.}},
  \bibinfo{year}{1980}, \bibinfo{journal}{Journal of Physics D: Applied
  Physics} \textbf{\bibinfo{volume}{13}}(\bibinfo{number}{5}),
  \bibinfo{pages}{L89},
  \urlprefix\url{http://stacks.iop.org/0022-3727/13/i=5/a=005}.

\bibitem[{\citenamefont{Kalmykov} \emph{et~al.}(2004)\citenamefont{Kalmykov,
  Coffey, Crothers, and Titov}}]{Kalmykov2004}
\bibinfo{author}{\bibnamefont{Kalmykov}, \bibfnamefont{Y.~P.}},
  \bibinfo{author}{\bibfnamefont{W.~T.} \bibnamefont{Coffey}},
  \bibinfo{author}{\bibfnamefont{D.~S.~F.} \bibnamefont{Crothers}}, and
  \bibinfo{author}{\bibfnamefont{S.~V.} \bibnamefont{Titov}},
  \bibinfo{year}{2004}, \bibinfo{journal}{Phys. Rev. E}
  \textbf{\bibinfo{volume}{70}}, \bibinfo{pages}{041103},
  \urlprefix\url{http://link.aps.org/doi/10.1103/PhysRevE.70.041103}.

\bibitem[{\citenamefont{Olmon} \emph{et~al.}(2012)\citenamefont{Olmon, Slovick,
  Johnson, Shelton, Oh, Boreman, and Raschke}}]{Olmon2012}
\bibinfo{author}{\bibnamefont{Olmon}, \bibfnamefont{R.~L.}},
  \bibinfo{author}{\bibfnamefont{B.}~\bibnamefont{Slovick}},
  \bibinfo{author}{\bibfnamefont{T.~W.} \bibnamefont{Johnson}},
  \bibinfo{author}{\bibfnamefont{D.}~\bibnamefont{Shelton}},
  \bibinfo{author}{\bibfnamefont{S.-H.} \bibnamefont{Oh}},
  \bibinfo{author}{\bibfnamefont{G.~D.} \bibnamefont{Boreman}}, and
  \bibinfo{author}{\bibfnamefont{M.~B.} \bibnamefont{Raschke}},
  \bibinfo{year}{2012}, \bibinfo{journal}{Phys. Rev. B}
  \textbf{\bibinfo{volume}{86}}, \bibinfo{pages}{235147},
  \urlprefix\url{http://link.aps.org/doi/10.1103/PhysRevB.86.235147}.

\bibitem[{\citenamefont{Palik}(1985)}]{Palik1985}
\bibinfo{author}{\bibnamefont{Palik}, \bibfnamefont{E.~D.}},
  \bibinfo{year}{1985}, \emph{\bibinfo{title}{Handbook of Optical Constants of
  Solids}} (\bibinfo{publisher}{San Diego: Academic Press}).

\bibitem[{\citenamefont{Raki\'c} \emph{et~al.}(1998)\citenamefont{Raki\'c,
  Djuri\v{s}i\'c, Elazar, and Majewski}}]{Rakic1998}
\bibinfo{author}{\bibnamefont{Raki\'c}, \bibfnamefont{A.~D.}},
  \bibinfo{author}{\bibfnamefont{A.~B.} \bibnamefont{Djuri\v{s}i\'c}},
  \bibinfo{author}{\bibfnamefont{J.~M.} \bibnamefont{Elazar}}, and
  \bibinfo{author}{\bibfnamefont{M.~L.} \bibnamefont{Majewski}},
  \bibinfo{year}{1998}, \bibinfo{journal}{Appl. Opt.}
  \textbf{\bibinfo{volume}{37}}(\bibinfo{number}{22}), \bibinfo{pages}{5271},
  \urlprefix\url{http://ao.osa.org/abstract.cfm?URI=ao-37-22-5271}.

\bibitem[{\citenamefont{Saslow and Wilkinson}(1971)}]{Saslow1971}
\bibinfo{author}{\bibnamefont{Saslow}, \bibfnamefont{W.~M.}}, and
  \bibinfo{author}{\bibfnamefont{G.}~\bibnamefont{Wilkinson}},
  \bibinfo{year}{1971}, \bibinfo{journal}{American Journal of Physics}
  \textbf{\bibinfo{volume}{39}}(\bibinfo{number}{10}), \bibinfo{pages}{1244},
  \urlprefix\url{http://scitation.aip.org/content/aapt/journal/ajp/39/10/10.11%
19/1.1976613}.

\bibitem[{\citenamefont{Sze}(1981)}]{Sze1981}
\bibinfo{author}{\bibnamefont{Sze}, \bibfnamefont{S.~M.}},
  \bibinfo{year}{1981}, \emph{\bibinfo{title}{Physics of Semiconductor
  Devices}} (\bibinfo{publisher}{New York: Wiley}).

\end{thebibliography}

\end{document}